\documentclass[aps,nofootinbib,preprintnumbers]{revtex4}
\usepackage{lipsum}
\usepackage{CJK}
\usepackage{amsfonts}
\usepackage{amsmath}
\usepackage{amssymb}
\usepackage[english]{babel}
\usepackage{graphicx}
\usepackage{epsfig}
\usepackage{bm}
\usepackage{verbatim}
\usepackage[utf8]{inputenc}
\usepackage{booktabs}
\usepackage{multirow}
\usepackage{subfig}
\usepackage{slashed}
\usepackage{xcolor}
\usepackage[colorlinks=true,urlcolor=red,citecolor=red]{hyperref}
\usepackage[font=small]{caption}
\usepackage{float}
\usepackage{blindtext}
\usepackage{placeins}
\usepackage{csquotes}
\usepackage[utf8]{inputenc}
\usepackage{bbm}
\usepackage{cancel}
\usepackage[colorinlistoftodos]{todonotes}
\usepackage{ulem}

\newenvironment{Eqnarray}{\arraycolsep 0.14em\begin{eqnarray}}{\end{eqnarray}}
\newcommand{\ba}{\begin{Eqnarray}}
\newcommand{\ea}{\end{Eqnarray}}
\newcommand{\be}{\begin{equation}}
\newcommand{\ee}{\end{equation}}
\newcommand{\bal}{\begin{aligned}}
\newcommand{\eal}{\end{aligned}}
\newcommand{\bea}{\begin{eqnarray}}
\newcommand{\eea}{\end{eqnarray}}
\newcommand{\ben}{\begin{enumerate}}
\newcommand{\een}{\end{enumerate}}
\newcommand{\bit}{\begin{itemize}}
\newcommand{\eit}{\end{itemize}}
\newcommand{\bde}{\begin{widetext}}
\newcommand{\ede}{\end{widetext}}
\newcommand{\nn}{\nonumber}

\renewcommand{\[}{\left[}

\def\nn{\nonumber}
\def\lsim{\mathrel{\rlap{\lower4pt\hbox{\hskip1pt$\sim$}}
    \raise1pt\hbox{$<$}}}
\def\gsim{\mathrel{\rlap{\lower4pt\hbox{\hskip1pt$\sim$}}
    \raise1pt\hbox{$>$}}}

\def\3211{$\mathrm{SU(3) \otimes SU(2)_L \otimes U(1)_R \otimes U(1)_{B-L}}$ }
\def\321{$\mathrm{SU(3) \otimes SU(2) \otimes U(1)}$ }
\def\422{$\mathrm{SU(4) \otimes SU(2) \otimes SU(2)_R}$ }

\newcommand{\mathsym}[1]{{}}

\definecolor{bostonuniversityred}{rgb}{0.8, 0.0, 0.0}

\input{tcilatex}
\begin{document}

\title{Dark matter and scalar sector in a novel two-loop scotogenic neutrino mass model}
\author{A. E. C\'{a}rcamo Hern\'{a}ndez}
\email{antonio.carcamo@usm.cl}
\affiliation{Universidad T\'ecnica Federico Santa Mar\'{\i}a, Casilla 110-V, Valpara\'{\i}%
so, Chile}
\affiliation{Centro Cient\'{\i}fico-Tecnol\'ogico de Valpara\'{\i}so, Casilla 110-V,
Valpara\'{\i}so, Chile}
\affiliation{Millennium Institute for Subatomic physics at high energy frontier - SAPHIR, Fernandez Concha 700, Santiago, Chile}
\author{Catalina Espinoza}
\email{m.catalina@fisica.unam.mx}
\affiliation{Cátedras SECIHTI,  Instituto de Física, Universidad Nacional Autónoma de México, Ciudad de México, C.P.~04510, Mexico}
\author{Juan Carlos G\'{o}mez-Izquierdo}
\affiliation{Instituto Polit\'ecnico Nacional, CECyT No.16, Carretera Pachuca-Actopan Kil\'ometro 1+500, Distrito de Educaci\'on, Salud, Ciencia, Tecnolog\'ia e
Innovaci\'on, San Agust\'in Tlaxiaca, Hidalgo, C.P. 42162, M\'exico.}
\affiliation{ Instituto Polit\'ecnico Nacional, UPIIH, Carretera Pachuca-Actopan Kil\'ometro 1+500, Distrito de Educaci\'on, Salud, Ciencia, Tecnolog\'ia e
Innovaci\'on, San Agust\'in Tlaxiaca, Hidalgo, C.P. 42162, M\'exico.}
\email{cizquierdo@ipn.mx}
\author{Juan Marchant Gonz\'{a}lez}
\email{juan.marchant@upla.cl}
\affiliation{Laboratorio de C\'omputo de F\'isica (LCF-UPLA), Facultad de Ciencias Naturales y Exactas, Universidad de Playa Ancha, Subida Leopoldo Carvallo 270, Valpara\'iso, Chile.}
\author{Myriam Mondrag\'on}
\email{myriam@fisica.unam.mx}
\affiliation{Instituto de Física, Universidad Nacional Autónoma de México, Ciudad de México, C.P.~04510, Mexico}
\date{\today }
   
\begin{abstract}
We propose an extended $3+1$ Higgs doublet model where the Standard Model (SM) gauge structure is enhanced by the discrete symmetry $Q_6 \times Z_2 \times Z_4$, and the fermion content is extended with right-handed Majorana neutrinos. The scalar sector, besides four $SU(2)$ doublets, incorporates multiple gauge-singlet scalars. In our model, the tiny active neutrino masses arise from a novel radiative seesaw mechanism at two-loop level and the leptonic mixing features the cobimaximal mixing pattern compatible with neutrino oscillation experimental data. Along with this,  the proposed model is consistent with SM quark masses and mixings as well as with the constraints arising from dark matter relic density and dark matter direct detection. Our analysis reveals that the best-fit point satisfying dark matter constraints yields a non-SM scalar with mass near $95$ GeV, which could be a possible candidate for the observed $95$ GeV diphoton excess. We further obtain other non SM scalars with masses at the subTeV scale which are within the LHC reach, while successfully complying with the experimental bounds arising from collider searches.
\end{abstract}

\maketitle

\section{Introduction}
Although the Standard Model (SM) has achieved remarkable success as a theory of strong and electroweak interactions, with its predictions experimentally verified to very high degree of accuracy, it still faces several unresolved issues. Some of these include, for instance, the smallness of neutrino masses, the hierarchy of SM charged fermion masses and mixing angles and the measured amount of dark matter in the Universe. These unresolved issues strongly motivate the development of extensions to the SM that incorporate an enlarged particle spectrum and extended symmetries. Among these, theories involving discrete flavor symmetries have acquired significant interest within the particle physics community. The spontaneous breaking of such symmetries can generate predictive and viable fermion mass matrix textures, which are essential for explaining the observed patterns of SM fermion masses and mixing angles. Comprehensive reviews of discrete flavor groups can be found in various works, including \cite{King:2013eh,Altarelli:2010gt,Ishimori:2010au,Grimus:2011fk,Fonseca:2014lfa, Chauhan:2022gkz,King:2015aea}. In particular, discrete flavor groups with a small number of doublets and singlets in their irreducible representations, such as for example $S_{3}$ \cite{Gerard:1982mm, Kubo:2003iw, Kubo:2003pd, Kobayashi:2003fh, Chen:2004rr, Mondragon:2007af, Mondragon:2008gm, Ma:2013zca, Kajiyama:2013sza, CarcamoHernandez:2015hjp, Emmanuel-Costa:2016vej, Arbelaez:2016mhg, CarcamoHernandez:2016pdu, Gomez-Izquierdo:2017rxi, Cruz:2017add, Ma:2017trv, Espinoza:2018itz, Garces:2018nar, CarcamoHernandez:2018vdj, CarcamoHernandez:2020pxw, Kuncinas:2020wrn, Khater:2021wcx,Kuncinas:2022whn,Kuncinas:2023ycz,Babu:2023oih,CarcamoHernandez:2024ycd}, $D_{4}$ \cite{Frampton:1994rk, Grimus:2003kq, Grimus:2004rj, Frigerio:2004jg, Blum:2007jz, Adulpravitchai:2008yp, Ishimori:2008gp, Hagedorn:2010mq, Meloni:2011cc, Vien:2013zra, Vien:2014ica, Vien:2014soa, CarcamoHernandez:2020ney, Vien:2020uzf, Vien:2021diw, Bonilla:2020hct},  $Q_{4}$~\cite{Aranda:2011dx,Lovrekovic:2012bz,Vien:2019lso},  $Q_{6}$~\cite{Frampton:1994rk,Babu:2004tn,Kajiyama:2005rk,Kajiyama:2007pr,Kifune:2007fj,Babu:2009nn,Kawashima:2009jv,Kaburaki:2010xc,Babu:2011mv,Araki:2011zg,Gomez-Izquierdo:2013uaa,Gomez-Izquierdo:2017med,Vien:2023zid} have been incorporated in extensions of the SM, as they offer an economical and straightforward approach for obtaining viable fermion mass matrix textures. This, in turn, allows for a successful explanation of the observed SM fermion masses and mixing patterns. 
In order to explain the tiny values of the active neutrino masses, very heavy right-handed Majorana neutrinos, singlets under the SM gauge symmetry, which mix with the active neutrinos, are added to the fermion spectrum of the SM, then allowing the implementation of the tree level type I seesaw mechanism. However, such mechanism despite being the most economical explanation for the smallness of the active neutrino masses, does not allow to successfully accommodate the current amount of dark matter relic density observed in the Universe and yields tiny rates for charged lepton flavor violating decays, too many orders of magnitude below the current experimental sensitivity, then making very limited the testability of theories having tree level type I seesaw mechanism. This motivates radiative seesaw models where a preserved discrete symmetry prevents the generation of tree level active neutrino masses and make them appearing at least at one-loop level. Theories based on radiative seesaw mechanisms allows to relate dark matter with active neutrino masses since the lightest of the electrically neutral seesaw messengers plays a crucial role in the generation of the observed dark matter relic abundance. In such theories the stability of the dark matter candidate is guaranteed by a preserved discrete symmetry which ensures the radiative nature of the seesaw mechanism responsible for producing tiny masses for active neutrinos. The most economical radiative seesaw models are the ones where active neutrino masses are produced at one-loop level; in such models to yield tiny values for active neutrino masses one has to rely either on very small neutrino Yukawa couplings or on unnaturally small value for the mass difference between the CP even and CP odd components of the electrically neutral scalar messengers. Theories where active neutrino masses arise at two-loop level yield a more natural explanation for the tiny neutrino masses than those where they arise at one loop level.
On the other hand, the cobimaximal pattern~\cite{Fukuura:1999ze,Miura:2000sx,Ma:2002ce,Grimus:2003yn,Chen:2014wxa,Ma:2015fpa,Joshipura:2015dsa,Li:2015rtz,He:2015xha,Chen:2015siy,Ma:2016nkf,Damanik:2017jar,Ma:2017trv,Grimus:2017itg,CarcamoHernandez:2017owh,CarcamoHernandez:2018hst,Ma:2019iwj,Hernandez:2021kju,Rivera-Agudelo:2022qpa,CarcamoHernandez:2024ycd, Rivera-Agudelo:2024vdn} for leptonic mixing provides a compelling explanation for the observed neutrino oscillation data. In the basis where the SM charged lepton mass matrix is diagonal, this pattern corresponds to a specific form of the neutrino mass matrix given by: 
\begin{equation}
\widetilde{M}_{\nu}=\left( 
\begin{array}{ccc}
A & C & C^{*} \\ 
C & B & D \\ 
C^{*} & D & B^{*}
\end{array}%
\right),
\label{X}
\end{equation}
It predicts a non-zero $\theta_{13}\neq 0$, $\theta_{23}=\frac{\pi}{4}$ and $\delta_{CP}=-\frac{\pi}{2}$, which is close to the current experimental results. The term `cobimaximal' reflects the fact that this pattern yields both maximal 2-3 mixing and a maximally CP-violating phase. Additionally, it arises from a generalized $\mu-\tau$ symmetry \cite{Babu:2002dz,Grimus:2003yn,King:2014nza,Xing:2022uax}
\begin{equation}
P^T\widetilde{M}_{\nu}P=\left(\widetilde{M}_{\nu}\right)^{*} 
\end{equation} 
with 
\begin{equation}
P=\left( 
\begin{array}{ccc}
1 & 0 & 0 \\ 
0 & 0 & 1 \\ 
0 & 1 & 0
\end{array}%
\right).
\label{X}
\end{equation}
To derive the cobimaximal leptonic mixing pattern, non-Abelian discrete groups with irreducible triplet representations such as $A_4$ \cite{Ma:2017moj,Ma:2021kfa} and $\Delta(27)$ \cite{Ma:2019iwj,CarcamoHernandez:2017owh,CarcamoHernandez:2018hst,CarcamoHernandez:2020udg} have been employed in extensions of the SM. Besides, discrete groups having doublets as irreducible representations such as $S_3$ \cite{Ma:2017trv,Gomez-Izquierdo:2023mph,Gomez-Izquierdo:2024apr,CarcamoHernandez:2024ycd} have also been used to derive the cobimaximal mixing pattern. In this work we demonstrate that the $Q_6$ flavor symmetry can successfully reproduce the cobimaximal leptonic mixing pattern within the framework of a two-loop level radiative seesaw neutrino mass model. To the best of our knowledge our model corresponds to the first implementation of the cobimaximal leptonic mixing pattern within the framework of a $Q_6$ discrete flavor group. 

Our model is based on the  $Q_{6}$ family group, which is supplemented by a $Z_2\times Z_4$ symmetry. The $Q_{6}$ and $Z_4$ symmetries are spontaneously broken, whereas the $Z_2$ symmetry is preserved. We assume that the spontaneous breaking of the $Z_4$ symmetry gives rise to a preserved $\widetilde{Z}_{2}$ symmetry, which  allows for three dark matter candidates. The model  has an extended $3+1$ Higgs doublet sector  featuring the cobimaximal mixing pattern for the lepton mixing. Then, the tiny active neutrino masses are radiatively generated at two-loop level, thanks to the preserved $Z_2$ and $\widetilde{Z}_{2}$ discrete symmetries, which guarantee the stability of the dark matter candidates as well as the radiative nature of the two-loop seesaw mechanism. The successful implementation of both the two-loop level radiative seesaw mechanism that generates the tiny active neutrino masses and the leptonic cobimaximal mixing pattern requires the inclusion of several scalar singlets, some of them acquiring complex vacuum expectation values (VEVs), those giving rise to geometrical CP violation arising from the spontaneous breaking of the discrete symmetries. Despite the large number of scalars, the effective number of parameters at low energies is greatly reduced due to the $Q_6$ flavor symmetry, rendering the model predictive.

The content of this paper is as follows. In section \ref{model} we explain the proposed model specifying its symmetry and particle content. The implications of the model in quark masses and mixing are described in section \ref{quarmassesandmixings}. In section \ref{leptonmassesandmixings} we discuss the consequences of the model in lepton masses and mixing. The low energy as well as the whole scalar potentials are analyzed in section \ref{scalarpotential}, considering two specific benchmark scenarios for the low energy case. We also provide a discussion about the quasialignment limit. The dark scalar sector and the consequences of the model for Dark Matter are analyzed in section \ref{DM}. We state our conclusions in section \ref{conclusions}.     

\section{The model}
\label{model}
We propose a novel two-loop level radiative seesaw mechanism to generate active neutrino masses, where the leptonic mixing is governed by the cobimaximal pattern. To this end, we consider an extended 4HDM theory where the SM gauge symmetry is enlarged by the inclusion of the $Q_{6}$ family symmetry~\cite{Babu:2004tn,Kajiyama:2005rk,Kajiyama:2007pr,Kifune:2007fj,Babu:2009nn,Kawashima:2009jv,Kaburaki:2010xc,Babu:2011mv,Araki:2011zg,Gomez-Izquierdo:2013uaa,Gomez-Izquierdo:2017med,Vien:2023zid} and the $Z_2\times Z_4$ discrete group. The SM particle content of the model under consideration is augmented by the inclusion of right-handed Majorana neutrinos and several electrically neutral gauge singlet scalars. We use the $Q_6$ flavor group as it has several doublet and singlet irreducible representations and allows the implementation of the cobimaximal mixing pattern with less amount of symmetries and fields than the non abelian $S_3$ discrete group. In our model, $Q_{6}$ is completely broken, $Z_2$ is preserved and the $Z_4$ symmetry is spontaneously broken down to a remnant conserved $\widetilde{Z}_2$ symmetry. The full symmetry of the model experiences the following spontaneous symmetry breaking scheme:
\begin{gather}
\mathcal{G}=SU(3)_{C}\times SU\left( 2\right) _{L}\times U\left( 1\right)_{Y}\times Q_6\times Z_{4}\times Z_{2}{\xrightarrow{v_{\sigma},v_{\rho},v_{\xi}}}
\notag \\
SU(3)_{C}\times SU\left( 2\right) _{L}\times U\left( 1\right) _{Y}\times
\widetilde{Z}_{2}\times Z_{2}{\xrightarrow{v_1,v_2,v_3}} \notag \\
SU(3)_{C}\times U\left( 1\right) _{Q}\times\widetilde{Z}_{2}\times Z_{2}.  \label{SB}
\end{gather}%
We assume that the $Z_{4}$ symmetry is spontaneously broken to a preserved
matter parity symmetry $\widetilde{Z}_2$ defined with charges given as $%
(-1)^{Q_{Z_{4}}+2s}$ where $Q_{Z_{4}}$ and $s$ are the $Z_{4}$ charge (in
additive notation) and spin of the particle under consideration,
respectively. The preserved $\widetilde{Z}_{2}\times Z_{2}$ symmetry ensures the radiative nature of the seesaw mechanism at two-loop level that generates the tiny masses of the active neutrinos.

\begin{figure}
\centering
\includegraphics[scale=0.25]{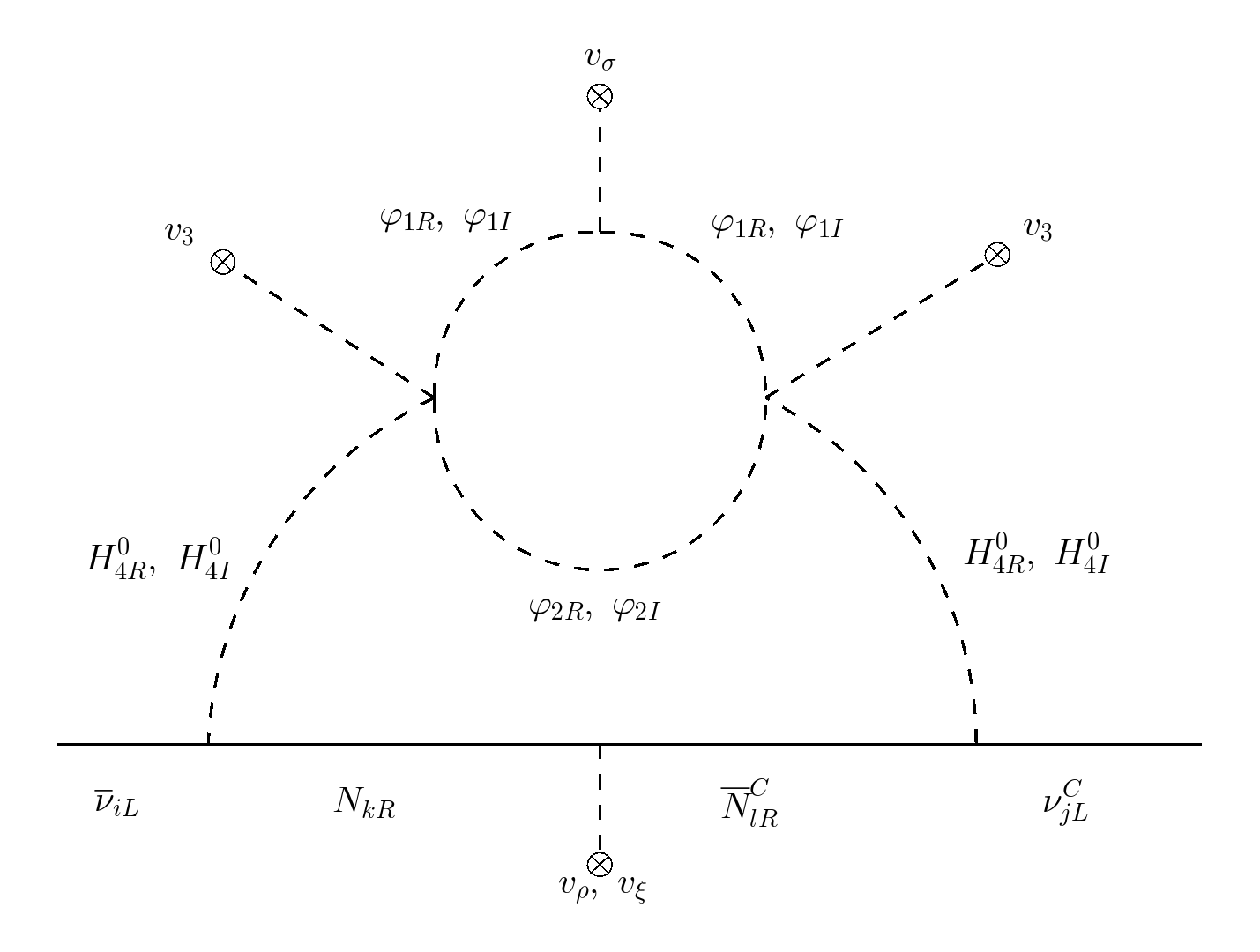}
\caption{two-loop Feynman diagram contribution to neutrino masses.}
\label{fig:neutrino-loop}
\end{figure}

In order to generate tree-level masses for the SM charged fermions and two-loop level masses for light active neutrinos, the scalar sector of our proposed model is composed of three active $SU(2)$ scalar doublets, namely $H_i$ ($i=1,2,3$), one inert $SU(2)$ scalar doublet $H_4$ and six electrically neutral scalar singlets $\varphi_{n}$, $\xi_n$ ($n=1,2$), $\sigma$, $\rho$. Moreover, the implementation of the radiative seesaw mechanism that produces the tiny active neutrino masses requires to extend the fermionic spectrum of the SM by including three right-handed Majorana neutrinos in singlet and doublet representations of the $Q_6$ discrete group, as shown in Table \ref{fermions}, which displays the fermionic particle content with their transformations under the $SU(3)_{C}\times SU\left( 2\right) _{L}\times U\left( 1\right)_{Y}\times Q_6\times Z_{4}\times Z_{2}$ group. It is worth mentioning that the scalar fields $H_n$ and $\xi_n$ ($n=1,2$) are grouped in the $Q_6$ doublets $H=\left(H_1,H_2\right)$, $\xi=\left(\xi_1,\xi_2\right)$, whereas the remaining scalar fields are assigned as $Q_6$ singlets.  The scalar particle content and their assignments under the $SU(3)_{C}\times SU\left( 2\right) _{L}\times U\left( 1\right)_{Y}\times Q_6\times Z_{4}\times Z_{2}$ group are displayed in Table \ref{scalars}. As shown in Table \ref{scalars}, the scalar fields $H_4$ and $\varphi_2$ are charged under the preserved $Z_2$ symmetry, whereas $H_4$ and $\varphi_1$ have $Z_4$ charges transforming non trivially under the remnant $\widetilde{Z}_{2}$ symmetry. Thus, the scalar fields $H_4$, $\varphi_1$ and $\varphi_2$ do not acquire VEV's forbidding tree-and one-loop level masses for active neutrinos, and then allowing these masses to be radiatively generated at two-loop level. These inert scalars together with the right-handed Majorana neutrinos mediate the two-loop level radiative seesaw mechanism that yields the tiny active neutrino masses, as indicated in the Feynman diagram of Fig. \ref{fig:neutrino-loop}. Furthermore, as follows from Table \ref{fermions}, the right-handed Majorana neutrinos are also charged under the preserved $Z_2\times\widetilde{Z}_{2}$ symmetry. Consequently, due to the preserved $Z_2\times \widetilde{Z}_{2}$ symmetry, our model has stable dark matter (DM) candidates, one will be the lightest among the $Z_{2}$ odd fields, the second one will be the lightest among the fields transforming non trivially under $\widetilde{Z}_{2}$ and the third one will correspond to the particle with non trivial $Z_2\times\widetilde{Z}_{2}$ charge and lowest mass. Thus, our model has a multicomponent dark matter which implies that the resulting relic density will be the sum of the relic densities generated by these three DM candidates. A detailed analysis of the consequences of the model for dark matter will be performed in section \ref{DM}. 
\begin{table}
\begin{tabular}{|c|c|c|c|c|c|c|c|c|c|c|c|c|}
\hline
& $q_{L}$ & $q_{3L}$ & $u_{R}$ & $u_{3R}$ & $d_{R}$ & $d_{3R}$ & $l_{1L}$ & $l_{L}$  & $e_{1R}$ & $e_{R}$
& $N_{1R}$ & $N_{R}$ \\ \hline
$SU(3)_{C}$ & $\mathbf{3}$ & $\mathbf{3}$ & $\mathbf{3}$ & $\mathbf{3}$
& $\mathbf{3}$ & $\mathbf{3}$ & $\mathbf{1}$ & $\mathbf{1}$ & 
$\mathbf{1}$ & $\mathbf{1}$ & $\mathbf{1}$ & $\mathbf{1}$ \\ \hline
$SU(2)_{L}$ & $\mathbf{2}$ & $\mathbf{2}$ & $\mathbf{1}$ & $\mathbf{1}$
& $\mathbf{1}$ & $\mathbf{1}$ & $\mathbf{2}$ & $\mathbf{2}$ & 
$\mathbf{1}$ & $\mathbf{1}$ & $\mathbf{1}$ & $\mathbf{1}$ \\ \hline
$U(1)_Y$ & $\frac{1}{6}$ & $\frac{1}{6}$ & $\frac{2}{3}$ & $\frac{2}{3}$ & $-\frac{1}{3}$ & $-\frac{1}{3}$ & $\frac{1}{2}$ & $\frac{1}{2}$ & $-1$ & $-1$ & $0$ & $0$ \\ \hline
$Q_{6}$ & $\mathbf{2}_{2}$ & $\mathbf{1}_{-+}$ & $\mathbf{2}_{2}$ & $\mathbf{1}_{-+}$
& $\mathbf{2}_{2}$ & $\mathbf{1}_{-+}$ & $\mathbf{1}_{-+}$ & $\mathbf{2}_{2}$ & 
$\mathbf{1}_{-+}$ & $\mathbf{2}_{2}$ & $\mathbf{1}_{-+}$ & $\mathbf{2}_{2}$ \\ \hline
$Z_{2}$ & $0$ & $0$ & $0$ & $0$ & $0$ & $0$ & $0$ & $0$ & $0$ & $0$ & $1$ & $1$ \\ \hline
$Z_{4}$ & $0$ & $0$ & $0$ & $0$ & $0$ & $0$ & $0$ & $0$ & $0$ & $0$ & $1$ & $1$ \\\hline
\end{tabular}
\caption{Fermion content with the $SU(3)_C\times SU(2)_L\times U(1)_Y\times Q_{6}\times Z_2\times Z_4$ assignments. }
\label{fermions}
\end{table}
\begin{table}[t]
\begin{tabular}{|c|c|c|c|c|c|c|c|c|}
\hline
& $H$ & $H_{3}$ & $H_{4}$ & $\varphi _{1}$ & $\varphi _{2}$ & $\sigma$ & $\xi$ & $\rho$ \\ \hline
$SU(3)_C$ & $\mathbf{1}$ & $\mathbf{1}$ & $\mathbf{1}$ & $\mathbf{1}$
& $\mathbf{1}$ & $\mathbf{1}$ & $\mathbf{1}$ & $\mathbf{1}$ \\ \hline
$SU(2)_L$ & $\mathbf{2}$ & $\mathbf{2}$ & $\mathbf{2}$ & $\mathbf{1}$
& $\mathbf{1}$ & $\mathbf{1}$ & $\mathbf{1}$ & $\mathbf{1}$ \\ \hline
$U(1)_Y$ & $\frac{1}{2}$ & $\frac{1}{2}$ & $\frac{1}{2}$ & $0$
& $0$ & $0$ & $0$ & $0$ \\ \hline
$Q_{6}$ & $\mathbf{2}_{1}$ & $\mathbf{1}_{++}$ & $\mathbf{1}_{++}$ & $\mathbf{1}_{++}$
& $\mathbf{1}_{++}$ & $\mathbf{1}_{++}$ & $\mathbf{2}_{1}$ & $\mathbf{1}_{--}$ \\ \hline
$Z_{2}$ & $0$ & $0$ & $1$ & $0$ & $1$ & $0$ & $0$ & $0$ \\ \hline
$Z_{4}$ & $0$ & $0$ & $1$ & $1$ & $2$ & $2$ & $2$ & $2$ \\\hline
\end{tabular}
\caption{Scalar content with the $SU(3)_C\times SU(2)_L\times U(1)_Y\times Q_{6}\times Z_2\times Z_4$ assignments. }
\label{scalars}
\end{table}
In order to get a nearly cobimaximal mixing pattern for lepton mixing, we consider the following VEV configuration for the $Q_{6}$ doublet scalar: 
\begin{equation}
\left\langle \xi \right\rangle =v_{\xi }\left( e^{i\theta },e^{-i\theta
}\right),
\label{eq:vev-xi}
\end{equation}
which is shown in Appendix~\ref{ScalarQ6doublet} to be consistent with the scalar potential minimization conditions for a large region of parameter space.

With the above specified particle content and symmetries, the following 
Yukawa terms arise: 
\begin{eqnarray}
\tciLaplace _{Y} &=&y_{1u}\left( \overline{q}_{L}\widetilde{H}%
_{3}u_{R}\right) _{\mathbf{1}_{++}}+y_{2u}\left( \overline{q}_{L}\widetilde{H%
}\right) _{\mathbf{1}_{-+}}u_{3R}+y_{3u}\overline{q}_{3L}\left( \widetilde{H}%
u_{R}\right) _{\mathbf{1}_{-+}}+y_{4u}\overline{q}_{3L}\widetilde{H}%
_{3}u_{3R}\notag \\
&&+y_{1d}\left( \overline{q}_{L}H_{3}d_{R}\right) _{\mathbf{1}%
_{++}}+y_{2d}\left( \overline{q}_{L}H\right) _{\mathbf{1}_{-+}}d_{3R}+y_{3d}%
\overline{q}_{3L}\left( Hd_{R}\right) _{\mathbf{1}_{-+}}+y_{4d}\overline{q}%
_{3L}H_{3}d_{3R}\notag \\
&&+y^{l}_{1} \overline{l}_{1 L}H_{3}e_{1 R}+ y^{l}_{2} \overline{l}_{1 L}\left(He_R\right)_{\mathbf{1}_{-+}}+y^{l}_{3} \left( \overline{l}_{L}H\right) _{\mathbf{1}_{-+}}e_{1 R}+ y^{l}_{4}\left(  \overline{l}_{L}H_{3}e_{R}\right)_{\mathbf{1}_{++}}\notag \\
&&+y_{1\nu }\overline{l}_{1L}\widetilde{H}_{4}N_{1R}+y_{2\nu }\left( 
\overline{l}_{L}\widetilde{H}_{4}N_{R}\right) _{\mathbf{1}_{++}}\notag \\
&&+y_{1N}N_{1R}\sigma \overline{N_{1R}^{C}}+y_{2N}\left( N_{R}\overline{%
N_{R}^{C}}\right) _{\mathbf{2}_{1}}\xi +y_{3N}\left( N_{R}\overline{N_{R}^{C}%
}\right) _{\mathbf{1}_{--}}\rho+y_{4N}\left[ \left( N_{R}\xi \right) _{\mathbf{1}_{-+}}\overline{%
N_{1R}^{C}}+h.c.\right] 
\end{eqnarray}
After the spontaneous breaking of the $Q_{6}\times Z_{4}$ symmetry, the above given Yukawa interactions take the following form:
\begin{eqnarray}
\tciLaplace _{Y} &=&y_{1u}\left[\overline{q}_{1L}\widetilde{H}_{3}u_{2R}-%
\overline{q}_{2L}\widetilde{H}_{3}u_{1R}\right] +y_{2u}\left[\overline{q}%
_{1L}\widetilde{H}_{1}u_{3R}+\overline{q}_{2L}\widetilde{H}_{2}u_{3R}\right]
+y_{3u}\overline{q}_{3L}\left[ \widetilde{H}_{1}u_{1R}+\widetilde{H}%
_{2}u_{2R}\right] +y_{4u}\overline{q}_{3L}\widetilde{H}_{3}u_{3R}\notag \\
&&+y_{1d}\left[ \overline{q}_{1L}H_{3}d_{2R}-\overline{q}_{2L}H_{3}d_{1R}%
\right] +y_{2d}\left[ \overline{q}_{1L}H_{1}d_{3R}+\overline{q}%
_{2L}H_{2}d_{3R}\right] 
+y_{3d}\overline{q}_{3L}\left[ H_{1}d_{1R}+H_{2}d_{2R}\right] +y_{4d}%
\overline{q}_{3L}H_{3}d_{3R}\notag \\
&&+y^{l}_{1} \overline{l}_{1 L}H_{3}e_{1 R}+   y^{l}_{2} \overline{l}_{1 L}\left[ H_{1}e_{2 R}+ H_{2}e_{3 R}\right]+y^{l}_{3} \left[ \overline{l}_{2 L}H_{1}+ \overline{l}_{3 L}H_{2}\right]e_{1 R}+ y^{l}_{4}\left[  \overline{l}_{2 L}H_{3}e_{3 R}-\overline{l}_{3 L}H_{3}e_{2 R}\right]
\notag\\
&&+y_{1\nu }\overline{l}_{1L}\widetilde{H}_{4}N_{1R}+y_{2\nu }\left[
\overline{l}_{2L}\widetilde{H}_{4}N_{3R}-\overline{l}_{3L}\widetilde{H}%
_{4}N_{2R}\right] +y_{1N}N_{1R}\sigma \overline{N_{1R}^{C}}+y_{2N}\left[ N_{2R}\overline{%
N_{2R}^{C}}\xi _{2}+N_{3R}\overline{N_{3R}^{C}}\xi _{1}\right]\notag\\
&&+y_{3N}\left[ N_{2R}\rho \overline{N_{3R}^{C}}+N_{3R}\rho \overline{%
N_{2R}^{C}}\right]+y_{4N}\left[ \left( N_{2R}\xi _{1}+N_{3R}\xi _{2}\right) \overline{%
N_{1R}^{C}}+h.c.\right],
\end{eqnarray}
To close this section, we provide a concise and qualitative discussion of the implications of our model in charged lepton flavor violation. Charged lepton flavor violating decays, like for instance $\mu\to e\gamma$, will receive radiative contributions at one-loop level mediated by neutral scalars and charged leptons as well as by charged scalars (arising from the inert doublet $H_4$) and right-handed neutrinos. For an appropriate region of parameter space, which implies small values of the flavor changing neutral Yukawa couplings involving electron and muon, not larger than about $10^{-6}$ \cite{Harnik:2012pb,Calibbi:2017uvl} and masses of the charged scalars arising from the inert doublet $H_4$ larger than several TeVs \cite{Hernandez:2021iss,Abada:2022dvm}, the charged lepton flavor violating decay $\mu\to e\gamma$ will acquire rates below its current experimental limit of $1.5\times 10^{-13}$ \cite{MEGII:2025gzr}. 
A detailed numerical analysis of the implications of the model in charged lepton flavor violation is beyond the scope of this work and will be presented elsewhere.
\section{Quark masses and mixing}
\label{quarmassesandmixings}
From the quark Yukawa interactions, we find that the up and down type quark mass matrices have the following form 
\begin{equation}
\mathbf{M}_{u}=%
\begin{pmatrix}
0 & y_{1}^{u}\langle \tilde{H}_{3}^{0}\rangle  & y_{2}^{u}\langle \tilde{H}%
_{1}^{0}\rangle  \\ 
-y_{1}^{u}\langle \tilde{H}_{3}^{0}\rangle  & 0 & y_{2}^{u}\langle \tilde{H}%
_{2}^{0}\rangle  \\ 
y_{3}^{u}\langle \tilde{H}_{1}^{0}\rangle  & y_{3}^{u}\langle \tilde{H}%
_{2}^{0}\rangle  & y_{4}^{u}\langle \tilde{H}_{3}^{0}\rangle 
\end{pmatrix}
,\qquad %
\mathbf{M}_{d}=%
\begin{pmatrix}
0 & y_{1}^{d}\langle H_{3}^{0}\rangle  & y_{2}^{d}\langle H_{1}^{0}\rangle 
\\ 
-y_{1}^{d}\langle H_{3}^{0}\rangle  & 0 & y_{2}^{d}\langle H_{2}^{0}\rangle 
\\ 
y_{3}^{d}\langle H_{1}^{0}\rangle  & y_{3}^{d}\langle H_{2}^{0}\rangle  & 
y_{4}^{d}\langle H_{3}^{0}\rangle 
\end{pmatrix}
\label{qm1}
\end{equation}%
The above mass matrices possess implicitly the nearest-neighbor interactions (NNI) textures, to show it, we take the VEV alignment 
$\langle
H_{1}^{0}\rangle =0$, $\langle H_{2}^{0}\rangle =\frac{v_{2}}{\sqrt{2}}$ and 
$\langle H_{3}^{0}\rangle =\frac{v_{3}}{\sqrt{2}}$, which is consistent with the scalar potential minimization conditions for a large region of parameter space, as shown in Appendix \ref{ScalarQ6doublet}. Then, the quark mass matrices are parameterized as follows
\begin{equation}
\mathbf{M}_{q}= 
\begin{pmatrix}
0 & A_{q} & 0 \\ 
-A_{a} & 0 & b_{q} \\ 
0 & c_{q} & F_{q}%
\end{pmatrix}%
,\label{qm2}
\end{equation}
where $q=u,d$. Both mass matrices are diagonalized by the $\mathbf{U}%
_{q(L,R)}$ unitary matrices such that $\mathbf{U}^{\dagger}_{q L}\mathbf{M}%
_{q} \mathbf{U}_{q R}=\hat{\mathbf{M}}_{q}$, with $\hat{\mathbf{M}}_{q}=\text{%
Diag}.\left(m_{q_{1}}, m_{q_{2}}, m_{q_{3}}\right)$ being the physical quark
masses. In order to obtain the CKM matrix, let us calculate the $\mathbf{U}_{q L}$
matrix by means of the bilineal form $\mathbf{\hat{M}}_{q} \mathbf{\hat{M}}%
^{\dagger}_{q}=\mathbf{U}^{\dagger}_{q L} \mathbf{M}_{q} \mathbf{M }%
^{\dagger}_{q} \mathbf{U}_{q L}$. As it is shown in the Appendix, $\mathbf{U}_{q L}=\mathbf{P}_{q}\mathbf{O}_{q L}$, where $\mathbf{P}_{q} = \text{diag.} \left( 1, e^{ i\eta_{q_{2}}}, e^{ i\eta_{q_{3}}
}\right)$ and the $\mathbf{O}_{qL}$ orthogonal matrix is parametrized as follows:
\begin{align}  \label{ortho}
\mathbf{O}_{q L}= 
\begin{pmatrix}
-\sqrt{\dfrac{\tilde{m}_{q_{2}} (\rho^{q}_{-}-R^{q}) K^{q}_{+}}{4 y_{q}
\delta^{q}_{1} \kappa^{q}_{1} }} & -\sqrt{\dfrac{\tilde{m}_{q_{1}}
(\sigma^{q}_{+}-R^{q}) K^{f}_{+}}{4 y_{q} \delta^{q}_{2} \kappa^{q}_{2} }} & 
\sqrt{\dfrac{\tilde{m}_{q_{1}} \tilde{m}_{q_{2}} (\sigma^{q}_{-}+R^{q})
K^{q}_{+}}{4 y_{q} \delta^{q}_{3} \kappa^{q}_{3} }} \\ 
-\sqrt{\dfrac{\tilde{m}_{q_{1}} \kappa^{q}_{1} K^{q}_{-}}{%
\delta^{q}_{1}(\rho^{q}_{-}-R^{q}) }} & \sqrt{\dfrac{\tilde{m}_{q_{2}}
\kappa^{q}_{2} K^{q}_{-}}{\delta^{q}_{2}(\sigma^{q}_{+}-R^{q}) }} & \sqrt{%
\dfrac{\kappa^{q}_{3} K^{q}_{-}}{\delta^{q}_{3}(\sigma^{q}_{-}+R^{q}) }} \\ 
\sqrt{\dfrac{\tilde{m}_{q_{1}} \kappa^{q}_{1}(\rho^{q}_{-}-R^{q})}{2
y_{q}\delta^{q}_{1}}} & -\sqrt{\dfrac{\tilde{m}_{q_{2}}
\kappa^{q}_{2}(\sigma^{q}_{+}-R^{q})}{2 y_{q}\delta^{q}_{2}}} & \sqrt{\dfrac{%
\kappa^{q}_{3}(\sigma^{q}_{-}+R^{q})}{2 y_{q}\delta^{q}_{3}}}%
\end{pmatrix}%
\end{align}
with 
\begin{align}  \label{defortho}
\rho^{q}_{\pm}&\equiv 1+\tilde{m}^{2}_{q_{2}}\pm\tilde{m}%
^{2}_{q_{1}}-y^{2}_{q},\quad \sigma^{q}_{\pm}\equiv 1-\tilde{m}%
^{2}_{q_{2}}\pm(\tilde{m}^{2}_{q_{1}}-y^{2}_{q}),\quad \delta^{q}_{(1,
2)}\equiv (1-\tilde{m}^{2}_{q_{(1, 2)}})(\tilde{m}^{2}_{q_{2}}-\tilde{m}%
^{2}_{q_{1}});  \notag \\
\delta^{q}_{3}&\equiv (1-\tilde{m}^{2}_{q_{1}})(1-\tilde{m}%
^{2}_{q_{2}}),\quad \kappa^{q}_{1} \equiv \tilde{m}_{q_{2}}-\tilde{m}%
_{q_{1}}y_{q},\quad \kappa^{q}_{2}\equiv \tilde{m}_{q_{2}}y_{q}-\tilde{m}%
_{q_{1}},\quad \kappa^{q}_{3}\equiv y_{q}-\tilde{m}_{q_{1}}\tilde{m}_{q_{2}};
\notag \\
R^{q}&\equiv \sqrt{\rho^{q 2}_{+}-4(\tilde{m}^{2}_{q_{2}}+\tilde{m}%
^{2}_{q_{1}}+\tilde{m}^{2}_{q_{2}}\tilde{m}^{2}_{q_{1}}-2\tilde{m}_{q_{1}}%
\tilde{m}_{q_{2}}y_{q})},\quad K^{q}_{\pm} \equiv y_{q}(\rho^{q}_{+}\pm
R^{q})-2\tilde{m}_{q_{1}}\tilde{m}_{q_{2}}.
\end{align}

We have to point out that the parameters have been normalized by  $%
m_{q_{3}}$, the heaviest physical quark mass. Additionally, there are two unfixed
parameters ($y_{q}\equiv \vert F_{q}\vert/m_{q_{3}}$) which are constrained
by the condition $1>y_{q}>\tilde{m}_{q_{2}}>\tilde{m}_{q_{1}}$.
Finally, the CKM mixing matrix is written as 
\begin{equation}
\mathbf{V}_{CKM}=\mathbf{O}^{T}_{u L}\bar{\mathbf{P}}_{q} \mathbf{O}_{d L},
\quad \mathbf{P}_{q}=\mathbf{P}^{\dagger}_{u}\mathbf{P}_{d}=\text{diag.}%
\left(1, e^{i\bar{\eta}_{q_{2}}}, e^{i\bar{\eta}_{q_{3}}} \right).
\end{equation}
This CKM mixing matrix has four free parameters namely $y_{u}$, $y_{d}$, and
two phases $\bar{\eta}_{q_{2}}$ and $\bar{\eta}_{q_{3}}$ which could be obtained
numerically. In addition to this, the expression for the mixing angles are given as follows:
\begin{eqnarray}
\sin^{2}{\theta^{q}_{13}}&=& \big| (\mathbf{V}_{CKM})_{13}\big|^{2}= \big |(\mathbf{O}_{u})_{11} (\mathbf{O}_{d})_{13}+(\mathbf{O}_{u})_{21} (\mathbf{O}_{d})_{23}e^{i\bar{\eta}_{q_{2}}}+(\mathbf{O}_{u})_{31} (\mathbf{O}_{d})_{33}e^{i\bar{\eta}_{q_{3}}}\big|^{2};\nn\\
\sin^{2}{\theta^{q}_{12}}&=&\frac{ \big| (\mathbf{V}_{CKM})_{12}\big|^{2}}{1- \big| (\mathbf{V}_{CKM})_{13}\big|^{2}}= \frac{\big |(\mathbf{O}_{u})_{11} (\mathbf{O}_{d})_{12}+(\mathbf{O}_{u})_{21} (\mathbf{O}_{d})_{22}e^{i\bar{\eta}_{q_{2}}}+(\mathbf{O}_{u})_{31} (\mathbf{O}_{d})_{32}e^{i\bar{\eta}_{q_{3}}}\big|^{2}}{1- \big| (\mathbf{V}_{CKM})_{13}\big|^{2}};\nn\\
\sin^{2}{\theta^{q}_{23}}&=&\frac{ \big| (\mathbf{V}_{CKM})_{23}\big|^{2}}{1- \big| (\mathbf{V}_{CKM})_{13}\big|^{2}}= \frac{\big |(\mathbf{O}_{u})_{12} (\mathbf{O}_{d})_{13}+(\mathbf{O}_{u})_{22} (\mathbf{O}_{d})_{23}e^{i\bar{\eta}_{q_{2}}}+(\mathbf{O}_{u})_{32} (\mathbf{O}_{d})_{33}e^{i\bar{\eta}_{q_{3}}}\big|^{2}}{1- \big| (\mathbf{V}_{CKM})_{13}\big|^{2}},
\end{eqnarray}
and the Jarlskog invariant takes the form:
\begin{equation}
\label{eq:JCP-1}
J_{CP}={\rm Im}\left[\left(\mathbf{V}_{CKM}\right)_{23}\left(\mathbf{V}_{CKM}^{\ast}\right)_{13}\left(\mathbf{V}_{CKM}\right)_{12}\left(\mathbf{V}_{CKM}^{\ast}\right)_{22}\right]=\frac{1}{8}\sin{2\theta^{q}_{12}}\sin{2\theta^{q}_{23}}\sin{2\theta^{q}_{13}}\cos{\theta^{q}_{13}}\sin{\delta^{q}_{CP}}
\end{equation}
\section{Lepton masses and mixing}
\label{leptonmassesandmixings}
\subsection{Charged lepton sector}
The charged lepton mass matrix is directly obtained from the leptonic Yukawa interactions and has the following form:
\begin{equation}
M_{l}=\left( 
\begin{array}{ccc}
y^{l}_{1}\frac{v_{3}}{\sqrt{2}} & y^{l}_{2}\frac{v_{1}}{\sqrt{2}} & y^{l}_{2}\frac{v_{2}}{\sqrt{2}} \\ 
y^{l}_{3}\frac{v_{1}}{\sqrt{2}} & 0 & y^{l}_{4}\frac{%
v_{3}}{\sqrt{2}} \\ 
y^{l}_{3}\frac{v_{2}}{\sqrt{2}} & -y^{l}_{4}\frac{%
v_{3}}{\sqrt{2}} & 0%
\end{array}%
\right),
\end{equation}
As one can notice, in the quark sector, the NNI textures were obtained by using the following VEV alignment $\left( \langle H^{0}_{1} \rangle , \langle H^{0}_{2} \rangle \right)=\left(0 , v_{2}/\sqrt{2}\right)$ and $ \langle H^{0}_{3} \rangle=v_{3}/\sqrt{2}$. This choice implies, in the charged lepton, the following textures
\begin{equation*}
M_{l}=\left( 
\begin{array}{ccc}
a_{l} & 0 & b_{l} \\ 
0 & 0 & d_{l} \\ 
c_{l} & -d_{l} & 0%
\end{array}%
\right),
\end{equation*}
where the matrix elements can be easily read off the above equation. Analogously to the quark sector, the aforementioned matrix is diagonalized by $\mathbf{U}^{\dagger}_{l L}\mathbf{M}_{l}\mathbf{U}_{l R}=\hat{\mathbf{M}}_{l}$ with $\hat{\mathbf{M}}_{l}=\textrm{Diag}.\left(m_{e},m_{\mu},m_{\tau}\right)$. Then,
we build the bilineal $\hat{\mathbf{M}}_{l}\hat{\mathbf{M}}^{\dagger}_{l}= \mathbf{U}^{\dagger}_{l L}\mathbf{M}_{l}\mathbf{M}^{\dagger}_{l} \mathbf{U}_{l L}$ in order to obtain the unitary matrix that appears in  the PMNS one. In the Appendix \ref{fermiondiagonalization}, we show that $\mathbf{U}_{l L}=\mathbf{P}_{l}\mathbf{O}_{l}$ where 
$\mathbf{P}_{l}=\textrm{Diag}.(1, e^{i\eta_{\mu}}, e^{i\eta_{\tau}} )$ and 
the latter matrix is real and orthogonal such that is parametrized as
\begin{equation}
\mathbf{O}_{l}= \left( X_{1}\quad X_{2} \quad X_{3}
\right),
\end{equation}
where the eigenvectors are written explicitly 
\begin{eqnarray}
X_{1}&=& \left( 
\begin{array}{c}
-\sqrt{\frac{2\vert m_{e}\vert\left(\vert m_{\tau}\vert\vert m_{\mu}\vert-\vert a_{l}\vert\vert m_{e}\vert\right)^{2}\left[\vert a_{l}\vert\left(\vert m_{\tau}  \vert^{2}+\vert m_{\mu}  \vert^{2}+\vert m_{e}  \vert^{2}-\vert a_{l}  \vert^{2}+R_{e}\right)-2\vert m_{\tau}  \vert \vert m_{\mu}  \vert \vert m_{e}  \vert\right]}{D_{e}}}\\
\sqrt{\frac{4\vert m_{\tau}\vert \vert m_{\mu}\vert \left(\vert m_{\tau}\vert\vert m_{\mu}\vert-\vert a_{l}\vert\vert m_{e}\vert\right) \left(\vert m_{\mu}\vert\vert a_{l}\vert-\vert m_{\tau}\vert\vert m_{e}\vert\right) \left(\vert m_{\tau}\vert\vert a_{l}\vert-\vert m_{\mu}\vert\vert m_{e}\vert\right)}{D_{e}}}\\
\sqrt{\frac{\vert a_{l}\vert \vert m_{e}\vert \left[2 \vert m_{\tau}\vert \vert m_{\mu}\vert \vert a_{l}\vert-\vert m_{e}\vert\left(\vert m_{\tau}  \vert^{2}+\vert m_{\mu}  \vert^{2}-\vert m_{e}  \vert^{2}+\vert a_{l}  \vert^{2}-R_{e}\right)\right]^{2}}{D_{e}}}
\end{array}%
\right);\nn\\
X_{2}&=&\left( 
\begin{array}{c}
\sqrt{\frac{\vert m_{\mu}\vert\left(\vert m_{\mu}\vert\vert a_{l}\vert-\vert m_{\tau}\vert\vert m_{e}\vert\right)\left(\vert m_{\tau}  \vert^{2}-\vert m_{\mu}  \vert^{2}+\vert m_{e}  \vert^{2}-\vert a_{l}  \vert^{2}+R_{e}\right)}{D_{\mu}}}\\  \sqrt{\frac{\vert m_{\tau}\vert \vert m_{e}\vert \left[ \vert m_{\mu}\vert\left(\vert m_{\tau}  \vert^{2}-\vert m_{\mu}  \vert^{2}+\vert m_{e}  \vert^{2}+\vert a_{l}  \vert^{2}+R_{e}\right) -  2 \vert m_{\tau}\vert \vert m_{e}\vert \vert a_{l}\vert\right]}{D_{\mu}}}\\ 
-\sqrt{\frac{\vert a_{l}\vert \vert m_{\mu}\vert \left[\vert m_{\mu}\vert\left(\vert m_{\tau}  \vert^{2}-\vert m_{\mu}  \vert^{2}+\vert m_{e}  \vert^{2}+\vert a_{l}  \vert^{2}-R_{e}\right) -  2 \vert m_{\tau}\vert \vert m_{e}\vert \vert a_{l}\vert\right]}{D_{\mu}}}
\end{array}%
\right);\nn\\
X_{3}&=&\left( 
\begin{array}{c}
\sqrt{\frac{2\vert m_{\tau}\vert\left(\vert m_{\tau}\vert\vert a_{l}\vert-\vert m_{\mu}\vert\vert m_{e}\vert\right)^{2}\left[\vert a_{l}\vert\left(\vert m_{\tau}  \vert^{2}+\vert m_{\mu}  \vert^{2}+\vert m_{e}  \vert^{2}-\vert a_{l}  \vert^{2}+R_{e}\right)-2\vert m_{\tau}  \vert \vert m_{\mu}  \vert \vert m_{e}  \vert\right]}{D_{\tau}}}\\  \sqrt{\frac{4\vert m_{\mu}\vert \vert m_{e}\vert \left(\vert m_{\tau}\vert\vert m_{\mu}\vert-\vert a_{l}\vert\vert m_{e}\vert\right) \left(\vert m_{\mu}\vert\vert a_{l}\vert-\vert m_{\tau}\vert\vert m_{e}\vert\right) \left(\vert m_{\tau}\vert\vert a_{l}\vert-\vert m_{\mu}\vert\vert m_{e}\vert\right)}{D_{\tau}}}\\ 
\sqrt{\frac{\vert a_{l}\vert \vert m_{\tau}\vert \left[\vert m_{\tau}\vert\left(\vert m_{\tau}  \vert^{2}-\vert m_{\mu}  \vert^{2}-\vert m_{e}  \vert^{2}-\vert a_{l}  \vert^{2}+R_{e}\right) + 2 \vert m_{\mu}\vert \vert m_{e}\vert \vert a_{l}\vert\right]^{2}}{D_{\tau}}}
\end{array}%
\right),
\end{eqnarray}
with
\begin{eqnarray}
D_{e}&=&2\vert a_{l}\vert \left(\vert m_{\tau}  \vert^{2}-\vert m_{e}  \vert^{2}\right)\left(\vert m_{\mu}  \vert^{2}-\vert m_{e}  \vert^{2}\right)\left[2 \vert m_{\tau}\vert \vert m_{\mu}\vert \vert a_{l}\vert-\vert m_{e}\vert\left(\vert m_{\tau}  \vert^{2}+\vert m_{\mu}  \vert^{2}-\vert m_{e}  \vert^{2}+\vert a_{l}  \vert^{2}-R_{e}\right)\right];\nn\\ 
D_{\mu}&=&2\vert a_{l}\vert \left(\vert m_{\tau}  \vert^{2}-\vert m_{\mu}  \vert^{2}\right)\left(\vert m_{\mu}  \vert^{2}-\vert m_{e}  \vert^{2}\right);\nn\\ D_{\tau}&=&2\vert a_{l}\vert \left(\vert m_{\tau}  \vert^{2}-\vert m_{e}  \vert^{2}\right)\left(\vert m_{\tau}  \vert^{2}-\vert m_{\mu}  \vert^{2}\right)\left[\vert m_{\tau}\vert\left(\vert m_{\tau}  \vert^{2}-\vert m_{\mu}  \vert^{2}-\vert m_{e}  \vert^{2}-\vert a_{l}  \vert^{2}+R_{e}\right) + 2 \vert m_{\mu}\vert \vert m_{e}\vert \vert a_{l}\vert\right].
\end{eqnarray}
In order to get the correct charged lepton masses, the unfixed parameter should  satisfy $\vert m_{\tau}\vert>\vert a_{l}\vert\approx(\vert m_{\tau}\vert/\vert m_{\mu}\vert)\vert m_{e}\vert$. As a result of this, $\mathbf{U}_{l L}$ must be almost
the identity matrix as one can verify in the Appendix \ref{fermiondiagonalization}.

\subsection{Neutrino sector}
Due to the preserved $\widetilde{Z}_{2}\times Z_{2}$ symmetry, the tiny masses of the active neutrinos are forbidden at tree as well as at one loop level. These masses are only generated at two-loop level. From the neutrino Yukawa interactions we find that the mass matrix for active neutrinos takes the form:
\begin{equation}
 M_{\nu }=\left( 
 \begin{array}{ccc}
 y_{1\nu }^{2}F\left( \left( M_{N}\right) _{11},m_{R},m_{I}\right) & y_{1\nu
 }y_{2\nu }F\left( \left( M_{N}\right) _{12},m_{R},m_{I}\right) & y_{1\nu
 }y_{2\nu }F\left( \left( M_{N}\right) _{13},m_{R},m_{I}\right) \\ 
 y_{1\nu }y_{2\nu }F\left( \left( M_{N}\right) _{12},m_{R},m_{I}\right) & 
 y_{2\nu }^{2}F\left( \left( M_{N}\right) _{22},m_{R},m_{I}\right) & -y_{2\nu
 }^{2}F\left( \left( M_{N}\right) _{23},m_{R},m_{I}\right) \\ 
 y_{1\nu }y_{2\nu }F\left( \left( M_{N}\right) _{13},m_{R},m_{I}\right) & 
 -y_{2\nu }^{2}F\left( \left( M_{N}\right) _{23},m_{R},m_{I}\right) & y_{2\nu
 }^{2}F\left( \left( M_{N}\right) _{33},m_{R},m_{I}\right)%
 \end{array}%
 \right) .
 \end{equation}
 The above given neutrino mass matrix can also be written as:
\begin{eqnarray}
M_{\nu } &=&\frac{1}{16\pi ^{2}}\widetilde{Y}_{\nu D}\left( 
\begin{array}{ccc}
m_{N_{1}}f_{1} & 0 & 0 \\ 
0 & m_{N_{2}}f_{2} & 0 \\ 
0 & 0 & m_{N_{3}}f_{3}%
\end{array}%
\right) \widetilde{Y}_{\nu D}^{T},\hspace{1.5cm}\widetilde{Y}_{\nu D}=Y_{\nu
D}R_{N},  \notag \\
Y_{\nu D} &=&\left( 
\begin{array}{ccc}
y_{_{1}}^{\left( \nu \right) } & 0 & 0 \\ 
0 & 0 & y_{2}^{\left( \nu \right) } \\ 
0 & -y_{2}^{\left( \nu \right) } & 0%
\end{array}%
\right) ,\hspace{1.5cm}\left( M_{N}\right) _{diag}=\left( 
\begin{array}{ccc}
m_{N_{1}} & 0 & 0 \\ 
0 & m_{N_{2}} & 0 \\ 
0 & 0 & m_{N_{3}}%
\end{array}%
\right) =R_{N}^{T}M_{N}R_{N}  \notag \\
f_{k} &=&\frac{m_{R}^{2}}{m_{R}^{2}-m_{N_{k}}^{2}}\ln \left( \frac{m_{R}^{2}%
}{m_{N_{k}}^{2}}\right) -\frac{m_{I}^{2}}{m_{I}^{2}-m_{N_{k}}^{2}}\ln \left( 
\frac{m_{I}^{2}}{m_{N_{k}}^{2}}\right) ,\hspace{1.5cm}k=1,2,3
\end{eqnarray}
where $m_{R}=m_{\func{Re}H_{4}^{0}}$, $m_{I}=m_{\func{Im}H_{4}^{0}}$ and $f_{k}$ is a loop function.

It is worth mentioning that the mass splitting between $\func{Re}H_{4}^{0}$ and $\func{Im}H_{4}^{0}$ is generated at one loop level. Furthermore, the Majorana neutrino mass matrix takes the form:
\begin{equation}
M_{N}=\left( 
\begin{array}{ccc}
y_{1N}\frac{v_{\sigma }}{\sqrt{2}} & y_{4N}\frac{v_{\xi }}{\sqrt{2}}%
e^{i\theta } & y_{4N}\frac{v_{\xi }}{\sqrt{2}}e^{-i\theta } \\ 
y_{4N}\frac{v_{\xi }}{\sqrt{2}}e^{i\theta } & y_{2N}\frac{v_{\xi }}{\sqrt{2}}%
e^{-i\theta } & y_{3N}\frac{v_{\rho }}{\sqrt{2}} \\ 
y_{4N}\frac{v_{\xi }}{\sqrt{2}}e^{-i\theta } & y_{3N}\frac{v_{\rho }}{\sqrt{2%
}} & y_{2N}\frac{v_{\xi }}{\sqrt{2}}e^{i\theta }%
\end{array}%
\right) .
\end{equation}
For the sake of simplicity, we consider the benchmark scenario where $\left( M_{N}\right) _{ij}<<m_{R}^{2}$, $m_{I}^{2}$. That scenario allows the cobimaximal pattern~\cite{Fukuura:1999ze,Miura:2000sx,Ma:2002ce,Grimus:2003yn,Chen:2014wxa,Ma:2015fpa,Joshipura:2015dsa,Li:2015rtz,He:2015xha,Chen:2015siy,Ma:2016nkf,Damanik:2017jar,Ma:2017trv,Grimus:2017itg,CarcamoHernandez:2017owh,CarcamoHernandez:2018hst,Ma:2019iwj,Hernandez:2021kju, Rivera-Agudelo:2022qpa, Gomez-Izquierdo:2023mph,Rivera-Agudelo:2024vdn}  of the light active neutrino mass matrix to be manifest,  since the mass matrix for active neutrinos takes the form:
\begin{equation}
M_{\nu }\simeq \frac{m_{R}^{2}-m_{I}^{2}}{8\pi ^{2}\left(
m_{R}^{2}+m_{I}^{2}\right) }\left( 
\begin{array}{ccc}
y_{1\nu }^{2}y_{1N}\frac{v_{\sigma }}{\sqrt{2}} & y_{1\nu }y_{2\nu }y_{4N}%
\frac{v_{\xi }}{\sqrt{2}}e^{i\theta } & y_{1\nu }y_{2\nu }y_{4N}\frac{v_{\xi
}}{\sqrt{2}}e^{-i\theta } \\ 
y_{1\nu }y_{2\nu }y_{4N}\frac{v_{\xi }}{\sqrt{2}}e^{i\theta } & y_{2\nu
}^{2}y_{2N}\frac{v_{\xi }}{\sqrt{2}}e^{-i\theta } & -y_{2\nu }^{2}y_{3N}\frac{%
v_{\rho }}{\sqrt{2}} \\ 
y_{1\nu }y_{2\nu }y_{4N}\frac{v_{\xi }}{\sqrt{2}}e^{-i\theta } & -y_{2\nu
}^{2}y_{3N}\frac{v_{\rho }}{\sqrt{2}} & y_{2\nu }^{2}y_{2N}\frac{v_{\xi }}{%
\sqrt{2}}e^{i\theta }%
\end{array}%
\right) ,
\end{equation}
As one can notice, the effective neutrino mass matrix can be parameterized as
\begin{equation}
\mathbf{M}_{\nu }=\left( 
\begin{array}{ccc}
A_{\nu } & \tilde{B}_{\nu } & \tilde{B}_{\nu }^{\ast } \\ 
\tilde{B}_{\nu } & \tilde{C}_{\nu }^{\ast } & D_{\nu } \\ 
\tilde{B}_{\nu }^{\ast } & D_{\nu } & \tilde{C}_{\nu }%
\end{array}%
\right), \label{Eq:NeutrMatParam}
\end{equation}%
where the cobimaximal pattern is clearly exhibited.  As it is well known, $\mathbf{M}_{\nu }$ is diagonalized by the mixing matrix $\mathbf{U}%
_{\nu }$, this is, $\mathbf{U}_{\nu }^{\dagger }\mathbf{M}_{\nu }\mathbf{U}%
_{\nu }^{\ast }=\hat{\mathbf{M}}_{\nu }$ with $\hat{\mathbf{M}}_{\nu }=\text{%
Diag.}(|m_{1}|,|m_{2}|,|m_{3}|)$. Explicitly, we have
\begin{eqnarray}
\mathbf{U}_{\nu}=%
\begin{pmatrix}
\cos{\gamma_{12}}\cos{\gamma_{13}} & \sin{\gamma_{12}}\cos{\gamma_{13}} & 
-\sin{\gamma_{13}} \\ 
-\frac{1}{\sqrt{2}}\left(\sin{\gamma_{12}}-i\cos{\gamma_{12}}\sin{\gamma_{13}}%
\right) & \frac{1}{\sqrt{2}}\left(\cos{\gamma_{12}}+i\sin{\gamma_{12}}\sin{%
\gamma_{13}}\right) & \frac{i\cos{\gamma_{13}}}{\sqrt{2}} \\ 
-\frac{1}{\sqrt{2}}\left(\sin{\gamma_{12}}+i\cos{\gamma_{12}}\sin{\gamma_{13}}%
\right) & \frac{1}{\sqrt{2}}\left(\cos{\gamma_{12}}-i\sin{\gamma_{12}}\sin{%
\gamma_{13}}\right) & -\frac{i\cos{\gamma_{13}}}{\sqrt{2}}%
\end{pmatrix} ~.%
\end{eqnarray}

\subsection{PMNS mixing matrix}

Once the lepton masses were calculated, the PMNS mixing matrix is given by $\mathbf{U}=\mathbf{U}^{\dagger}_{l}\mathbf{U}_{\nu}=\mathbf{O}^{T}_{l}\mathbf{P}^{\dagger}_{l}\mathbf{U}_{\nu}$. Consequently, the reactor, solar and atmospheric angles are given as follows
\begin{eqnarray}
\sin^{2}{\theta}_{13}&=& \big| \left(\mathbf{U}\right)_{13}\vert^{2}=\vert \left(\mathbf{O}_{l}\right)_{11}\left(\mathbf{U}_{\nu}\right)_{13}+\left(\mathbf{O}_{l}\right)_{21}\left(\mathbf{U}_{\nu}\right)_{23}e^{-i\eta_{\mu}}+\left(\mathbf{O}_{l}\right)_{31}\left(\mathbf{U}_{\nu}\right)_{33}e^{-i\eta_{\tau}}\big|^{2};\nn\\
\sin^{2}{\theta}_{12}&=&\frac{\big| \left(\mathbf{U}\right)_{12}\big|^{2}}{1-\big|\left(\mathbf{U}\right)_{13}\big|^{2}}=\frac{\big|\left(\mathbf{O}_{l}\right)_{11}\left(\mathbf{U}_{\nu}\right)_{12}+\left(\mathbf{O}_{l}\right)_{21}\left(\mathbf{U}_{\nu}\right)_{22}e^{-i\eta_{\mu}}+\left(\mathbf{O}_{l}\right)_{31}\left(\mathbf{U}_{\nu}\right)_{32}e^{-i\eta_{\tau}}\big|^{2}}{1-\big|\left(\mathbf{U}\right)_{13}\big|^{2}};\nn\\
\sin^{2}{\theta}_{23}&=&\frac{\big| \left(\mathbf{U}\right)_{12}\big|^{2}}{1-\big|\left(\mathbf{U}\right)_{13}\big|^{2}}=\frac{\big|\left(\mathbf{O}_{l}\right)_{12}\left(\mathbf{U}_{\nu}\right)_{13}+\left(\mathbf{O}_{l}\right)_{22}\left(\mathbf{U}_{\nu}\right)_{23}e^{-i\eta_{\mu}}+\left(\mathbf{O}_{l}\right)_{32}\left(\mathbf{U}_{\nu}\right)_{33}e^{-i\eta_{\tau}}\big|^{2}}{1-\big| \left(\mathbf{U}\right)_{13}\big|^{2}}. \label{Eq:mixPMNS}
\end{eqnarray}

Besides this, we can obtain the $\delta_{CP}$ Dirac CP-violating phase which comes from the Jarlskog invariant, 
\begin{equation}
\sin{\delta_{CP}} = \frac{\textrm{Im}\left[(\mathbf{U})_{23}(\mathbf{U})^{\ast}_{13}(\mathbf{U})_{12}(\mathbf{U})^{\ast}_{22}\right]}{\frac{1}{8}\sin{2\theta_{12}}\sin{2\theta_{23}}\sin{2\theta_{13}}\cos{\theta_{13}}} ~.    
\end{equation}

Notice that there are still  free parameters in the PMNS matrix, these are $\gamma_{12}$, $\gamma_{13}$ and $\vert a_{l} \vert$. In addition to those, two phases $\eta_{\mu}$ and $\eta_{\tau}$. Nevertheless, these might be irrelevant because there is a region in parameter space where  $\mathbf{U}_{l}$ is close to the identity matrix (See Appendix \ref{fermiondiagonalization}). Consequently, the PMNS matrix is controlled mainly by the cobimaximal one. 

Let us calculate the mixing angles and the Dirac CP violating phases in the limit $\vert a_{l}\vert=(\vert m_{\tau} \vert/ \vert m_{\mu} \vert) \vert m_{e} \vert $, then,
\begin{equation}
\mathbf{U}_{l}\approx \left( 
\begin{array}{ccc}
-1   & 0 & \vert \bar{m}_{e}\vert e^{i\eta_{\tau}}\\ 
0 & e^{i\eta_{\mu}} & 0 \\ 
\vert \bar{m}_{e}\vert & 0 & e^{i\eta_{\tau}} %
\end{array}%
\right),
\end{equation}
where $\vert \bar{m}_{e}\vert= \vert m_{e}\vert/\vert m_{\mu}\vert\sim \mathcal{O}\left(10^{
-3}\right)$. In consequence, the involved matrix elements are
\begin{eqnarray}
\left(\mathbf{U}\right)_{13}&\approx& \sin{\gamma_{13}}-\frac{i}{\sqrt{2}}\vert \bar{m}_{e}\vert \cos{\gamma_{13}}e^{-i\eta_{\tau}};\nn\\
\left(\mathbf{U}\right)_{12}&\approx& -\sin{\gamma_{12}}\cos{\gamma_{13}};\nn\\
\left(\mathbf{U}\right)_{23}&\approx& \frac{i}{\sqrt{2}}\cos{\gamma_{13}}e^{-i\eta_{\mu}};\nn\\
\left(\mathbf{U}\right)_{22}&\approx&\frac{1}{\sqrt{2}}\left(\cos{\gamma_{12}}-i\sin{\gamma_{12}}\sin{\gamma_{13}}\right)e^{-i\eta_{\mu}}.
\end{eqnarray}
As noticed, in this limit, the $\eta_{\mu}$ phase does not play an important role in the mixing parameters and the Dirac CP-violating phase, as one can verify by using the above expressions and Eq.~(\ref{Eq:mixPMNS}). Then, we obtain
\begin{eqnarray}
\sin{\theta}_{13}&\approx& \sin{\gamma_{13}}\left[1-\frac{\vert \bar{m}_{e}\vert}{\sqrt{2}}\cot{\gamma_{13}}\sin{\eta_{\tau}}\right];\nn\\
\sin{\theta}_{12}&\approx&\sin{\gamma_{12}}\left[1+\frac{\vert \bar{m}_{e}\vert}{\sqrt{2}}\tan{\gamma_{13}}\sin{\eta_{\tau}}\right];\nn\\
\sin{\theta}_{23}&\approx&\frac{1}{\sqrt{2}}\left[1+\frac{\vert \bar{m}_{e}\vert}{\sqrt{2}}\tan{\gamma_{13}}\sin{\eta_{\tau}}\right];\nn\\
\sin{\delta_{CP}}&\approx&-1+\frac{\vert \bar{m}_{e}\vert}{\sqrt{2}}\tan{\gamma_{13}}\sin{\eta_{\tau}}.
\end{eqnarray}
Therefore, we realized that the charged lepton sector modifies the cobimaximal predictions such that the solar angle and Dirac CP-violating phase are deviated from $\pi/4$ and $3\pi/2$, respectively. This deviation is tiny in this limit, as a result of this $\theta_{13} \approx \gamma_{13}$ and $\theta_{12}\approx \gamma_{12}$. In short, this brief analytical study exhibits that the current model might fit quite well the PMNS matrix. To finish this section, a $\chi^{2}$ analysis was performed to scan the allowed region for the free parameters.
\begin{figure}
\centering
\subfloat[]{\includegraphics[scale=0.32]{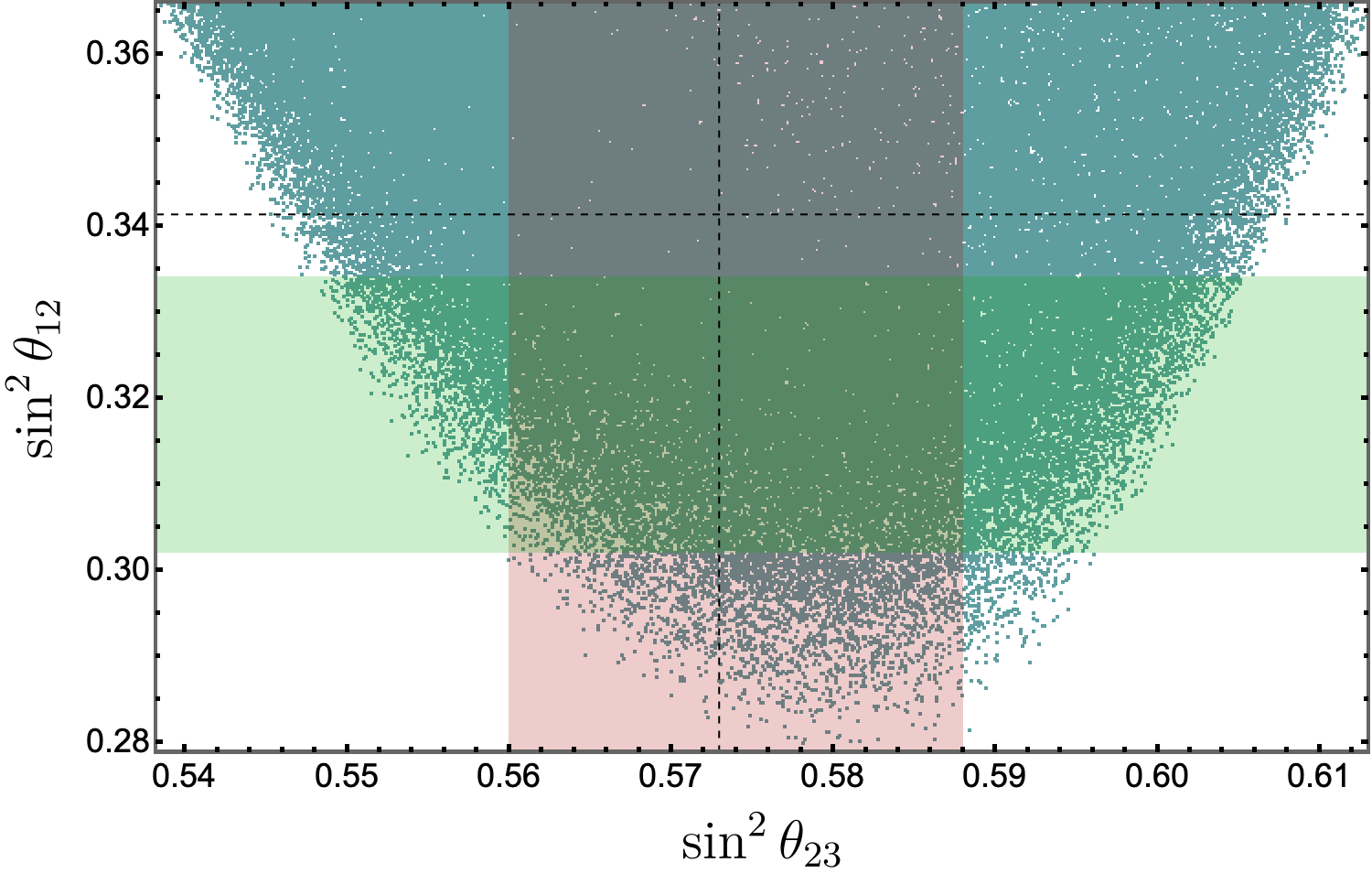}}\quad 
\subfloat[]{\includegraphics[scale=0.3]{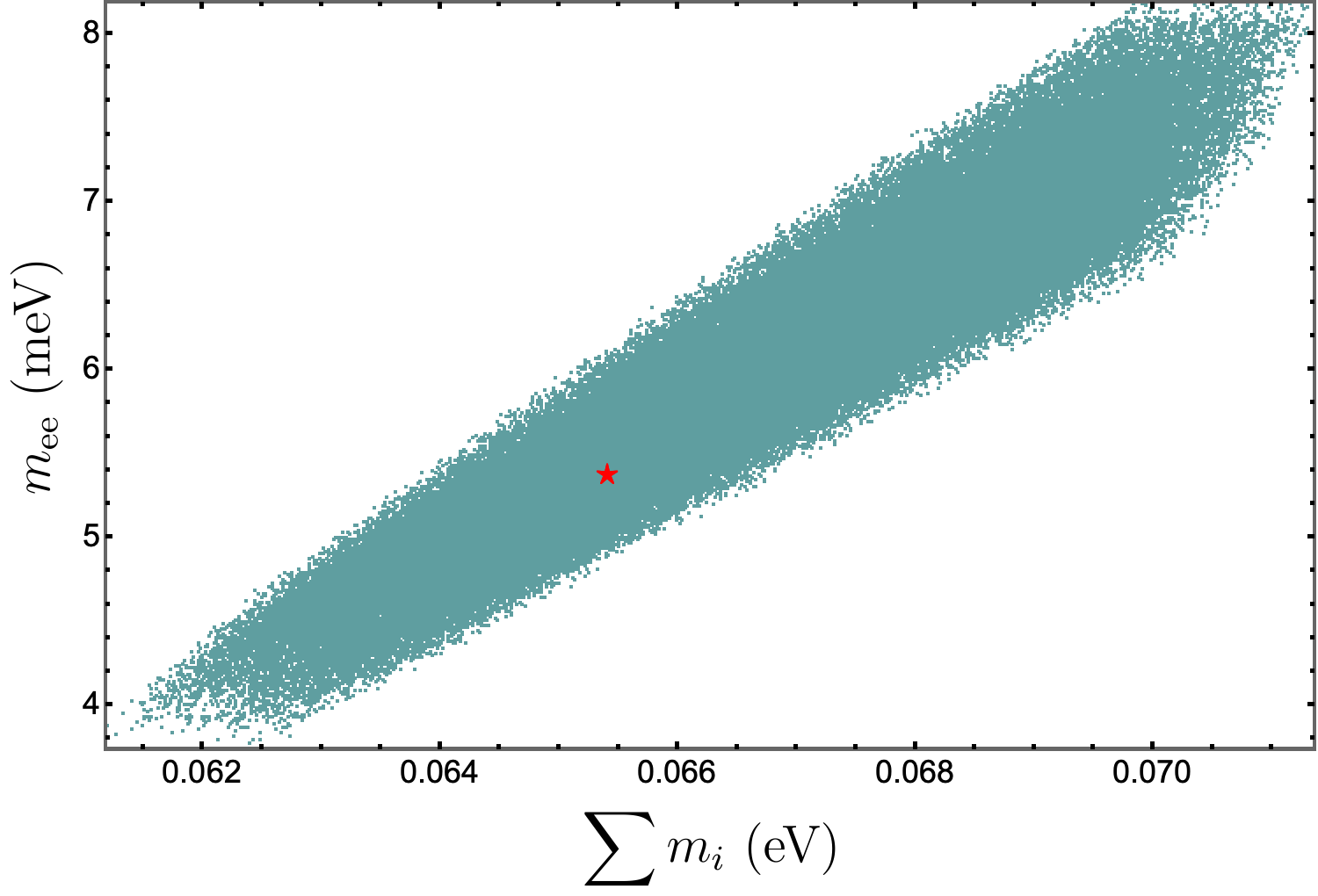}}\quad 
\caption{Correlation plots between mixing angles, effective Majorana mass, and sum of lightest neutrino masses.}
\label{fig:corr-neut}
\end{figure}
\begin{figure}
\centering
\includegraphics[scale=0.4]{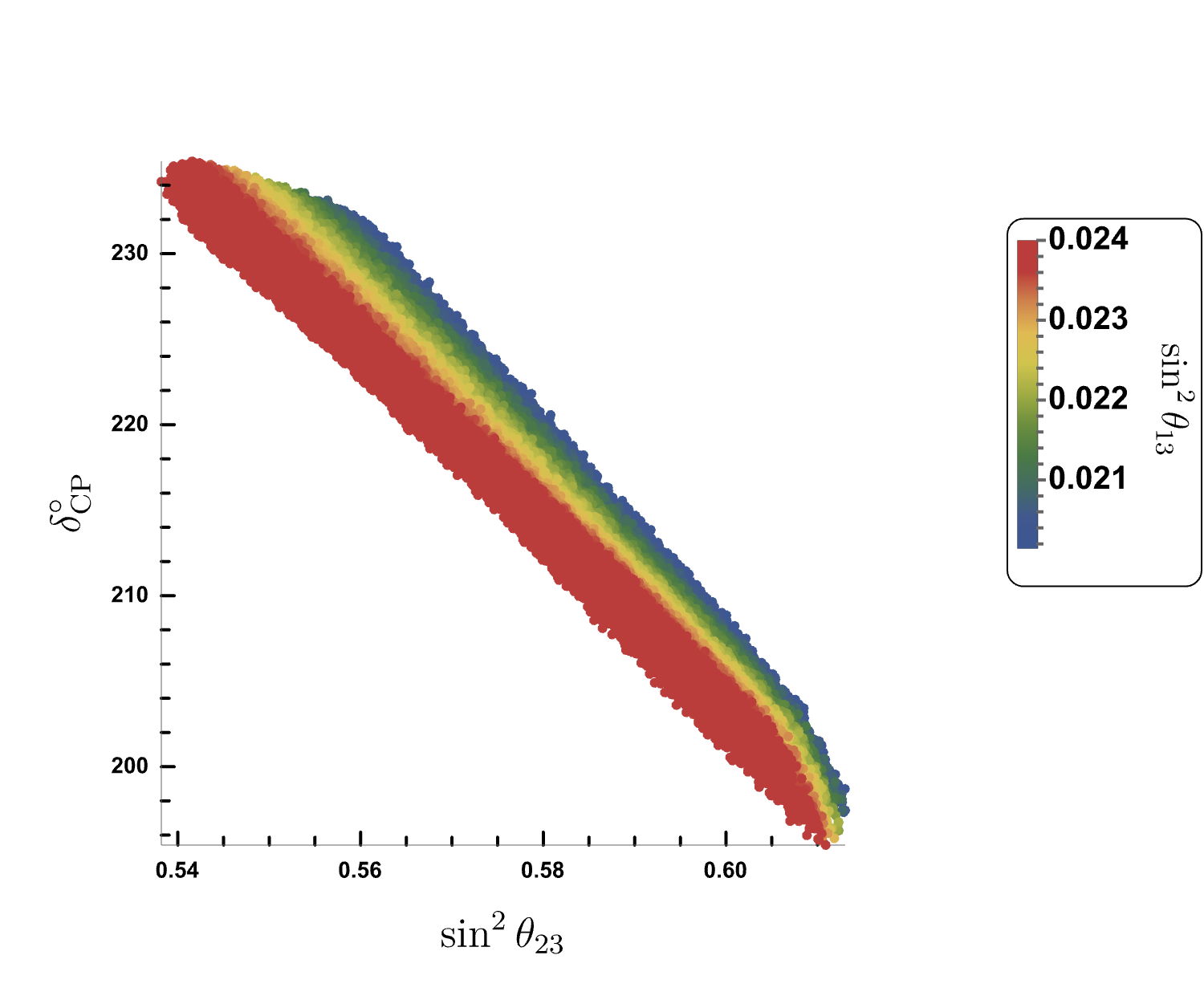}
\caption{Correlation plot between mixing angles and CP violation phase, for different values of $\sin^2\theta_{13}$.}
\label{fig:mixangles-CP}
\end{figure}

\begin{table}[tp]
\begin{tabular}{c|c|cccccc}
\toprule[0.13em] Observable & range & $\Delta m_{21}^{2}$ [$10^{-5}$eV$^{2}$]
& $\Delta m_{31}^{2}$ [$10^{-3}$eV$^{2}$] & $\sin^2\theta^{(l)}_{12}/10^{-1}$
& $\sin^2\theta^{(l)}_{13}/10^{-2}$ & $\sin^2\theta^{(l)}_{23}/10^{-1}$ & $%
\delta^{(l)}_{\text{CP}} (^{\circ })$ \\ \hline
Experimental & $1\sigma$ & $7.50_{-0.20}^{+0.22}$ & $2.55_{-0.03}^{+0.02}$ & 
$3.18\pm 0.16$ & $2.200_{-0.062}^{+0.069}$ & $5.74\pm 0.14$ & $%
194_{-22}^{+24}$ \\ 
Value~\cite{deSalas:2020pgw} & $3\sigma$ & $6.94-8.14$ & $2.47-2.63 $ & $2.71-3.69$ & $2.000-2.405$
& $4.34-6.10$ & $128-359$ \\ \hline
Experimental & $1\sigma$ & $7.49\pm 0.19$ & $2.513_{-0.019}^{+0.021}$ & $3.08_{-0.11}^{+0.12}$ & $2.215_{-0.056}^{+0.058}$ & $4.7_{-0.13}^{+0.17}$ & $212_{-41}^{+26}$ \\ 
Value~\cite{Esteban:2024eli} & $3\sigma$ & $6.92-8.05$ & $2.451-2.578 $ & $2.75-3.45$ & $2.03-2.388$
& $4.35-5.85$ & $124-364$ \\ \hline
Fit & $1\sigma-3\sigma$ & $7.69$ & $2.54$ & $3.41$ & $2.24$ & $5.73$ & $219.7$\\ \hline
\end{tabular}
\caption{Model predictions for the scenario of normal order (NO) neutrino mass.}
\label{Table:neutrinos_value}
\end{table}

To fit the parameters of the effective neutrino sector and successfully reproduce the experimental values of the neutrino mass-squared splittings, the leptonic mixing angles, and the leptonic Dirac CP phase, we minimize the following $\chi^2$ function:
\begin{eqnarray}
\chi ^{2} &= & \frac{\left( \Delta m_{21}^{2\ \exp }-\Delta m_{21}^{2\ \text{th}}\right) ^{2}}{\sigma
_{\Delta m^2_{21}}^{2}}+\frac{\left( \Delta m_{31}^{2\ \exp }-\Delta m_{31}^{2\ \text{th}}\right) ^{2}}{\sigma
_{\Delta m^2_{31}}^{2}}+
\frac{\left( \sin^2\theta_{12}^{(l)\exp }-\sin^2\theta
_{12}^{(l)\text{th}}\right) ^{2}}{\sigma _{\sin^2 \theta^{(l)}_{12}}^{2}}\label{ec:error-neu}\\
\nn & & + \frac{\left( \sin^2\theta_{23}^{(l)\exp}-\sin^2\theta_{23}^{(l)\text{th}}\right) ^{2}}{\sigma_{\sin^2\theta^{(l)}_{23}}^{2}}
 + \frac{\left( \sin^2\theta_{13}^{(l)\exp }-\sin^2\theta_{13}^{(l)\text{th}}\right) ^{2}}{\sigma_{\sin^2\theta^{(l)}_{13}}^{2}}
+\frac{\left( \delta_{\text{CP}}^{\exp }-\delta_{\text{CP}}^{\text{th}}\right) ^{2}}{\sigma _{\delta_{\text{CP}}}^{2}}\;,  \notag
\end{eqnarray}

where $\Delta m_{i1}^2$ (with $i= 2, 3)$ are the neutrino mass squared differences, $\sin\theta^{(l)}_{jk}$ is the sine function of the mixing angles (with $j,k=1,2,3$) and $\delta_{\text{CP}}$ is the CP violation phase. The supra indices represent the experimental (\enquote{exp}) and theoretical (\enquote{$\text{th}$}) values, and the $1\sigma$ are the experimental errors. By performing the numerical analysis of our model, randomly varying the magnitude of each free parameter between $[10^{-4}, 2.5]$ eV, while the phase was varied between $[0, 2\pi]$ rad, the $\chi^2$ function was minimized, obtaining the following value,
\begin{equation}
\chi^2= 0.497
\end{equation}

On the other hand, the value of our free parameters that minimize $\chi^2$, as given in Eq.~(\ref{Eq:NeutrMatParam}), which represent our best-fit point are
\begin{align}
A&=4.86\times 10^{-2}\ \text{eV} & B&= -7.03\times 10^{-3}\ \text{eV} & C&= -7.37\times 10^{-3}\ \text{eV} \notag\\
D&= 2.81\times 10^{-3}\ \text{eV} & \theta&= -1.94\ \text{rad}
\label{eq:bestPara}
\end{align}

After performing the fit of the effective parameters and obtaining the best-fit point, we obtained the values shown in Table \ref{Table:neutrinos_value}, alongside the experimental values of neutrino oscillation parameters within the $1\sigma$ and $3\sigma$ ranges, as reported in Refs.~\cite{deSalas:2020pgw, Esteban:2024eli}. In Table \ref{Table:neutrinos_value}, we see that the neutrino mass-squared differences ($\Delta m_{21}^2$, $\Delta m_{31}^2$) and the solar and reactor mixing angles ($\sin^2\theta_{12}^{(l)}$, $\sin^2\theta_{13}^{(l)}$) lie within the $1\sigma$ range. The atmospheric mixing angle ($\sin^2\theta_{23}^{(l)}$) and the leptonic Dirac CP-violating phase ($\delta_{\text{CP}}$) are within the $2\sigma$ range.

Fig.~\ref{fig:corr-neut}a shows the correlation between the neutrino mixing angles, where the green and pink background fringes represent the $1\sigma$ range of the experimental values and the intersection of the dotted lines represent our best-fit point for each observable. In Fig.~\ref{fig:mixangles-CP}, we see that for the mixing angles, we can get values in the $1\sigma$ range, while for the CP violating phase, we obtain values up to $3\sigma$, where each lepton sector observable is obtained in the following range of values: $0.279\leq \sin^2\theta^{(l)}_{12}\leq 0.366$, $0.538\leq \sin^2\theta^{(l)}_{23}\leq 0.613$.\\

In addition to the previously discussed observables from the neutrino sector, our model also predicts another observable, the effective Majorana neutrino mass parameter relevant for neutrinoless double beta decay, which serves as a probe of the Majorana nature of neutrinos. This effective mass parameter is defined as follows:
\begin{equation}
m_{ee}=\left| \sum_i \mathbf{U}_{ei}^2m_{\nu i}\right|\;,
\label{ec:mee}
\end{equation}
where $\mathbf{U}_{ei}$ and $m_{\nu i}$ are the matrix elements of the PMNS leptonic mixing matrix and the light active neutrino masses, respectively. From  Eq. \eqref{ec:mee}, we can see that the neutrinoless double beta ($0\nu\beta\beta$) decay amplitude is proportional to $m_{ee}$. Fig.~\ref{fig:corr-neut}b shows the correlation between the effective Majorana neutrino mass parameter $m_{ee}$ and the sum of the masses of the active neutrinos $\sum m_i$, where the neutrino sector model parameters were randomly generated in a range of values where the neutrino mass squared splittings and the mixing parameters are inside the $3\sigma$ experimentally allowed range, consistent with the above mentioned $\chi^2$ analysis. As
seen from Fig.~\ref{fig:corr-neut}b, our model predicts an effective Majorana neutrino mass parameter in the range $3.73\  \text{meV}\lesssim m_{ee}\lesssim 8.19\ \text{meV}$, while the star point in the figure represents the value of $m_{ee}$ corresponding to the best-fit point of the model, according to the values of the parameters of Eq.~\eqref{eq:bestPara}, whose value is $m_{ee}\simeq 5.38\ \text{meV}$ for the scenario of normal neutrino mass hierarchy. The current most stringent experimental upper bound on the effective Majorana neutrino mass parameter, i.e., $m_{ee}\leq 50\ \text{meV}$ arises from the KamLAND-Zen limit on the $^{136}X_e\; 0\nu\beta\beta$ decay half-life $T_{1/2}^{0\nu\beta\beta}(^{136}X_e) >2.0\times 10^{26}$ yr \citep{KamLAND-Zen:2022tow}. As for the sum of the neutrino masses $\sum m_i$, it can also be seen from Fig.~\ref{fig:corr-neut}b that the value lies in the $0.061-0.0715$ eV range, while of the value $\sum m_i$ for the best-fit point is $\sum m_i\simeq 6.54\times 10^{-2}\ \text{eV}$, well within the recent bounds from refs.~\cite{Jiang:2024viw,Naredo-Tuero:2024sgf}, $\sum m_{i(cosmo)} \lesssim 0.04–0.3$ eV.

\section{Scalar potential}
\label{scalarpotential}
\subsection{Scalar spectrum}
The scalar potential of the model invariant under the symmetries takes the form: 
\begin{eqnarray}\label{eq:potencialQ6}
V &=& -\mu_1^2 \left(H_1^{\dagger}H_1\right)
-\mu_2^2 \left(H_2^{\dagger}H_2\right) -\mu_3^2 \left(H_3^{\dagger}H_3\right) -\mu_{13}^2 \left(H_3^{\dagger}H_1+ H_1^{\dagger}H_3\right)-\mu_{23}^2\left(H_2^{\dagger}H_3 + H_3^{\dagger}H_2\right)-\mu_{12}^2 \left(H_1^{\dagger}H_2+ H_2^{\dagger}H_1\right) \notag\\
&&-\mu_4^2 \left(\sigma^*\sigma\right) 
-\mu_5^2 \left(\xi_1^*\xi_1\right)-\mu_6^2 \left(\xi_2^*\xi_2\right) -\mu_7^2 \left(\rho^*\rho\right)  +\lambda_1 \left(H^{\dagger}H\right)^2_{\mathbf{1_{++}}} +\lambda_2 \left(H_3^{\dagger}H_3\right)^2 +\lambda_3 \left(\sigma^*\sigma\right)^2 +\lambda_4 \left(\xi^{\dagger}\xi\right)^2_{\mathbf{1_{++}}} +\lambda_5 \left(\rho^*\rho\right)^2 \notag\\
&&+ \lambda_6 \left(H^{\dagger}H\right)_{\mathbf{1_{++}}}\left(H_3^{\dagger}H_3\right) + \lambda_7 \left(H^{\dagger}H\right)_{\mathbf{1_{++}}}\left(\sigma^*\sigma\right) + \lambda_8 \left(H^{\dagger}H\right)_{\mathbf{1_{++}}}\left(\rho^*\rho\right) + \lambda_9 \left(\xi^{\dagger}\xi\right)_{\mathbf{1_{++}}}\left(H_3^{\dagger}H_3\right)+ \lambda_{10}\left(\xi^{\dagger}\xi\right)_{\mathbf{1_{++}}}\left(\sigma^*\sigma\right) \notag\\
&&+ \lambda_{11} \left(\xi^{\dagger}\xi\right)_{\mathbf{1_{++}}}\left(\rho^*\rho\right) +\lambda_{12} \left(H^{\dagger}H\right)_{\mathbf{1_{+}}}\left(\xi^{\dagger}\xi\right)_{\mathbf{1_{++}}} +\lambda_{13}\left(H^{\dagger}H\right)_{\mathbf{1_{--}}}\left(\xi^{\dagger}\xi\right)_{\mathbf{1_{--}}} +\lambda_{14}\left(H^{\dagger}H\right)_{\mathbf{1_{--}}}\left(\sigma^*\rho\right)\notag\\
&&+\lambda_{15}\left(\xi^{\dagger}\xi\right)_{\mathbf{1_{--}}}\left(\sigma^*\rho\right)
+\lambda_{16} \left(H_3^{\dagger}H_3\right)\left(\sigma^*\sigma\right) +\lambda_{17} \left(H_3^{\dagger}H_3\right)\left(\rho^*\rho\right)+\lambda_{18} \left(\sigma^*\sigma\right)\left(\rho^*\rho\right) +h.c.
\end{eqnarray}
where $\lambda_6=\lambda_7=\lambda_8=\lambda_{12}=0$ as required by CP conservation. Having $\lambda_{12}$ real will yield mixing between CP even and CP odd scalar states. Here we include soft-breaking mass terms in order to keep consistency of the VEV configurations of the $Q_6$ scalar doublets with the scalar potential minimization conditions in the whole region of parameter space. 
The condition $\langle H_{1}^{0}\rangle =0$ imposes a constraint on the potential, which yields a nontrivial relation among the soft-breaking mass parameters. Specifically, we find:
\begin{equation}
\mu_{12}^2 = \lambda _{13} \cos (2 \theta ) v_{\xi }^2+\frac{1}{2} \lambda _{14}
   v_{\rho } v_{\sigma }-\frac{\mu _{13}^2 v_3}{v_2} ~.
\end{equation}

The minimization conditions of the scalar potential are given by:
\begin{eqnarray}
\mu_2^2 &=& -\frac{\mu _{23}^2 v_3}{v_2}\; , \\
\mu_3^2 &=& \frac{1}{2} \lambda _{16}v_{\sigma }^2+\lambda _2 v_3^2-i \lambda _9 \sin (2 \theta ) v_{\xi }^2-\frac{\mu _{23}^2 v_2}{v_3}\; ,\\
\mu_4^2 &=& \frac{1}{2} \left(\frac{\lambda _{15} \cos (2 \theta ) v_{\xi }^2
   v_{\rho }}{v_{\sigma }}-2 i \lambda _{10} \sin (2 \theta ) v_{\xi
   }^2+\lambda _{18} v_{\rho }^2+2 \lambda _3 v_{\sigma }^2+\lambda
   _{16} v_3^2\right)\; , \\
\mu_5^2 &=& \mu_6^2= \frac{\lambda_{15} v_{\rho}v_{\sigma}}{2 \cos(2\theta)}\; , \label{eq:mu5}\\
\mu_7^2 &=& \frac{1}{2} \left(\frac{\lambda _{15} \cos (2 \theta ) v_{\xi }^2
   v_{\sigma }}{v_{\rho }}-2 i \lambda _{11} \sin (2 \theta ) v_{\xi
   }^2+2 \lambda _5 v_{\rho }^2+\lambda _{18} v_{\sigma }^2\right).
\end{eqnarray}

The minimization conditions are derived in the standard manner by imposing that the first derivatives of the scalar potential with respect to all field VEVs vanish. 
However, as indicated in Eq. \eqref{eq:vev-xi}, the VEV of the field $\xi$ is, in general, complex. Consequently, the minimization of the scalar potential with respect to $\xi$ is not carried out directly in terms of its VEV, but rather with respect to its real and imaginary components, following the procedure discussed in Ref. \cite{Kuncinas:2023ycz}. Therefore, the minimization with respect to the field $\xi$ 
yields the following relations:
\begin{align}
\frac{\partial V}{\partial ({\text{Re}}\xi_1)} &= \frac{1}{2} v_{\xi } \left(\cos (\theta ) \left(\lambda _{15} v_{\rho } v_{\sigma
   }-2 \mu _5^2\right)-8 \lambda _4 \sin ^2(\theta ) \cos (\theta ) v_{\xi }^2-i
   \sin (\theta ) \left(\lambda _{11} v_{\rho }^2+\lambda _{10} v_{\sigma
   }^2+\lambda _9 v_3^2\right)\right)= 0 ~, \label{eq:xi1re} \\
\frac{\partial V}{\partial ({\text{Im}}\xi_1)} &=  -\frac{1}{2} v_{\xi } \left(\sin (\theta ) \left(2 \mu _5^2+\lambda _{15} v_{\rho }
   v_{\sigma }\right)+8 \lambda _4 \sin (\theta ) \cos ^2(\theta ) v_{\xi }^2+i \cos
   (\theta ) \left(\lambda _{11} v_{\rho }^2+\lambda _{10} v_{\sigma }^2+\lambda _9
   v_3^2\right)\right) = 0~, \label{eq:xi1im}\\[10pt]
\frac{\partial V}{\partial ({\text{Re}}\xi_2)} &=  \frac{1}{2} v_{\xi } \left(\cos (\theta ) \left(\lambda _{15} v_{\rho } v_{\sigma
   }-2 \mu _6^2\right)-8 \lambda _4 \sin ^2(\theta ) \cos (\theta ) v_{\xi }^2-i
   \sin (\theta ) \left(\lambda _{11} v_{\rho }^2+\lambda _{10} v_{\sigma
   }^2+\lambda _9 v_3^2\right)\right)= 0 ~, \label{eq:xi2re}\\
\frac{\partial V}{\partial ({\text{Im}}\xi_2)} &=  \frac{1}{2} v_{\xi } \left(\sin (\theta ) \left(2 \mu _6^2+\lambda _{15} v_{\rho }
   v_{\sigma }\right)+8 \lambda _4 \sin (\theta ) \cos ^2(\theta ) v_{\xi }^2+i \cos
   (\theta ) \left(\lambda _{11} v_{\rho }^2+\lambda _{10} v_{\sigma }^2+\lambda _9
   v_3^2\right)\right) = 0~. \label{eq:xi2im}
\end{align}
From the above given equations we find: 
combine Eqs. \eqref{eq:xi1re} with \eqref{eq:xi1im} and \eqref{eq:xi2re} with \eqref{eq:xi2im}, we obtain,
\begin{align}
\lambda _{15} v_{\rho } v_{\sigma }-2 \mu _5^2 \cos (2 \theta ) &= 0 ~,\notag\\
\lambda _{15} v_{\rho } v_{\sigma }-2 \mu _6^2 \cos (2 \theta ) &= 0 ~.
\end{align}
Then, it follows that $\mu_5$ and $\mu_6$ are equal, obtaining the relationship provided by 
Eq. \eqref{eq:mu5}.

Therefore, the squared scalar mass matrices of the CP-even neutral, CP-odd neutral and electrically charged fields are given by:
\begin{align}
&\mathbf{M}_{CP-\text{even}}^{2}=\begin{pmatrix}
A_{3\times 3} & B_{3\times 4} \\
B_{4\times 3}^T & C_{4\times 4}
\end{pmatrix},\qquad 
\mathbf{M}_{CP-\text{odd}}^{2}=\begin{pmatrix}
X_{3\times 3} & 0_{3\times 4} \\
0_{4\times 3} & Y_{4\times 4}
\end{pmatrix}, \label{eq:Meven_Modd}\\[5pt]
&\mathbf{M}_{\text{charged}}^{2} =
\begin{pmatrix}
-\mu_1^2 & \frac{\mu _{13}^2 v_3}{v_2} & -\mu _{13}^2 \\
 \frac{\mu _{13}^2 v_3}{v_2} & \frac{\mu _{23}^2 v_3}{v_2}
   & -\mu _{23}^2 \\
 -\mu _{13}^2 & -\mu _{23}^2 & \frac{\mu _{23}^2 v_2}{v_3}
\end{pmatrix}. \label{eq:Mch}
\end{align}
The submatrices of Eq.~\eqref{eq:Meven_Modd} are detailed in Appendix \ref{ap:scalar}.

\subsection{Low energy scalar mass spectrum}
As a first approximation, we analyze the low-energy scalar sector. A complete analysis is presented in the next section. Therefore, at low-energies, the squared mass matrices of the CP-even, CP-odd, and charged scalar sectors, transforming trivially under the $\tilde{Z}_2\times Z_2$ symmetry, are given by,
\begin{align}
\mathbf{M}_{CP-\text{even}}^{2} &= \begin{pmatrix}
-\mu_1^2 & \frac{\mu _{13}^2 v_3}{v_2} & -\mu _{13}^2 \\
 \frac{\mu _{13}^2 v_3}{v_2} & \frac{\mu _{23}^2 v_3}{v_2} & -\mu
   _{23}^2 \\
 -\mu _{13}^2 & -\mu _{23}^2 & 2 \lambda _2 v_3^2+\frac{\mu _{23}^2
   v_2}{v_3}
\end{pmatrix}  \\
\mathbf{M}_{CP-\text{odd}}^{2} &= \begin{pmatrix}
-2 \lambda _1 v_2^2- \mu_1^2 & \frac{\mu _{13}^2 v_3}{v_2} & -\mu _{13}^2 \\
 \frac{\mu _{13}^2 v_3}{v_2} & \frac{\mu _{23}^2 v_3}{v_2} & -\mu
   _{23}^2 \\
 -\mu _{13}^2 & -\mu _{23}^2 & \frac{\mu _{23}^2 v_2}{v_3}
\end{pmatrix}
\\
\mathbf{M}_{CP-\text{charged}}^{2} &= \begin{pmatrix}
-\mu_1^2 & \frac{\mu _{13}^2 v_3}{v_2} & -\mu _{13}^2 \\
 \frac{\mu _{13}^2 v_3}{v_2} & \frac{\mu _{23}^2 v_3}{v_2}
   & -\mu _{23}^2 \\
 -\mu _{13}^2 & -\mu _{23}^2 & \frac{\mu _{23}^2 v_2}{v_3}
\end{pmatrix}
\end{align}
From the squared scalar mass matrices given above, we find that the physical $\tilde{Z}_2\times Z_2$ even low energy scalar mass spectrum is composed of three massive CP even neutral scalars, two CP odd scalars and two electrically charged scalar fields. Out of the three CP even scalar states, one corresponds to the $125$ GeV SM like Higgs boson, whereas the remaining two are non SM scalar fields having masses at the subTeV scale. Furthermore, we have one massless CP odd neutral scalar state as well as an electrically charged scalar field, which correspond to the SM Goldstone bosons associated with the longitudinal components of the $Z$ and $W$ gauge bosons. 

\begin{figure}
\centering
\subfloat[]{\includegraphics[scale=0.3]{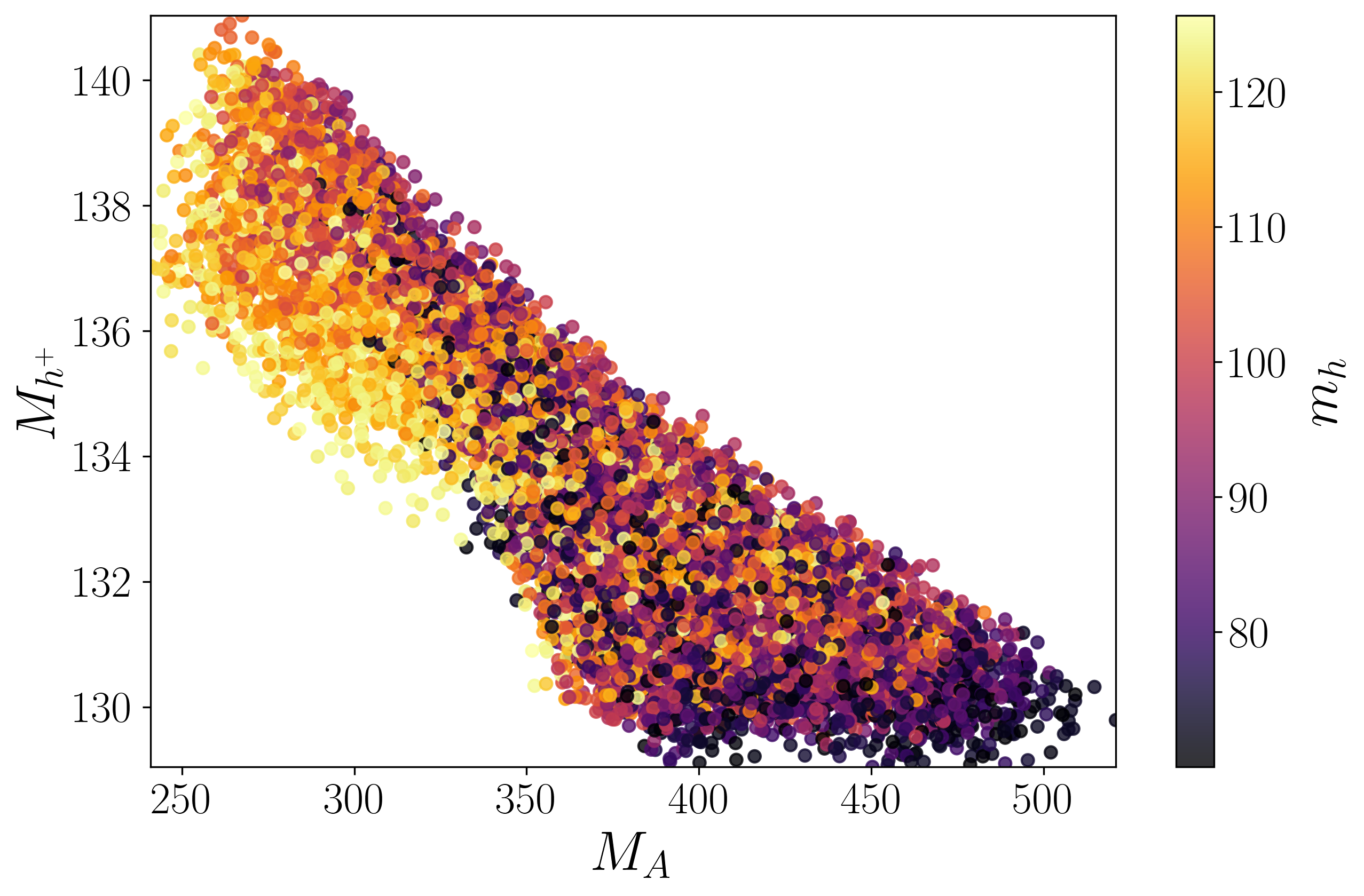}}\quad
\subfloat[]{\includegraphics[scale=0.3]{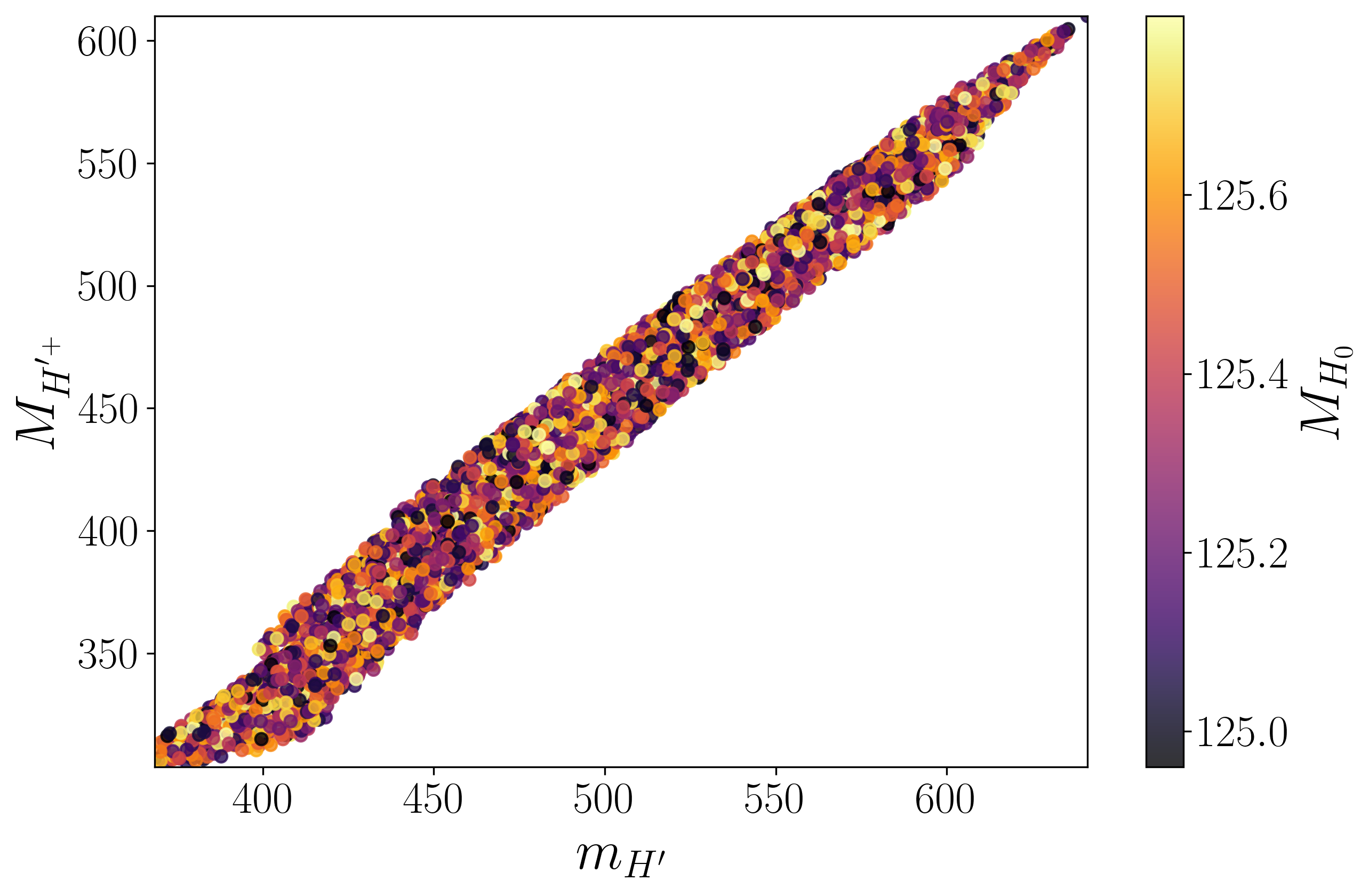}}\quad
\subfloat[]{\includegraphics[scale=0.3]{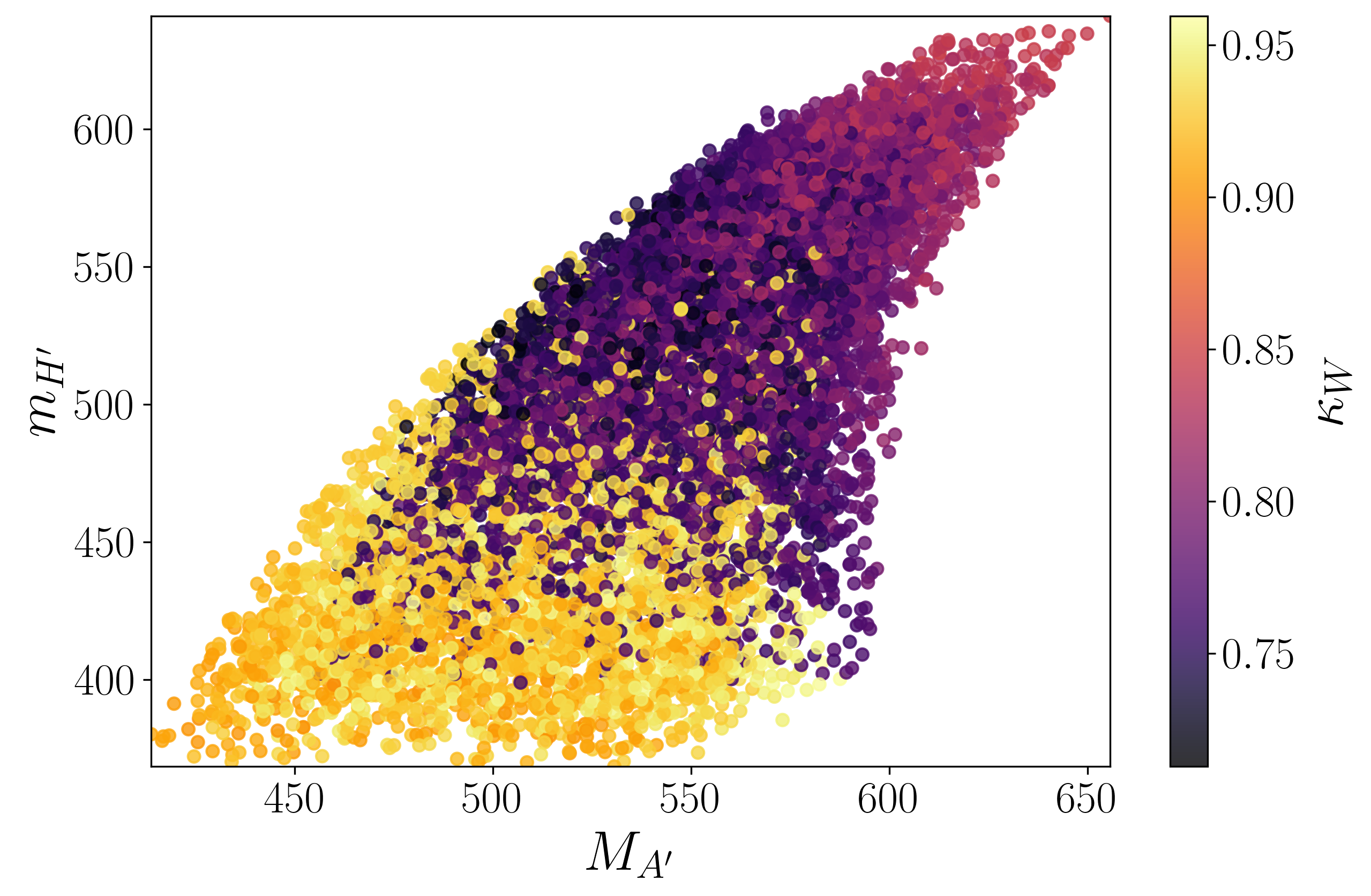}}\quad
\subfloat[]{\includegraphics[scale=0.3]{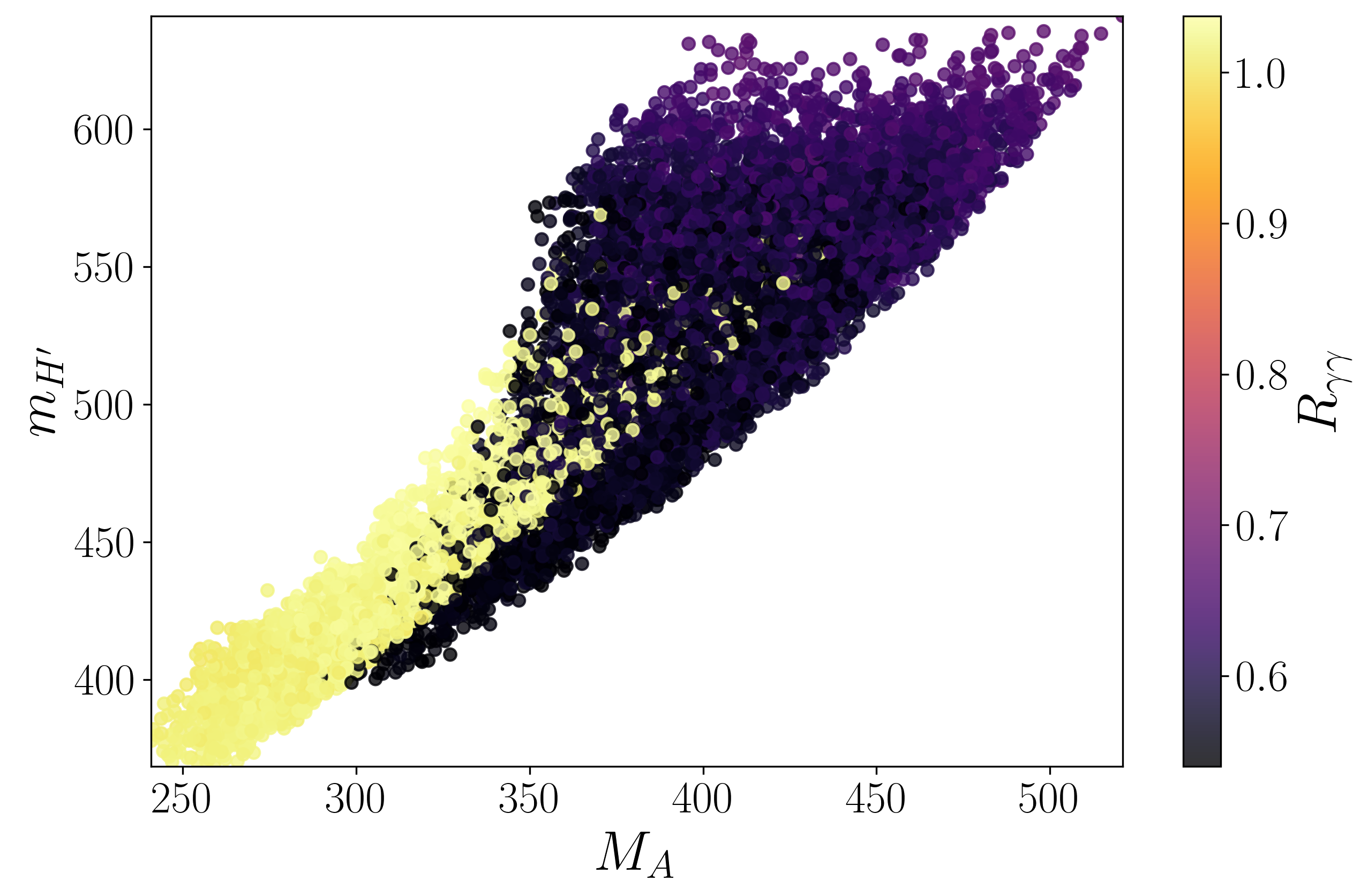}}
\caption{Correlation plots between the masses of the scalar sector, $R_{\gamma\gamma}$ and $\kappa_W$, considering low energies, considering $\mu_1^2 =0$.}
\label{fig:corr_scalar}
\end{figure}

\begin{figure}
\centering
\subfloat[]{\includegraphics[scale=0.3]{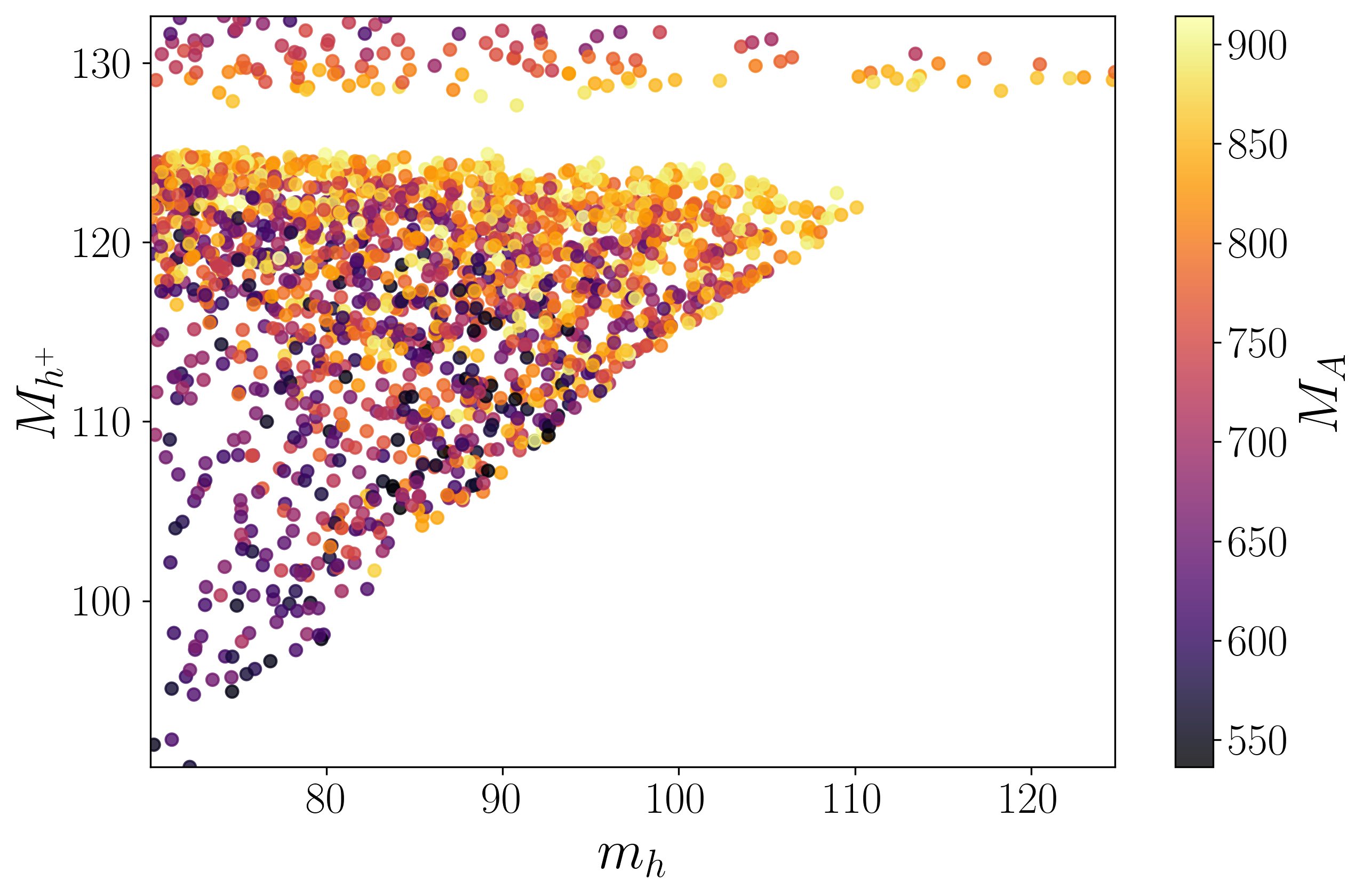}}\quad
\subfloat[]{\includegraphics[scale=0.3]{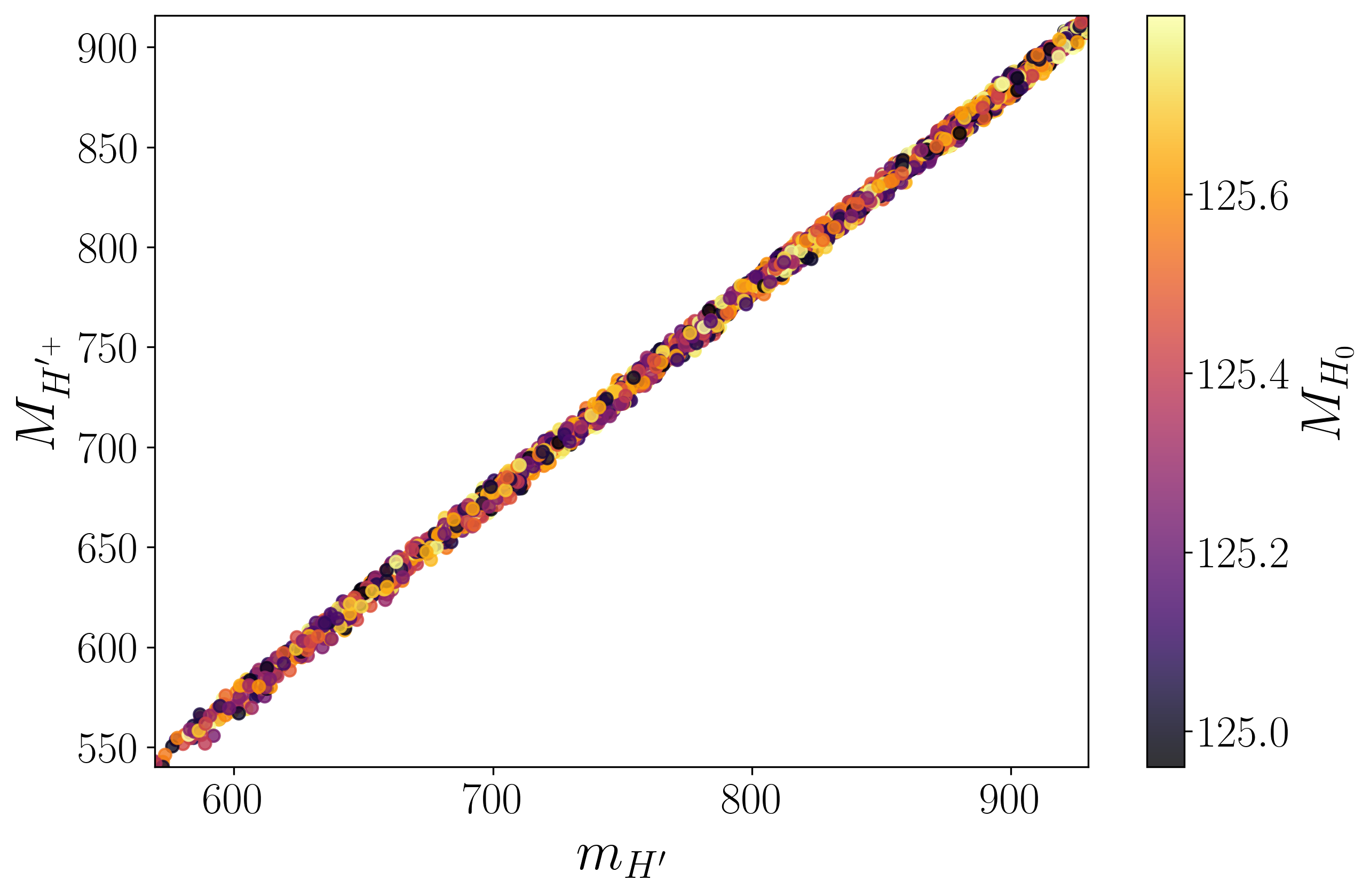}}\quad
\subfloat[]{\includegraphics[scale=0.3]{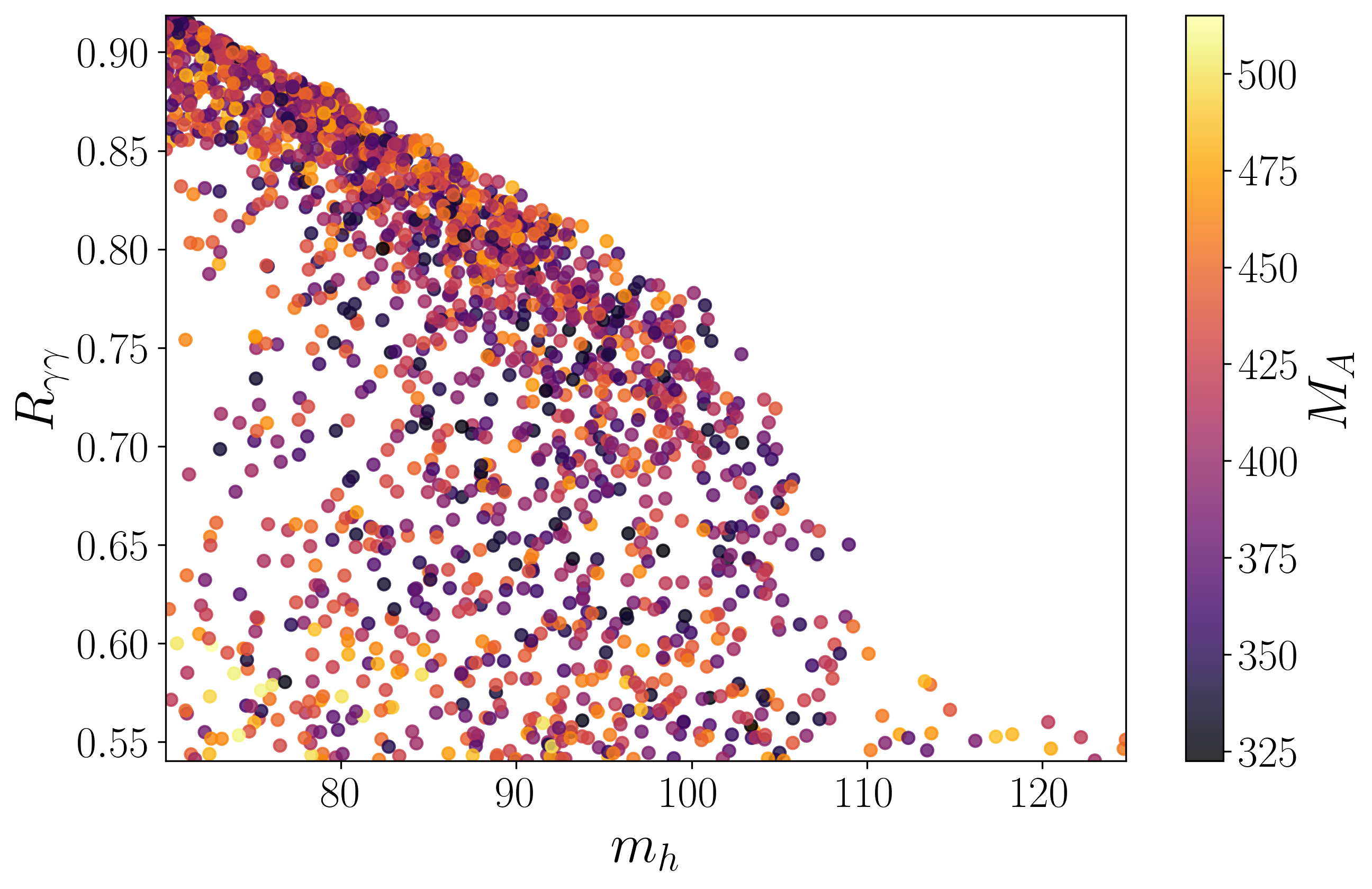}}\quad
\subfloat[]{\includegraphics[scale=0.3]{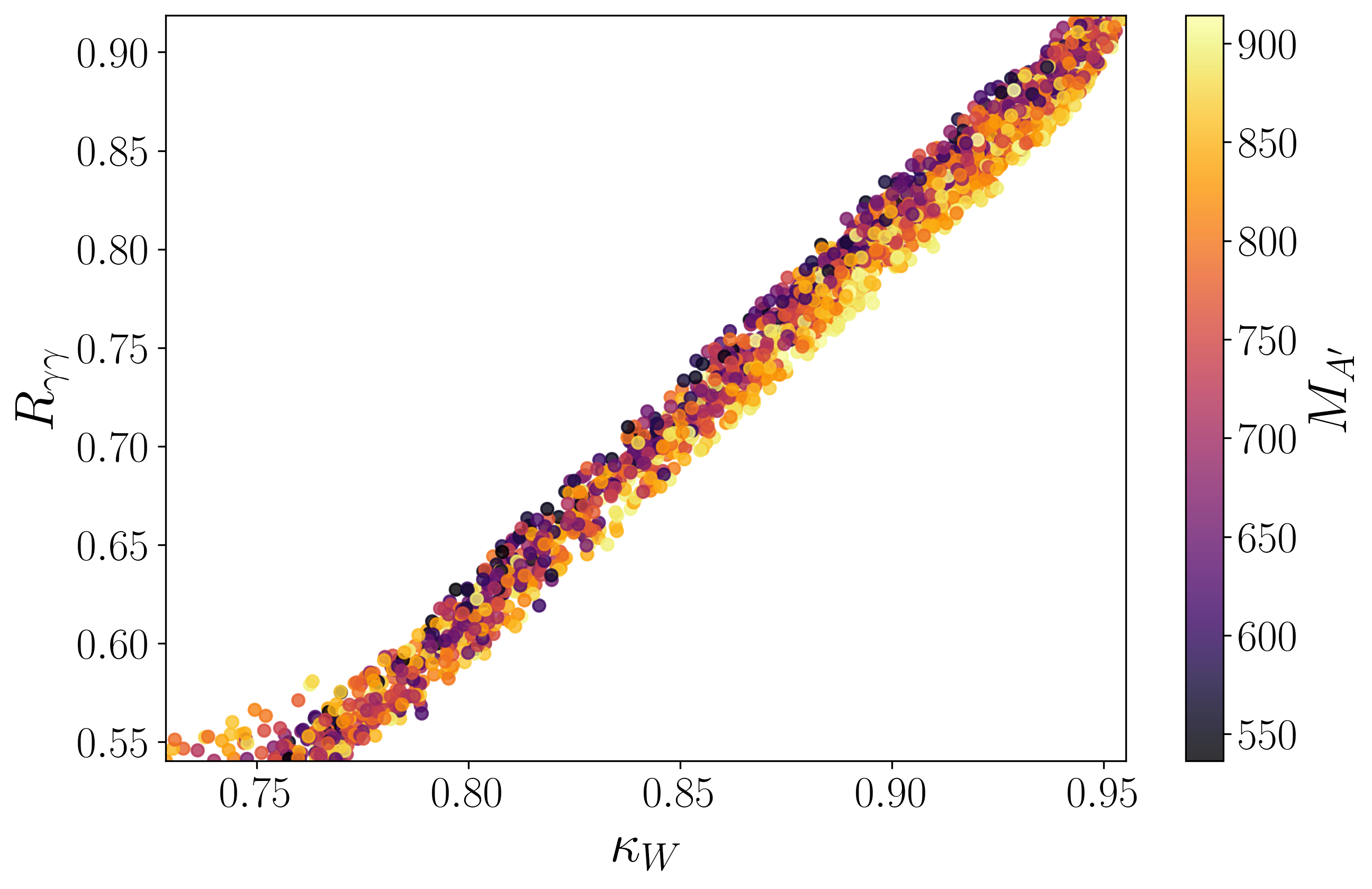}}
\caption{Correlation plots between the masses of the scalar sector, $R_{\gamma\gamma}$ and $\kappa_W$, considering low energies, considering $\mu_1^2 \not =0$.}
\label{fig:corr_scalar_mu}
\end{figure}

Fig.s \ref{fig:corr_scalar} and \ref{fig:corr_scalar_mu} show different correlations between the scalar sector masses and the $R_{\gamma\gamma}$ and $\kappa_W$ observables, considering two particular benchmarks 
corresponding to $\mu_1^2=0$ and $\mu_1^2\ \not=0$. It is worth mentioning that $R_{\gamma\gamma}$ is the Higgs diphoton signal strength where $\kappa_W$ parameterizes the deviation of the $125$ GeV Higgs boson's coupling to $W$ bosons from the Standard Model value. For the scalar sector masses and considering $\mu_1^2=0$, the light non SM CP-even scalar values are obtained in the range $70\ \text{GeV} \lesssim m_h\lesssim 125.7\ \text{GeV}$, with a central value of $m_h\simeq 119.8\ \text{GeV}$, while for the benchmark $\mu_1^2 \not =0$, we obtain, $70.02\ \text{GeV} \lesssim m_h\lesssim 124.7\ \text{GeV}$, with a central value of $m_h\simeq 86.1.8\ \text{GeV}$. In section \ref{DM}, a different value is obtained because it considers a complete analysis of the scalar potential, allowing for a higher value for this mass. Looking at the remaining two CP-even scalar masses, 
we obtain values in the ranges $124.96\ \text{GeV} \lesssim m_{H_0}\lesssim 125.7\ \text{GeV}$ and $368.4\ \text{GeV} \lesssim m_{H'}\lesssim 641.2\ \text{GeV}$, whereas for our second benchmark, we obtain, $124.96\ \text{GeV} \lesssim m_{H_0}\lesssim 125.8\ \text{GeV}$ and $569.4\ \text{GeV} \lesssim m_{H'}\lesssim 929.8\ \text{GeV}$. For the case of the CP-odd scalar sector, we get 
the following ranges of values for each mass: $241\ \text{GeV}\lesssim M_A\lesssim 521\ \text{GeV}$ and $413.8\ \text{GeV}\lesssim M_{A'}\lesssim 655.7\ \text{GeV}$ and for the case of $\mu_1^2 \not =0$, $322.5\ \text{GeV}\lesssim M_A\lesssim 515\ \text{GeV}$ and $536.4\ \text{GeV}\lesssim M_{A'}\lesssim 914\ \text{GeV}$, 
whereas for the charged scalar masses we find $129\ \text{GeV}\lesssim M_{h^+}\lesssim 141 \ \text{GeV}$ and $303.7\ \text{GeV}\lesssim M_{H^{'+}}\lesssim 610\ \text{GeV}$, but when considering the benchmark in Fig. \ref{fig:corr_scalar_mu}, we find $90.7\ \text{GeV}\lesssim M_{h^+}\lesssim 132.6 \ \text{GeV}$ and $540.1\ \text{GeV}\lesssim M_{H^{'+}}\lesssim 915.8\ \text{GeV}$. Furthermore, we can also observe that we get values for $R_{\gamma\gamma}$ and $\kappa_W$ compatible with the corresponding experimental bounds for both benchmarks \cite{ATLAS:2022vkf,CMS:2022dwd,ParticleDataGroup:2024cfk}, obtaining in this approximation $0.540\lesssim R_{\gamma\gamma}\lesssim 1.04$ ($\mu_1^2= 0$), $0.540\lesssim R_{\gamma\gamma}\lesssim 0.918$ ($\mu_1^2 \not= 0$), $0.713\lesssim\kappa_W\lesssim 0.959$ ($\mu_1^2= 0$) and $0.728\lesssim\kappa_W\lesssim 0.955$ ($\mu_1^2\not= 0$).

\subsection{Quasialignment limit}
As shown in Section \ref{quarmassesandmixings}, where $v_1 = 0$, we can achieve the alignment limit in a 2HDM in a general way according to ~\cite{Das:2019yad}. 
Let us perform a rotation from the basis of the interaction states $\Psi_i$ to an intermediate basis formed by the states $h_i$ ($i=2,3$) through an orthogonal rotation~\cite{Das:2019yad}.
\begin{equation}
\begin{pmatrix}
h_2\\
h_3
\end{pmatrix}= \mathcal{O}_{\Psi} \begin{pmatrix}
\Psi_2\\
\Psi_3
\end{pmatrix}=\begin{pmatrix}
\cos\theta_1 & \sin\theta_1 \\
-\sin\theta_1 & \cos\theta_1 \\
\end{pmatrix}\begin{pmatrix}
\Psi_2\\
\Psi_3
\end{pmatrix}.
\end{equation}

The physical basis $(h,H_0)$ can be obtained using another orthogonal rotation,:
\begin{equation}
\mathcal{O}_r= \begin{pmatrix}
\cos\theta_2 & \sin\theta_2 \\
-\sin\theta_2 & \cos\theta_2 \\
\end{pmatrix}.\label{eq:rotationCPeven}
\end{equation}

Therefore,
\begin{equation}
\begin{pmatrix}
h\\
H_0
\end{pmatrix}= \mathcal{O}_r \begin{pmatrix}
\Psi_2\\
\Psi_3
\end{pmatrix}=\mathcal{O}_r \mathcal{O}_{\Psi}^T\begin{pmatrix}
h_2\\
h_3
\end{pmatrix} ~.
\end{equation}

The alignment boundary will be when $h_2$ overlaps with $h$, i.e. $\mathcal{O}_{11}=1$, with,
\begin{equation}
\mathcal{O}=\mathcal{O}_r \mathcal{O}_{\Psi}^T = \begin{pmatrix}
\cos \left(\theta_2- \theta_1\right) & \sin \left(\theta_2- \theta_1\right)\\
-\sin \left(\theta_2- \theta_1\right) & \cos\left(\theta_2- \theta_1\right)
\end{pmatrix}~.
\end{equation}

The alignment limit will be given by the conditions on  the quartic couplings of the potential that reduce them to the SM Higgs coupling,  plus small deviations \cite{Das:2019yad}, so we  first look at the mass matrix of the CP-even sector to low-energy, which we can diagonalize with the rotation matrix~\eqref{eq:rotationCPeven}:
\begin{equation}
\mathcal{O}_r\cdot m_{\text{CP-even}}^2\cdot \mathcal{O}_r^T=\begin{pmatrix}
m_h^2 & 0 \\
0 & m_{H_0}^2 \\
\end{pmatrix} ~.
\end{equation}
Inverting the relationship, we obtain
\begin{equation}
m_{\text{CP-even}}^2 = \mathcal{O}_r^T\cdot \begin{pmatrix}
m_h^2 & 0 \\
0 & m_{H_0}^2 \\
\end{pmatrix}\cdot\mathcal{O}_r
\end{equation}
where,
\begin{equation}
m_{\text{CP-even}}^2 = \begin{pmatrix}
\frac{\mu _{23}^2 v_3}{2 v_2} &
   -\frac{\mu _{23}^2}{2} \\
 -\frac{\mu _{23}^2}{2} & \lambda _2
   v_3^2+\frac{\mu _{23}^2 v_2}{2 v_3}
\end{pmatrix}
\end{equation}
and  can get the following results: 
\begin{equation}
\lambda_2= \frac{2 v_3 \left(m_h^2 \sin
   ^2\left(\theta _2\right)+\cos
   ^2\left(\theta _2\right)
   m_H^2\right)-\mu _{23}^2 v_2}{2
   v_3^3} ~,
\end{equation}
so that when $\theta_1 = \theta_2$ we recover exactly the coupling for the SM Higgs boson.

\section{Dark matter phenomenology}
\label{DM}
\subsection{Dark matter sector}
\label{DM1}
In the dark sector, the scalar potential contains the following terms:

\begin{eqnarray}
	V &\supset& \mu_{8}^2 H_4^\dagger H_4 + \mu_{9}^2 \varphi_1^* \varphi_1 + \mu_{10}^2 \varphi_2^* \varphi_2 
	            + \mu_{\textrm{SB}}^2 ( \varphi_1^2 + \textrm{h.c.} ) + 
	            \kappa_{1} (\varphi_1^* \varphi_1)^2 +  \kappa_{2} (\varphi_2^* \varphi_2)^2 +  \kappa_{3} (\varphi_1^* \varphi_1)(\varphi_2^* \varphi_2) \nonumber \\ 
	&& + \left[ \kappa_{4}(\varphi_1^2 \varphi_2^2) + \kappa_{5}(H_4^\dagger H_3)(\varphi_1^* \varphi_2) 
	   + \kappa_{6}(H_3^\dagger H_4)(\varphi_1 \varphi_2)  + \textrm{h.c.}  \right] \\
	&& + \sum_{i=1}^{2} (\varphi_i^* \varphi_i) \left[ \kappa_{6+i} ( H_4^\dagger H_4 )
	     + \kappa_{8+i} ( H_3^\dagger H_3 ) + \kappa_{10+i} ( H^\dagger H )_{1++} 
	 \right] \nonumber \\
	&& + \kappa_{13} (H_4^\dagger H_4) ( H^\dagger H )_{1++} +  \kappa_{14} (H_4^\dagger H_4) ( H_3^\dagger H_3 ), \nonumber
\end{eqnarray}
note the soft symmetry breaking term driven by the parameter $\mu_{\textrm{SB}}$, which induces a mass gap between the components of the field $\varphi_1$ in order to have non-zero neutrino masses.
As before, we consider the VEV alignment $v_1=0$ and define $\tan\beta = v_2/v_3$. 
We keep assuming the masses of the components of the scalar doublet $H_4$ to be greater than the right-handed neutrino masses, so that the lightest of these is a DM candidate. The other DM candidates are one component of $\varphi_{1}$ and one of $\varphi_{2}$, which we denote by $\phi_{1}$ and $\phi_{2}$ respectively. We analyze the DM phenomenology in the region of masses of the DM candidates where the standard cold DM freeze-out scenario describes the DM abundance.

Concerning direct detection (DD), the scattering amplitudes of right-handed neutrinos off nucleons vanish at the leading order, so that this DM candidate is out of the reach of current DD experiments and we analyze only the constraints on the scalar DM candidates in this respect.

\subsection{Numerical results}
\label{DM2}
We implement the model in 
\texttt{SARAH}~\cite{Staub:2013tta,Staub:2009bi,Staub:2010jh,Staub:2012pb},
for which we first find the analytical expressions for the left and right mixing matrices of charged leptons and quarks following a similar procedure to the outlined previously. This in order to write the Yukawa lagrangian in the mass eigenstate basis. To simplify the calculations, we neglect off-diagonal terms and also the masses of the first and second generation of fermions. No other simplifications are made in the implementation,
from which we generate corresponding model files for some of the other tools using the
\texttt{SARAH-\allowbreak SPheno} framework~\cite{Staub:2015kfa,Porod:2003um,Porod:2011nf}.

The theoretical and experimental constraints are divided into two categories: hard cuts and likelihoods.
When testing a given point of parameter space, for positivity
and stability of the scalar potential 
we
employ the public tool \texttt{EVADE} \cite%
{Hollik:2018wrr,Ferreira:2019iqb},
which features the minimization of the scalar potential through 
polynomial homotopy continuation \cite{Maniatis:2012ex}, and an estimation of the decay rate of
a false vacuum \cite{Coleman:1977py,Callan:1977pt}.
Tree level large energy LQT~\cite{Lee:1977eg} unitarity conditions over the quartic couplings and conditions at finite energy $\sqrt{s}$  over the trilinear scalar couplings \cite{Goodsell:2018fex,Krauss:2018orw} are calculated numerically with \texttt{SPheno}.
Exclusion limits from scalar searches at Tevatron, LEP
and the LHC are implemented with the aid of \texttt{HiggsTools/HiggsPredictions/HiggsBounds} \cite%
{Bahl:2022igd,Bechtle:2020pkv}. 
To generate the input needed by \texttt{HiggsTools} we employ the 
\texttt{CalcHEP/Micromegas} \cite%
{Belyaev:2012qa,Alguero:2023zol}
framework.

\begin{figure}[!h]
	\includegraphics[width=8.5cm, height=7cm]{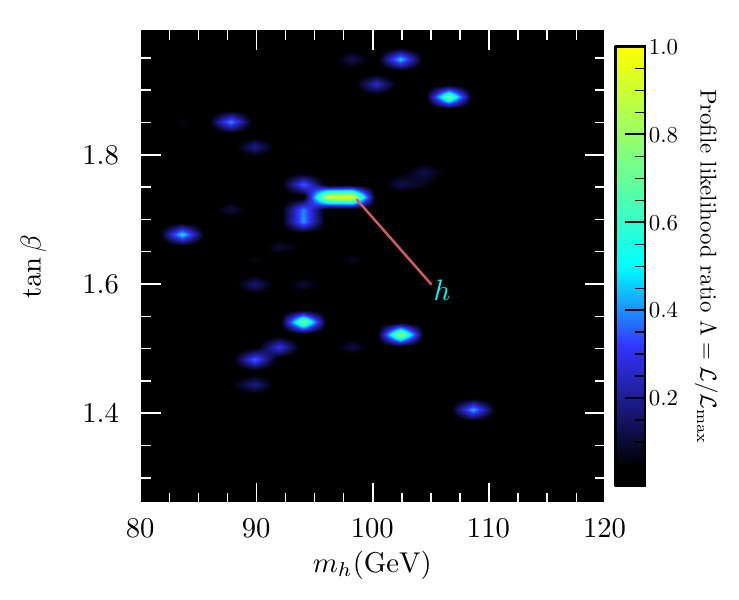}\newline
	\includegraphics[width=8.5cm, height=7cm]{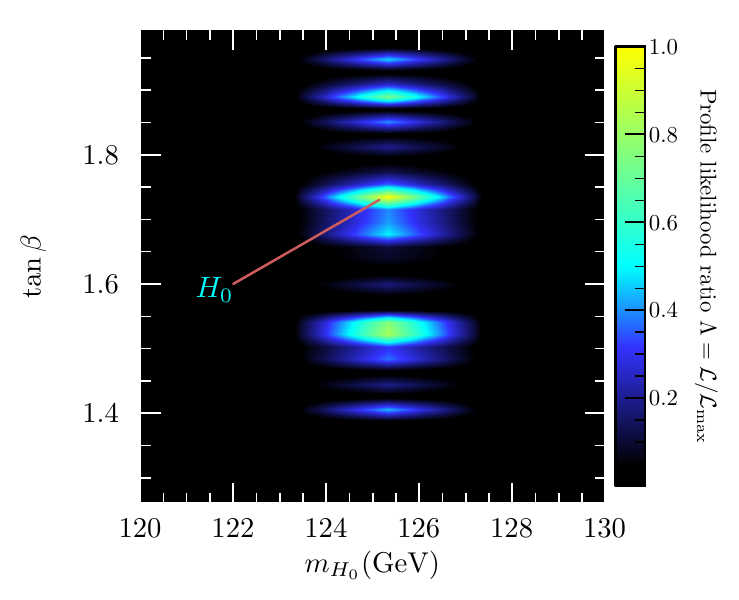}\newline
	\includegraphics[width=8.5cm, height=7cm]{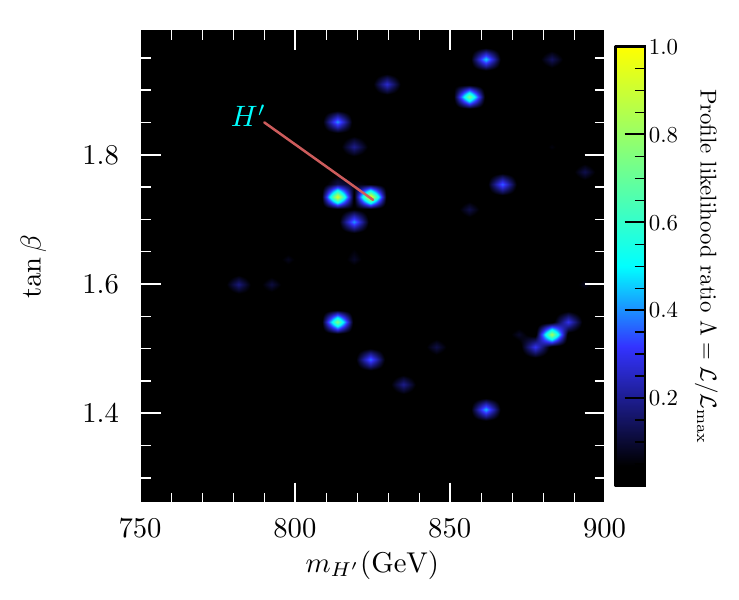}\newline
	\caption{Mass spectra of CP-even Higgs scalars. The best fit point (BFP) is signaled by the respective tags with masses $(m_h,m_{H_0},m_{H^\prime})=(98.7, 125.1, 825)$ GeV and $\tan\beta=1.73$.}
    \label{scalar_spectra}
\end{figure}

\begin{figure}[!h]
	\includegraphics[width=8.5cm, height=7cm]{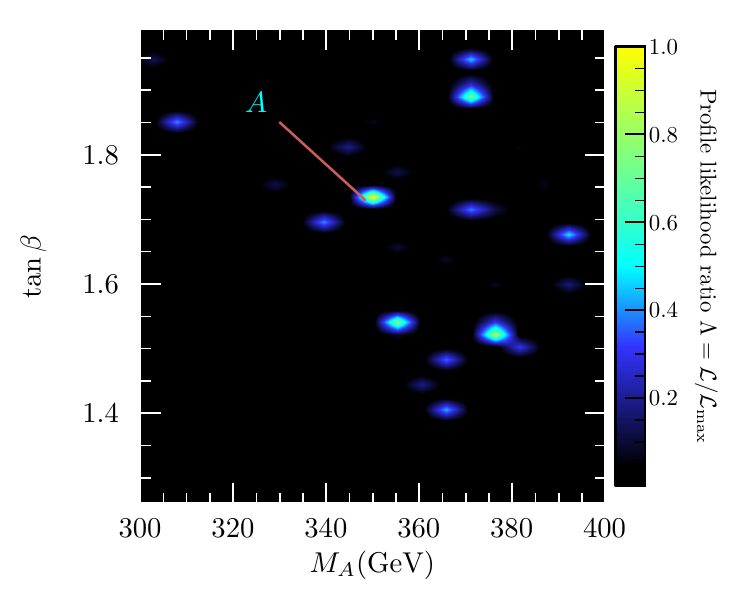}\includegraphics[width=8.5cm, height=7cm]{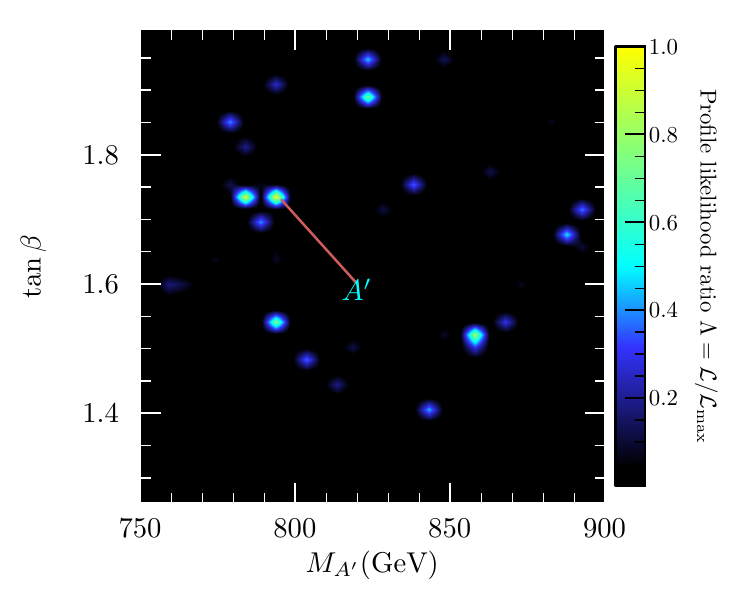}\newline
	\includegraphics[width=8.5cm, height=7cm]{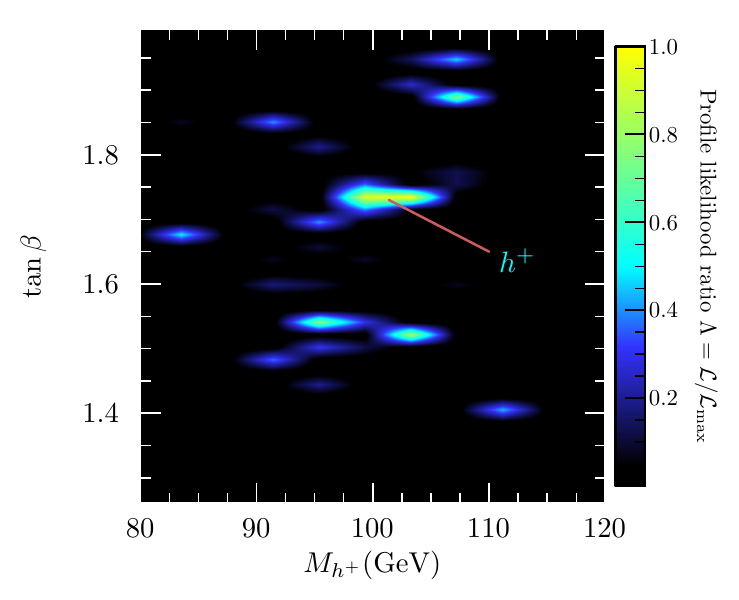}\includegraphics[width=8.5cm, height=7cm]{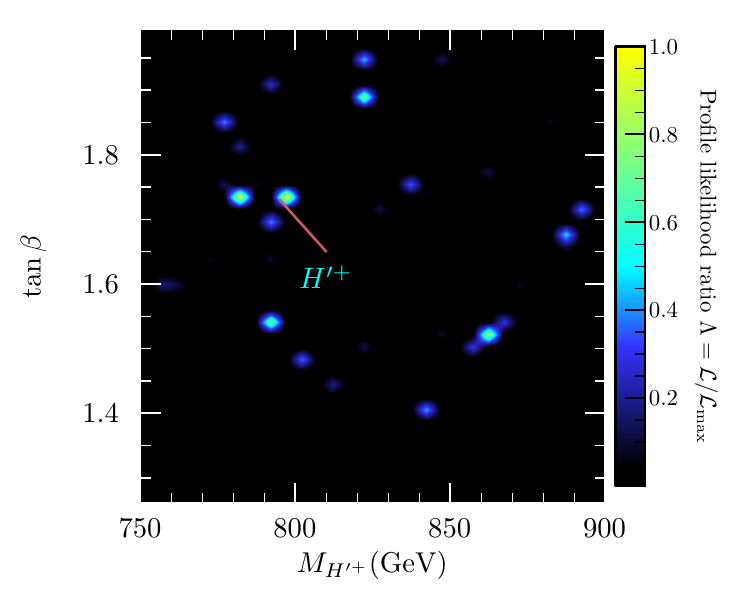}
	\caption{Mass spectra of CP-odd Higgs scalars (top panel) and the corresponding spectra for the charged Higgses (bottom panel). The best fit point (BFP) is signaled by the respective tags with masses $(m_A,m_{A^\prime},m_{h^+},m_{H^{\prime +}})=(348.3, 795.6, 101.4,795)$ GeV and $\tan\beta=1.73$.}
    \label{scalar_spectra_2}
\end{figure}

We impose hard cuts discarding points not complying with the above constraints.
For points not filtered by the previous hard cuts we calculate numerically the model predicted observables
that are used to construct a composite likelihood function. 
We calculate the couplings and decay branching ratios of the scalars with the help of the \texttt{HiggsTools/HiggsPredictions} code. 
We use the above predictions of the model to construct the composite likelihood function:
\begin{equation}
	\label{likeScalar}
	\log {\mathcal L}_{\text{scalar}} = \log {\mathcal L}_{\text{Higgs}} + \log {\mathcal L}_{H_0\rightarrow\gamma\gamma}
\end{equation}
The likelihood $\log {\mathcal L}_{H_0\rightarrow\gamma\gamma}$ regarding the branching ratio of the $125$ GeV SM-like Higgs into two photons
is constructed using the experimental value~\cite{ParticleDataGroup:2024cfk}:
\begin{equation}
	\text{BR}^\text{exp}_{h\rightarrow\gamma\gamma} = (2.5\pm 0.20)\times 10^{-3}
\end{equation}
to construct a simple chi-square function
$-2 \log \left( {\mathcal L}_{H_0\rightarrow\gamma\gamma} / {\mathcal L}^{\text{max}}_{H_0\rightarrow\gamma\gamma}  \right) = \chi^2_{H_0\rightarrow\gamma\gamma}$.
The likelihood $\log {\mathcal L}_{\text{Higgs}}$ that measures how well the couplings of $H_0$ resemble that of the already discovered SM Higgs is computed  through the equation:

\begin{equation}
	-2\log \left( {\mathcal L}_{\text{Higgs}} /  {\mathcal L}^{\text{max}}_{\text{Higgs}}\right) = 
	\chi^2_{\text{Higgs}}
\end{equation}
where $\chi^2_{\text{Higgs}}$ is constructed to minimize the quantity:
\begin{equation}
	\left| \chi^2_{\text{SM}} - \chi^2_{\text{Q6}} \right| ~,
\end{equation}
here $\chi^2_{\text{SM}}$ refers to the total chi-square of the LHC rate measurements of the observed Higgs boson
while $\chi^2_{\text{Q6}}$ is the prediction of the model under study here, both of these quantities are calculated with
\texttt{HiggsTools/HiggsPredictions/HiggsSignals}~\cite{Bechtle:2020uwn}.
In this manner, the scan of the parameter space yields model predictions that are ensured to
be contained mostly on an interval close to the SM prediction which is well in agreement with 
the LHC measurements.
Fig. \ref{scalar_spectra} shows the result of the numerical scan concerning the mass spectra of the CP-even scalars, where $\cal{L}$
is defined below.
The corresponding mass spectra for the pseudo-scalars and the charged scalars is shown in Fig. \ref{scalar_spectra_2}.
From the numerical analysis we are able to find a relatively small region of parameter space where the model correctly predicts a SM-like Higgs satisfying all the aforementioned constraints. The mass spectra resulting from these findings contains one light CP-even scalar of mass $m_h \approx 98$ GeV and one light charged scalar of mass $m_{h^+} \approx 101$ GeV. The rest of the scalars are heavier than $\sim 340$ GeV but up to $825$ GeV.

To proceed with the DM sector, we construct a log-likelihood function involving the observables in the (visible) scalar sector
and the DD and relic abundance observables:
\begin{equation}
	\log {\mathcal L} = \log {\mathcal L}_{\text{scalar}} + \log {\mathcal L}_{\text{DD}} + \log {\mathcal L}_{\Omega h^2}
	\label{likeTot} ~.
\end{equation}
For the numerical calculation of the relic density, as well as the DM-nucleon scattering cross sections, we use the capabilities of \texttt{Micromegas} \cite%
{Belanger:2013oya,Belanger:2014vza,Barducci:2016pcb,Belanger:2018ccd}. We construct ${\mathcal L}_{\Omega h^2}$ as a basic Gaussian likelihood with respect
to the PLANCK \cite%
{Planck:2018vyg} measured value, while the likelihood ${\mathcal L}_{\text{DD}}$ involves publicly available data from
the direct detection experiment LZ \cite{LZ:2022lsv}. We use
the numerical tool \texttt{DDCalc}
to compute the Poisson likelihood given by
\begin{equation}
	\mathcal{L}_\text{DD} = \frac{(b+s)^o \exp{\{-(b+s)\}}}{o!} ~,
\end{equation}
where $o$ is the number of observed events in the detector and $b$ is the
expected background count. From the model's predicted DM-nucleon scattering cross
sections as input, \texttt{DDCalc} computes the number
of expected signal events $s$ for given DM local halo and velocity
distribution models (we take the tool's default ones, for specific details
on the implementation such as simulation of the detector efficiencies and
acceptance rates, possible binning etc., see \cite%
{GAMBITDarkMatterWorkgroup:2017fax,GAMBIT:2018eea}).
Finally, we perform the scan of the
parameter space and construct the likelihood profiles using \texttt{Diver} \cite{Martinez:2017lzg,DarkMachinesHighDimensionalSamplingGroup:2021wkt,Scott:2012qh} (in standalone
mode).

\begin{figure}[!h]
	\includegraphics[width=8.5cm, height=7cm]{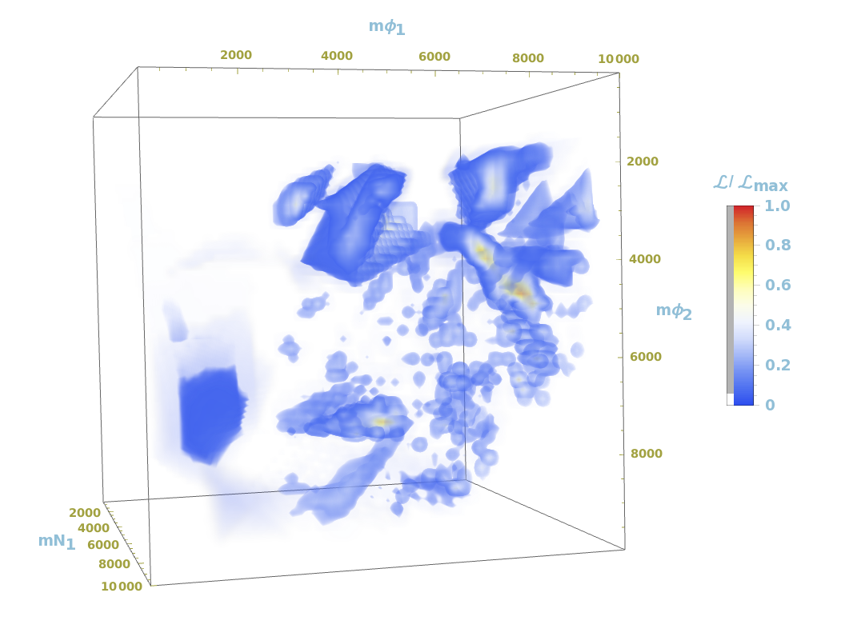}\newline
	\includegraphics[width=5.0cm, height=3.5cm]{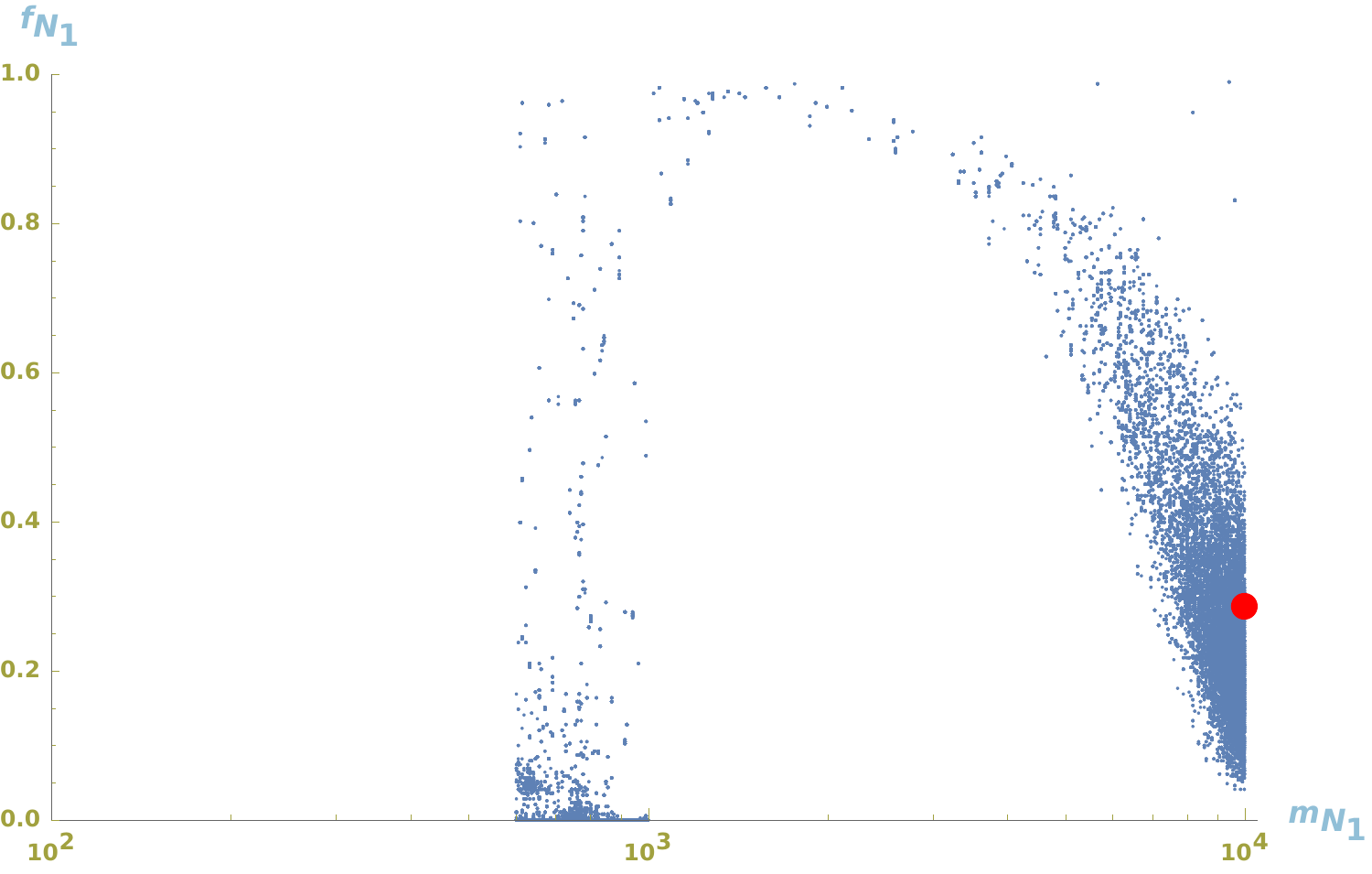}\qquad
	\includegraphics[width=5.0cm, height=3.5cm]{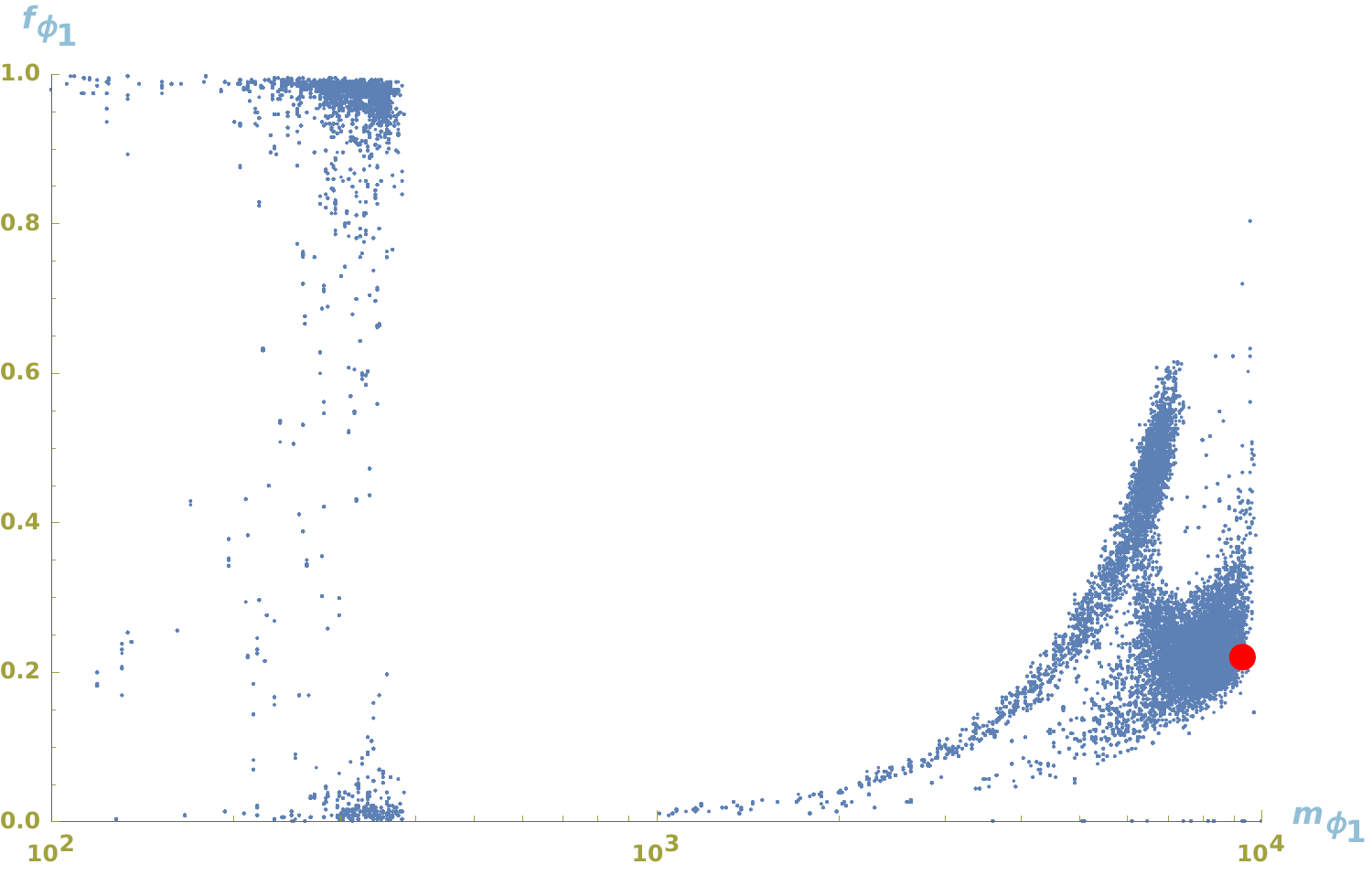}\qquad
	\includegraphics[width=5.0cm, height=3.5cm]{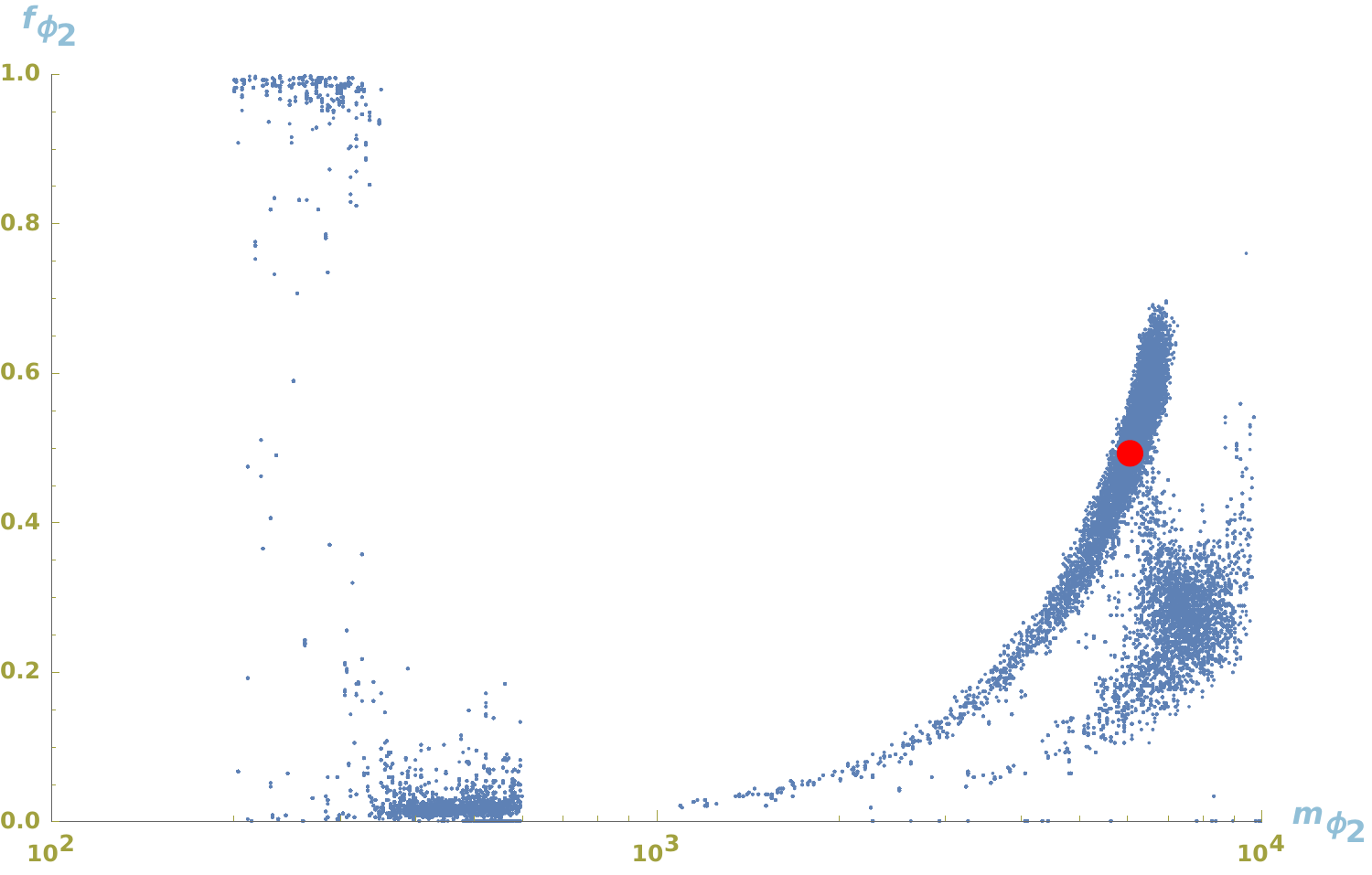}
	\caption{Top panel: Scattered plot of points in parameter space that lie inside the experimental Planck interval for the DM abundance, bright/red points are most consistent with the global constraints (all masses are in GeV). Bottom panel: The fraction of the relic density contributed by each DM candidate. The corresponding masses and fractions for the best fit point (BFP) are marked in red and have values $(m_{N_1}, m_{\phi_1}, m_{\phi_2})=(9991, 9323, 6045)$ GeV and $(f_{N_1}, f_{\phi_1}, f_{\phi_2})=(0.2867077, 0.2200167, 0.4932755)$.}
    \label{relics}
\end{figure}

\begin{figure}[!h]
	\includegraphics[width=8.5cm, height=7cm]{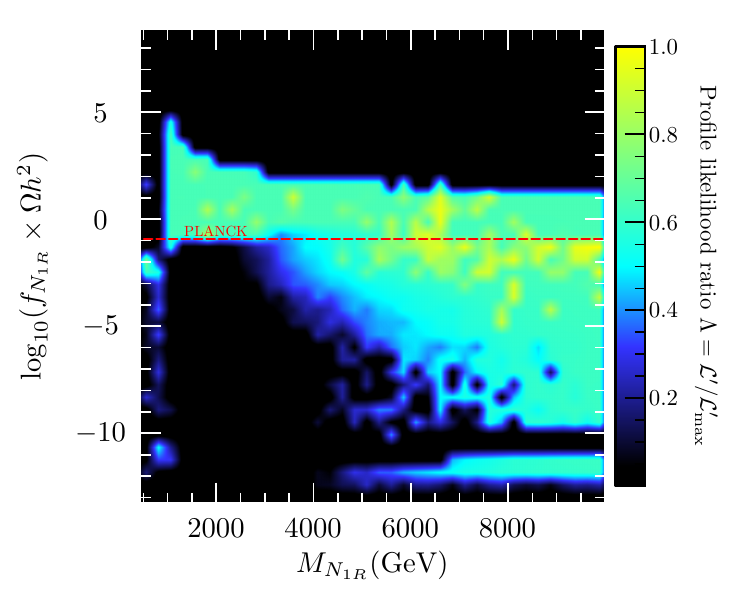}\newline
	\includegraphics[width=8.5cm, height=7cm]{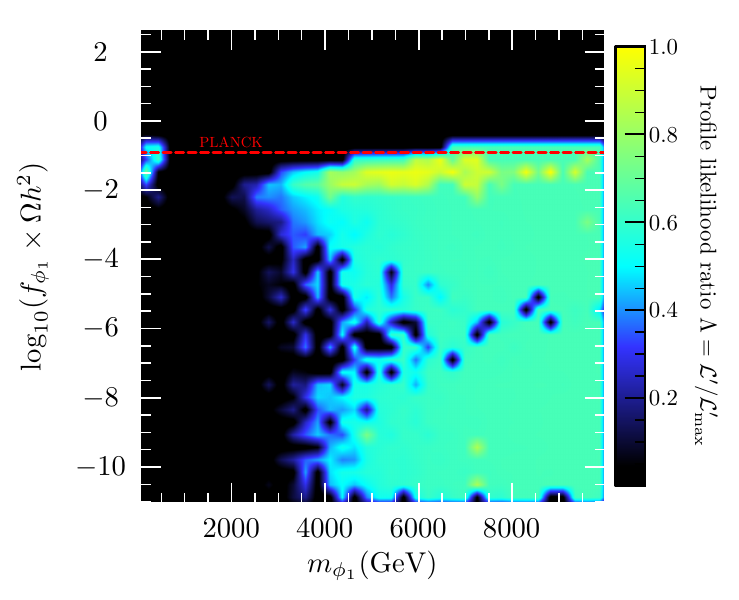}\newline
	\includegraphics[width=8.5cm, height=7cm]{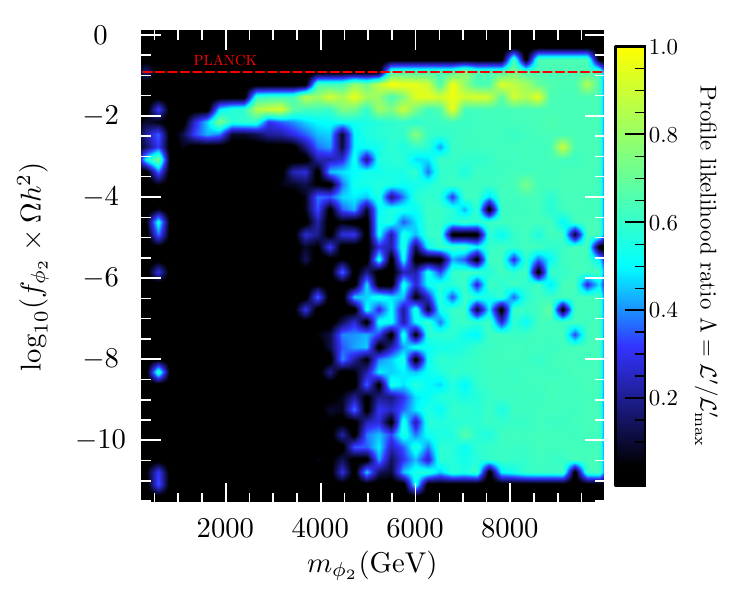}\newline
	\caption{DM relic abundance weighted by the respective DM fractions as a function of the masses of the DM candidates. The profiles are with respect to the partial likelihood ${\mathcal L}^\prime$ in Eq. (\ref{likePartial})}
    \label{rels}
\end{figure}

Fig. \ref{relics} shows the values of the masses of the DM candidates for which the model predicts a DM abundance
within the experimental PLANCK interval. Also shown are the corresponding fractions per DM candidate with which
each of them contribute to the total abundance.
We observe from the bottom panel of this figure that for masses of the right handed neutrino DM candidate below $\sim 600$ GeV the model is not capable to account for the observed DM abundance. This desert region also corresponds to the intervals of the scalar DM candidates around $m_{\phi_1} \sim (200 - 1000)$ GeV, $m_{\phi_2} \sim (600 - 1000)$ GeV and $m_{\phi_2} \lsim 200$ GeV.
This is also seen in Fig. \ref{rels} which shows the likelihood profiles for the three DM candidates with respect to the predicted fraction of DM abundance of each of the candidates and their masses. For visual aid%
\footnote{
In this case the profiles with respect to the total likelihood
${\mathcal L}$ which includes the relic density
constraint is of course just a horizontal slim bright band
around the PLANCK experimental value.
}
these profiles are shown with respect to the likelihood defined by:
\begin{equation}
	\log {\mathcal L}^\prime = \log {\mathcal L} - \log {\mathcal L}_{\Omega h^2}
	\label{likePartial} ~.
\end{equation}
The panels of this figure also portrait that in the interval of masses below $\sim 600$ GeV
the DM candidates $N_{1R}$ and $\phi_2$
are underabundant for the most part of the region while $\phi_1$ is slightly both underabundant and overproduced.
Other characteristics that can be inferred from these plots are, for instance, that the fermion DM candidate is almost entirely overproduced in the mass region below $\sim 3$ TeV (but above $600$ GeV). In this same mass region the scalar $\phi_1$ appears to have been annihilated out of existence.
For masses of the DM candidates above $3$ TeV all three of them contribute to the DM abundance but the scalar ones are mostly underproduced while the fermion one can also be overproduced some $\sim 3$ orders of magnitude above the measured value of the DM abundance.
\begin{figure}[!h]
	\includegraphics[width=8.5cm, height=7cm]{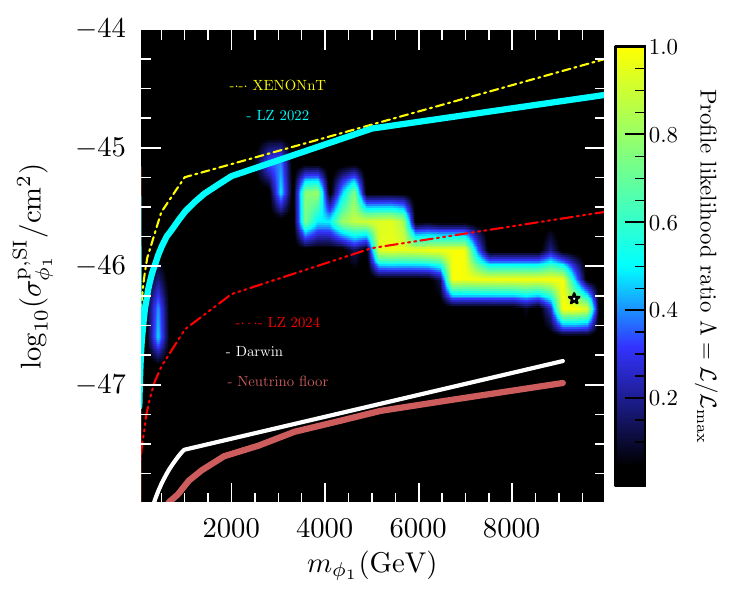}\includegraphics[width=8.5cm, height=7cm]{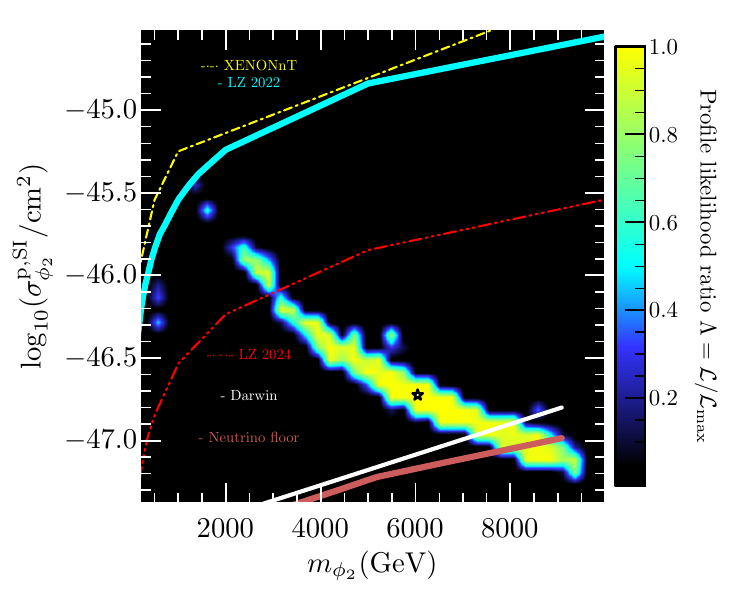}
	\caption{DM-proton spin-independent elastic scattering cross section as a function of the masses of the scalar DM candidates. Brightest areas are most consistent with all imposed constraints, dark areas are excluded by the analysis. The best fit point (BFP) is marked with a small star. For comparison, exclusion limits of the XENONnT, LZ 2022 and LZ 2024 experiments are shown, alongside with the projection of the DARWIN experiment and the neutrino floor. }
    \label{sigs}
\end{figure}

Finally, Fig. \ref{sigs} shows the likelihood profiles concerning the values of the spin independent scattering 
cross section consistent with all constraints in the model.
The plots in both panels show the dependence
of the likelihood on the DM mass and the DM-proton spin independent (SI)
cross section, for each of the scalar DM candidates.
We also depict the 90\% CL upper limit on the SI cross section from
the XENONnT \cite{XENON:2023cxc} and the 
LZ \cite{LZ:2022lsv,LZ:2024zvo}
experiments, alongside with the DARWIN experiment
from the projections of reference
\cite%
{DARWIN:2016hyl} \footnote{
	For better comparison with the other curves we extrapolated linearly the
	data available from this reference from 1 TeV up to $\sim$ 10 TeV}
and an estimation of the neutrino floor \cite%
{Billard:2013qya}.
We observe that the LZ experiment improved considerably their limits in just 2 years since the release of their first results. LZ is already able to exclude about half of the allowed parameter space for the case of the $\phi_1$ DM candidate, but still is far from excluding a sizable portion for the case of the $\phi_2$ DM candidate.
On the other hand, the capabilities of the DARWIN experiment will be able to probe the entire region for $\phi_1$ and around 80\% of the respective region for $\phi_2$, setting strong constrains on the model.

\section{Conclusions}
\label{conclusions}
We have proposed an extended $3+1$ extended Higgs doublet model where the tiny masses of active neutrinos are radiatively generated at two-loop level, and has three viable dark matter candidates plus a phenomenologically rich scalar sector. In the model under consideration, the SM gauge symmetry is enlarged by the inclusion of the $Q_6\times Z_2\times Z_4$ discrete group, whereas the SM fermionic spectrum is augmented by the inclusion of right-handed Majorana neutrinos. In addition to the four $SU(2)$ scalar doublets, the scalar sector also includes six gauge singlet scalars. Such extended particle content and symmetries allows for a successful implementation of the two-loop level radiative seesaw mechanism that yields the tiny active neutrino masses. In addition, it also generates a predictive cobimaximal pattern for the leptonic mixing, which successfully complies with current neutrino oscillation experimental data. Despite the extended scalar particle content, the number of low energy effective parameters is significantly reduced, thus rendering the model predictible. 
In our proposed model, the $Q_6$ symmetry is spontaneously broken, whereas the $Z_4$ symmetry breaks spontaneously down to a residual preserved $\tilde{Z}_2$ symmetry. Furthermore, the $Z_2$ symmetry is preserved. The preserved $Z_2$ and $\tilde{Z}_2$ discrete symmetries ensure two-loop induced masses for active neutrinos and also allow for stable dark matter candidates. We have analyzed in detail the implications of our model for  fermion masses and mixings, scalar sector and dark matter. We have found that our model successfully reproduces the low energy SM fermion flavor data and is compatible with current dark matter constraints. In particular we found that our model is compatible with lepton masses and mixings for normal neutrino mass ordering, and the inverted neutrino mass hierarchy is disfavored. Besides that, we found that the sum of the neutrino masses are located in the $0.061-0.0715$ eV range, while the value $\sum m_i$ for the best-fit point is $\sum m_i\simeq 6.54\times 10^{-2}\ \text{eV}$, consistent with current experimental bound $\sum m_{i(cosmo)} \lesssim 0.04–0.3$ eV arising from cosmological observations.

Furthermore,  our model successfully complies with the alignment limit constraints. A remarkable feature of the model is that after requiring its  consistency with all LHC constraints, we find a non-SM scalar with mass close to $95$ GeV, which could provide a possible explanation for the $95$ GeV diphoton excess. Additionally, we find several subTeV non SM scalars within the LHC reach, rendering our model testable at colliders. Furthermore, the model has three viable dark matter candidates, two scalars and one right-handed neutrino or three scalars, whose combined dark matter relic density is compatible with cosmological observations of the dark matter relic abundance.   
Regarding direct detection dark matter experiments, we found that the model could be strongly constrained by the Darwin experiment.

\section*{Acknowledgments}

This research has received funding from Chilean grants ANID-Chile FONDECYT 1241855, ANID CCTVal CIA250027, ANID – Millennium Science Initiative Program code ICN2019$\_$%
044; and
Mexican grants UNAM PAPIIT IN111224 and SECHITI grant CBF2023-2024-548. Red de Altas Energ\'{\i}as-CONACYT for the
financial support. C.E. acknowledges the support of SECIHTI (M\'{e}xico) C%
\'{a}tedra no. 341. JCGI is supported by SIP Project 20254394 , PAPIIT IN111224 and PAPIIT IA105025.

\appendix

\appendix

\section{$\mathcal{Q}_6$ multiplication rules}
\label{app} The $\mathcal{Q}_6$ has four singlets, $\mathbf{1}_{++}$, $%
\mathbf{1}_{+-}$, $\mathbf{1}_{-+}$, and $\mathbf{1}_{--}$, and two
doublets, $\mathbf{2}_1$ and $\mathbf{2}_2$. The tensor products for the $%
\mathcal{Q}_6$ representations are given by~\cite{Ishimori:2010au}
\begin{eqnarray}
\left( 
\begin{array}{c}
a \\ 
b%
\end{array}
\right)_{\mathbf{2}_2} \otimes \left( 
\begin{array}{c}
c \\ 
d%
\end{array}
\right)_{\mathbf{2}_{1}} = \left(a c- b d \right)_{\mathbf{1}_{+-}} \oplus
\left(a c+ b d \right)_{\mathbf{1}_{-+}} \oplus \left( 
\begin{array}{c}
a d \\ 
b c%
\end{array}
\right)_{\mathbf{2}_{1}},
\end{eqnarray}
\vspace{-1.5em} 
\begin{eqnarray}
\left( 
\begin{array}{c}
a \\ 
b%
\end{array}
\right)_{\mathbf{2}_k} \otimes \left( 
\begin{array}{c}
c \\ 
d%
\end{array}
\right)_{\mathbf{2}_{k}} = \left(a d- b c \right)_{\mathbf{1}_{++}} \oplus
\left(a d+ b c \right)_{\mathbf{1}_{--}} \oplus \left( 
\begin{array}{c}
a c \\ 
- b d%
\end{array}
\right)_{\mathbf{2}_{k^{\prime }}},
\end{eqnarray}
for $k,k^{\prime }=1,2$ and $k^{\prime }\neq k$, 
\begin{eqnarray}
\hspace{-3mm} & & \left( w\right)_{\mathbf{1}_{++}} \otimes \left( 
\begin{array}{c}
a_k \\ 
b_{-k}%
\end{array}
\right)_{\mathbf{2}_k}= \left( 
\begin{array}{c}
wa_k \\ 
wb_{-k}%
\end{array}
\right)_{\mathbf{2}_k}, \quad \left( w\right)_{\mathbf{1}_{--}} \otimes
\left( 
\begin{array}{c}
a_k \\ 
b_{-k}%
\end{array}
\right)_{\mathbf{2}_k}= \left( 
\begin{array}{c}
wa_k \\ 
-w b_{-k}%
\end{array}
\right)_{\mathbf{2}_k},  \notag \\
\hspace{-3mm} & & \left( w\right)_{\mathbf{1}_{+-}} \otimes \left( 
\begin{array}{c}
a_k \\ 
b_{-k}%
\end{array}
\right)_{\mathbf{2}_k}= \left( 
\begin{array}{c}
wb_{-k} \\ 
wa_{k}%
\end{array}
\right)_{\mathbf{2}_k}, \quad \left( w\right)_{\mathbf{1}_{-+}} \otimes
\left( 
\begin{array}{c}
a_{k} \\ 
b_{-k}%
\end{array}
\right)_{\mathbf{2}_k}= \left( 
\begin{array}{c}
wb_{-k} \\ 
- wa_{k}%
\end{array}
\right)_{\mathbf{2}_k},
\end{eqnarray}
\begin{eqnarray}
\mathbf{1}_{s_1s_2} \otimes \mathbf{1}_{s^{\prime }_1s^{\prime }_2} = 
\mathbf{1}_{s^{\prime \prime }_1s^{\prime \prime }_2},
\end{eqnarray}
where $s^{\prime \prime }_1=s_1s^{\prime }_1$ and $s^{\prime \prime
}_2=s_2s^{\prime }_2$.

\section{The scalar potential for a $Q_6$ doublet}
\label{ScalarQ6doublet}
The relevant terms for the scalar potential of $Q_6$ doublets are:
\begin{eqnarray}\label{VD}
V_D&=&-g_{\chi}^2 \left( \chi \chi^*\right)_{\mathbf{1}_{++}}+k_1 \left( \chi \chi^*\right)_{\mathbf{1}_{++}}\left( \chi \chi^*\right)_{\mathbf{1}_{++}} +k_2 \left( \chi \chi^*\right)_{\mathbf{1}_{--}}\left( \chi \chi^*\right)_{\mathbf{1}_{--}} + k_3   \left( \chi \notag\chi^*\right)_{\mathbf{2}_2} \left( \chi \chi^*\right)_{\mathbf{2}_2} \notag\\
&&-g_{H\xi}^2\left(H\xi^*\right)_{\mathbf{1}_{++}}+k_4 \left(H \xi^*\right)_{\mathbf{1}_{++}}\left( \xi H^*\right)_{\mathbf{1}_{++}} +k_5 \left(H \xi^*\right)_{\mathbf{1}_{--}}\left(\xi H^*\right)_{\mathbf{1}_{--}} + k_6   \left(H \xi^*\right)_{\mathbf{2}_2} \left(\xi H^*\right)_{\mathbf{2}_2},
\end{eqnarray}

where $\chi= H, \xi$. We obtain four unrestricted parameters: one bilinear term and three quadratic terms. From the minimization condition of the scalar potential: 
\begin{eqnarray}
 \frac{\partial  \langle V_D \rangle}{\partial v_2} &=& 0 \notag\\
 &=& v_{\xi _1} \left(2 v_2 (k_4+k_5) v_{\xi
   _1}-g_{H\xi}^2\right)  \label{eq:vevQ6H2}\\
 \frac{\partial  \langle V_D \rangle}{\partial v_{\xi_1}} &=& 0 \notag\\
 &=& -g_{H\xi}^2 v_2+2 v_2^2 (k_4+k_5)
   v_{\xi _1}+8 k_2 v_{\xi _1} v_{\xi _2}^2 \label{eq:vevQ6}\\
 \frac{\partial  \langle V_D \rangle}{\partial v_{\xi_2}} &=& 0 \notag\\
 &=& 8 k_2 v_{\xi _1}^2 v_{\xi _2},\label{eq:vevQ6xi2}
\end{eqnarray}

from Eq.~\eqref{eq:vevQ6xi2}, we can see that,
\begin{equation}
k_2=0.
\end{equation}

Therefore, we obtain the parameter $g_{H\xi}$ as a function of the other parameters, i.e.
\begin{equation}
    \begin{aligned}
        g_{H\xi}^2 = & \ \frac{2 \left(k_4
   v_2 v_{\xi_1}^2 -k_4
   v_{H_1} v_{\xi_1} v_{\xi_2}+k_5 v_{H_1}
   v_{\xi_1} v_{\xi_2}+k_5 v_2
   v_{\xi_1}^2\right)}{v_{\xi_1}},
    \end{aligned}
\end{equation}
with $k_4, \ k_5 \in \mathbb{R}$.
Furthermore, from the global minimum conditions, we obtain the following inequalities.:
\begin{eqnarray}
\frac{\partial^2 \langle V_D \rangle}{\partial v_2^2}  &&> 0 \notag\\
&& 2 (k_4+k_5) v_{\xi _1}^2 > 0 \notag\\
\frac{\partial^2 \langle V_D \rangle}{\partial v_{ \xi_1}^2}  &&> 0 \notag\\
&& 2 v_2^2 (k_4+k_5) > 0 
\end{eqnarray}

From this, we see that the VEV configuration of the $Q_6$ doublet $\xi$, given in Eq.~\eqref{eq:vev-xi}, is consistent with the scalar potential minimization condition in Eqs.~\eqref{eq:vevQ6H2}, \eqref{eq:vevQ6} and \eqref{eq:vevQ6xi2}. These results show that the VEV directions of the $Q_6$ scalar doublets $H$ and $\xi$ correspond to a global minimum of the scalar potential for a broad region of parameter space.

\section{Diagonalization of fermion mass matrices}
\label{fermiondiagonalization}
In this section, we will describe with detail the diagonalization procedure for the quark and lepton sectors.
\subsection{Quark sector}
Going back to Eq.(\ref{qm2}), we have
\begin{equation}
\mathbf{M}_{q}= 
\begin{pmatrix}
0 & A_{q} & 0 \\ 
-A_{a} & 0 & b_{q} \\ 
0 & c_{q} & F_{q}%
\end{pmatrix}%
.
\end{equation}

That mass matrices are diagonalized by the $\mathbf{U}%
_{q(L,R)}$ unitary matrices such that $\mathbf{U}^{\dagger}_{q L}\mathbf{M}%
_{q} \mathbf{U}_{q R}=\hat{\mathbf{M}}_{q}$ with $\hat{\mathbf{M}}_{q}=\text{%
Diag}.\left(m_{q_{1}}, m_{q_{2}}, m_{q_{3}}\right)$ being the physical quark
masses. In order to obtain the CKM matrix, let us calculate the $\mathbf{U}_{q L}$
matrix by means the bilineal form $\mathbf{\hat{M}}_{q} \mathbf{\hat{M}}%
^{\dagger}_{q}=\mathbf{U}^{\dagger}_{q L} \mathbf{M}_{q} \mathbf{M }%
^{\dagger}_{q} \mathbf{U}_{q L}$. Then, the hermitian matrix is written in
the polar form $\mathbf{M}_{q} \mathbf{M }^{\dagger}_{q}= \mathbf{P}_{q}%
\mathbf{m}_{q} \mathbf{m }^{\dagger}_{q} \mathbf{P}^{\dagger}_{q}$ with $%
\mathbf{P}_{q} = \text{diag.} \left( 1, e^{ i\eta_{q_{2}}}, e^{ i\eta_{q_{3}}
}\right)$ and the phases are given by 
\begin{equation}
\eta_{q_{2}}=\text{arg.} [\left( \mathbf{M}_{q} \mathbf{M }%
^{\dagger}_{q}\right)_{23}]+ \text{arg.} [\left( \mathbf{M}_{q} \mathbf{M }%
^{\dagger}_{q}\right)_{13}],\qquad \eta_{q_{3}}=-\text{arg.} [\left( \mathbf{%
M}_{q} \mathbf{M }^{\dagger}_{q}\right)_{13}].
\end{equation}

Therefore, $\mathbf{U}_{qL}=\mathbf{P}_{q}\mathbf{O}_{q L}$ so the $\mathbf{%
m}_{q} \mathbf{m }^{\dagger}_{q}$ real symmetric matrix is diagonalized by
the $\mathbf{O}_{q L}$ orthogonal one, this means, $\mathbf{\hat{M}}_{q} \mathbf{\hat{M}}%
^{\dagger}_{q}=\mathbf{O}^{T}_{q L} \mathbf{m}_{q} \mathbf{m }%
^{\dagger}_{q} \mathbf{O}_{q L}$. This last expression allows to fix three free parameters in terms of the physical masses and one unfixed parameter. To to do that, we use the invariants:  the trace ($Tr[\mathbf{\hat{M}}_{q} \mathbf{\hat{M}}%
^{\dagger}_{q}]$), determinant ($Det[\mathbf{\hat{M}}_{q} \mathbf{\hat{M}}%
^{\dagger}_{q}]$) and $(Tr[(\mathbf{\hat{M}}_{q} \mathbf{\hat{M}}%
^{\dagger}_{q})^{2}]-(Tr[\mathbf{\hat{M}}_{q} \mathbf{\hat{M}}%
^{\dagger}_{q}])^{2}/2$. As a result, we obtain
\begin{eqnarray}
\vert m_{q_{1}}\vert^{2}+\vert m_{q_{2}}\vert^{2}+\vert m_{q_{3}}\vert^{2}&=&2\vert A_{q}\vert^{2}+\vert B_{q}\vert^{2}+\vert C_{q}\vert^{2}+\vert F_{q}\vert^{2};\nn\\
 \vert m_{q_{1}}\vert^{2}\vert m_{q_{2}}\vert^{2}\vert m_{q_{3}}\vert^{2}&=&\vert A_{q}\vert^{4}\vert F_{q}\vert^{2};\nn\\
\vert m_{q_{1}}\vert^{2}(\vert m_{q_{2}}\vert^{2}+\vert m_{q_{3}}\vert^{2})+ \vert m_{q_{2}}\vert^{2} \vert m_{q_{3}}\vert^{2}&=&2\vert A_{q}\vert^{2}\vert F_{q}\vert^{2}+ (\vert A_{q}\vert^{2}+\vert B_{q}\vert^{2})(\vert A_{q}\vert^{2}+\vert C_{q}\vert^{2}).
\end{eqnarray}

Then, let $\vert F_{q}\vert$ the unfixed parameter so that
\begin{eqnarray}
\vert A_{q}\vert&=&\sqrt{\frac{\vert m_{q_{1}}\vert \vert m_{q_{2}}\vert \vert m_{q_{3}}\vert}{\vert F_{q}\vert}};\nn\\
\vert B_{q}\vert&=&\sqrt{\frac{\vert F_{q}\vert(\vert m_{q_{3}}\vert^{2}+\vert m_{q_{2}}\vert^{2}+\vert m_{q_{1}}\vert^{2}-\vert F_{q}\vert^{2}-R_{q})-2\vert m_{q_{3}}\vert \vert m_{q_{2}}\vert \vert m_{q_{1}}\vert}{2 \vert F_{q}\vert}};\nn\\
\vert C_{q}\vert&=&\sqrt{\frac{\vert F_{q}\vert(\vert m_{q_{3}}\vert^{2}+\vert m_{q_{2}}\vert^{2}+\vert m_{q_{1}}\vert^{2}-\vert F_{q}\vert^{2}+R_{q})-2\vert m_{q_{3}}\vert \vert m_{q_{2}}\vert \vert m_{q_{1}}\vert}{2 \vert F_{q}\vert}},
\end{eqnarray}
where $R_{q}=\sqrt{((\vert m_{q_{3}}\vert^{2}+\vert m_{q_{2}}\vert^{2}+\vert m_{q_{1}}\vert^{2}-\vert F_{q}\vert^{2})^{2}-4(\vert m_{q_{1}}\vert^{2}(\vert m_{q_{2}}\vert^{2}+\vert m_{q_{3}}\vert^{2})+ \vert m_{q_{2}}\vert^{2} \vert m_{q_{3}}\vert^{2}-2\vert m_{q_{1}}\vert \vert m_{q_{2}}\vert \vert m_{q_{3}}\vert \vert F_{q}\vert)}$.

Having done that, the orthogonal real matrix is given explicitly by
\begin{align}  \label{ortho}
\mathbf{O}_{q L}= 
\begin{pmatrix}
-\sqrt{\dfrac{\tilde{m}_{q_{2}} (\rho^{q}_{-}-R^{q}) K^{q}_{+}}{4 y_{q}
\delta^{q}_{1} \kappa^{q}_{1} }} & -\sqrt{\dfrac{\tilde{m}_{q_{1}}
(\sigma^{q}_{+}-R^{q}) K^{f}_{+}}{4 y_{q} \delta^{q}_{2} \kappa^{q}_{2} }} & 
\sqrt{\dfrac{\tilde{m}_{q_{1}} \tilde{m}_{q_{2}} (\sigma^{q}_{-}+R^{q})
K^{q}_{+}}{4 y_{q} \delta^{q}_{3} \kappa^{q}_{3} }} \\ 
-\sqrt{\dfrac{\tilde{m}_{q_{1}} \kappa^{q}_{1} K^{q}_{-}}{%
\delta^{q}_{1}(\rho^{q}_{-}-R^{q}) }} & \sqrt{\dfrac{\tilde{m}_{q_{2}}
\kappa^{q}_{2} K^{q}_{-}}{\delta^{q}_{2}(\sigma^{q}_{+}-R^{q}) }} & \sqrt{%
\dfrac{\kappa^{q}_{3} K^{q}_{-}}{\delta^{q}_{3}(\sigma^{q}_{-}+R^{q}) }} \\ 
\sqrt{\dfrac{\tilde{m}_{q_{1}} \kappa^{q}_{1}(\rho^{q}_{-}-R^{q})}{2
y_{q}\delta^{q}_{1}}} & -\sqrt{\dfrac{\tilde{m}_{q_{2}}
\kappa^{q}_{2}(\sigma^{q}_{+}-R^{q})}{2 y_{q}\delta^{q}_{2}}} & \sqrt{\dfrac{%
\kappa^{q}_{3}(\sigma^{q}_{-}+R^{q})}{2 y_{q}\delta^{q}_{3}}}%
\end{pmatrix}%
\end{align}
with 
\begin{align}  \label{defortho}
\rho^{q}_{\pm}&\equiv 1+\tilde{m}^{2}_{q_{2}}\pm\tilde{m}%
^{2}_{q_{1}}-y^{2}_{q},\quad \sigma^{q}_{\pm}\equiv 1-\tilde{m}%
^{2}_{q_{2}}\pm(\tilde{m}^{2}_{q_{1}}-y^{2}_{q}),\quad \delta^{q}_{(1,
2)}\equiv (1-\tilde{m}^{2}_{q_{(1, 2)}})(\tilde{m}^{2}_{q_{2}}-\tilde{m}%
^{2}_{q_{1}});  \notag \\
\delta^{q}_{3}&\equiv (1-\tilde{m}^{2}_{q_{1}})(1-\tilde{m}%
^{2}_{q_{2}}),\quad \kappa^{q}_{1} \equiv \tilde{m}_{q_{2}}-\tilde{m}%
_{q_{1}}y_{q},\quad \kappa^{q}_{2}\equiv \tilde{m}_{q_{2}}y_{q}-\tilde{m}%
_{q_{1}},\quad \kappa^{q}_{3}\equiv y_{q}-\tilde{m}_{q_{1}}\tilde{m}_{q_{2}};
\notag \\
R^{q}&\equiv \sqrt{\rho^{q 2}_{+}-4(\tilde{m}^{2}_{q_{2}}+\tilde{m}%
^{2}_{q_{1}}+\tilde{m}^{2}_{q_{2}}\tilde{m}^{2}_{q_{1}}-2\tilde{m}_{q_{1}}%
\tilde{m}_{q_{2}}y_{q})},\quad K^{q}_{\pm} \equiv y_{q}(\rho^{q}_{+}\pm
R^{q})-2\tilde{m}_{q_{1}}\tilde{m}_{q_{2}}.
\end{align}

We have to point out that the parameters have been normalized by the $%
m_{q_{3}}$ heaviest physical quark mass. Additionally, there are two unfixed
parameters ($y_{q}\equiv \vert F_{q}\vert/m_{q_{3}}$) which are constrained
by the condition $1>y_{q}>\tilde{m}_{q_{2}}>\tilde{m}_{q_{1}}$. Then, the
relevant matrices that take place in the CKM matrix are given by $\mathbf{U}%
_{q L}= \mathbf{P}_{q}\mathbf{O}_{q L}$ where $q=u, d$.
Finally, the CKM mixing matrix is written as 
\begin{equation}
\mathbf{V}_{CKM}=\mathbf{O}^{T}_{u L}\bar{\mathbf{P}}_{q} \mathbf{O}_{d L},
\quad \mathbf{P}_{q}=\mathbf{P}^{\dagger}_{u}\mathbf{P}_{d}=\text{diag.}%
\left(1, e^{i\bar{\eta}_{q_{2}}}, e^{i\bar{\eta}_{q_{3}}} \right).
\end{equation}
This CKM mixing matrix has four free parameters namely $y_{u}$ $y_{d}$, and
two phases $\eta_{q_{1}}$ and $\eta_{q_{2}}$ which could be obtained
numerically. In addition, the expression for the mixing angles are given as follows:
\begin{eqnarray}
\sin^{2}{\theta^{q}_{13}}&=& \big| (\mathbf{V}_{CKM})_{13}\big|^{2}= \big |(\mathbf{O}_{u})_{11} (\mathbf{O}_{d})_{13}+(\mathbf{O}_{u})_{21} (\mathbf{O}_{d})_{23}e^{i\bar{\eta}_{q_{2}}}+(\mathbf{O}_{u})_{31} (\mathbf{O}_{d})_{33}e^{i\bar{\eta}_{q_{3}}}\big|^{2};\nn\\
\sin^{2}{\theta^{q}_{12}}&=&\frac{ \big| (\mathbf{V}_{CKM})_{12}\big|^{2}}{1- \big| (\mathbf{V}_{CKM})_{13}\big|^{2}}= \big |(\mathbf{O}_{u})_{11} (\mathbf{O}_{d})_{12}+(\mathbf{O}_{u})_{21} (\mathbf{O}_{d})_{22}e^{i\bar{\eta}_{q_{2}}}+(\mathbf{O}_{u})_{31} (\mathbf{O}_{d})_{32}e^{i\bar{\eta}_{q_{3}}}\big|^{2};\nn\\
\sin^{2}{\theta^{q}_{23}}&=&\frac{ \big| (\mathbf{V}_{CKM})_{23}\big|^{2}}{1- \big| (\mathbf{V}_{CKM})_{13}\big|^{2}}= \big |(\mathbf{O}_{u})_{12} (\mathbf{O}_{d})_{13}+(\mathbf{O}_{u})_{22} (\mathbf{O}_{d})_{23}e^{i\bar{\eta}_{q_{2}}}+(\mathbf{O}_{u})_{32} (\mathbf{O}_{d})_{33}e^{i\bar{\eta}_{q_{3}}}\big|^{2}.
\end{eqnarray}

\subsection{Lepton sector}
As it was shown before, the charged lepton mass matrix has the following textures
\begin{equation}
M_{l}=\left( 
\begin{array}{ccc}
y^{l}_{1}\frac{w_{3}}{\sqrt{2}} & y^{l}_{2}\frac{w_{1}}{\sqrt{2}} & y^{l}_{2}\frac{w_{2}}{\sqrt{2}} \\ 
y^{l}_{3}\frac{w_{1}}{\sqrt{2}} & 0 & y^{l}_{4}\frac{%
w_{3}}{\sqrt{2}} \\ 
y^{l}_{3}\frac{w_{2}}{\sqrt{2}} & -y^{l}_{4}\frac{%
w_{3}}{\sqrt{2}} & 0%
\end{array}%
\right),
\end{equation}
The aforementioned matrix is diagonalized by $\mathbf{U}^{\dagger}_{l L}\mathbf{M}_{l}\mathbf{U}_{l R}=\hat{\mathbf{M}}_{l}$ with $\hat{\mathbf{M}}_{l}=\textrm{Diag}.\left(m_{e},m_{\mu},m_{\tau}\right)$. Then,
we build the bilineal $\hat{\mathbf{M}}_{l}\hat{\mathbf{M}}^{\dagger}_{l}= \mathbf{U}^{\dagger}_{l L}\mathbf{M}_{l}\mathbf{M}^{\dagger}_{l} \mathbf{U}_{l L}$ in order to obtain the relevant mixing matrix that takes places in the PMNS one. To do so, the CP-violating phases are factorized as follows: $\mathbf{M}_{l}\mathbf{M}^{\dagger}_{l}=\mathbf{P}_{l} \mathbf{m}_{l}\mathbf{m}^{\dagger}_{l}\mathbf{P}^{\dagger}_{l}$ where $\mathbf{P}_{l}=\textrm{Diag}.\left(e^{i\eta_{e}},e^{i\eta_{\mu}},e^{i\eta_{\tau}} \right)$. These phases must satisfy the following conditions
\begin{equation}
 \eta_{e}- \eta_{\mu}=\textrm{arg}(b_{l}) - \textrm{arg}(d_{l}),\qquad \eta_{e}- \eta_{\tau}=\textrm{arg}(a_{l}) - \textrm{arg}(c_{l}).
\end{equation}
Without of losing of generality, we take $\eta_{e}=0$. Then, $\mathbf{U}_{l L}=\mathbf{P}_{l}\mathbf{O}_{l}$ where the latter matrix is real and orthogonal such that  $\hat{\mathbf{M}}_{l}\hat{\mathbf{M}}^{\dagger}_{l}= \mathbf{O}^{T}_{l}\mathbf{m}_{l}\mathbf{m}^{\dagger}_{l} \mathbf{O}_{l}$
\begin{equation}
 \mathbf{m}_{l}\mathbf{m}^{\dagger}_{l}=  \left( 
\begin{array}{ccc}
\vert a_{l}  \vert^{2}+ \vert b_{l}  \vert^{2}& \vert b_{l}  \vert \vert d_{l}  \vert  & \vert a_{l}  \vert \vert c_{l}  \vert \\ 
\vert b_{l}  \vert \vert d_{l}  \vert & \vert d_{l}  \vert^{2} & 0 \\ 
\vert a_{l}  \vert \vert c_{l}  \vert & 0 & \vert c_{l}  \vert^{2}+ \vert d_{l}  \vert^{2}%
\end{array}%
\right),
\end{equation}
Given the $\mathbf{m}_{l}\mathbf{m}^{\dagger}_{l}$ real matrix, three free parameters can be fixed in terms of the charged lepton masses. This is realized by means the following invariant: the  $Tr[\hat{\mathbf{M}}_{l}\hat{\mathbf{M}}^{\dagger}_{l}]$ trace,  the $Det[\hat{\mathbf{M}}_{l}\hat{\mathbf{M}}^{\dagger}_{l}]$ determinant and $(Tr[(  \hat{\mathbf{M}}_{l}\hat{\mathbf{M}}^{\dagger}_{l})^{2}] -(Tr[\hat{\mathbf{M}}_{l}\hat{\mathbf{M}}^{\dagger}_{l}])^{2}/2$
\begin{eqnarray}
\vert a_{l}  \vert^{2}+ \vert b_{l} \vert^{2}+\vert c_{l} \vert^{2}+2 \vert d_{l} \vert^{2}&=& \vert m_{e}  \vert^{2}+\vert m_{\mu}  \vert^{2}+\vert m_{\tau}  \vert^{2};\nn\\
\vert a_{l}  \vert^{2} \vert d_{l}  \vert^{4}&=& \vert m_{e}  \vert^{2}\vert m_{\mu}  \vert^{2}\vert m_{\tau}  \vert^{2};\nn\\
 2\vert a_{l}  \vert^{2} \vert d_{l} \vert^{2} +\left(\vert b_{l}  \vert^{2}  +\vert d_{l}  \vert^{2} \right)\left(\vert c_{l}  \vert^{2} \vert +\vert d_{l}  \vert^{2} \right)&=&\vert m_{e}  \vert^{2}\left( \vert m_{\mu}  \vert^{2}+\vert m_{\tau}  \vert^{2}\right)+\vert m_{\mu}  \vert^{2}\vert m_{\tau}  \vert^{2}
\end{eqnarray}
In this case, there is an unfixed parameter ($\vert a_{l}\vert$) and the rest of them are written in terms of it and the charged lepton masses. This is
\begin{eqnarray}
\vert b_{l}\vert&=&\sqrt{\frac{\vert a_{l}  \vert\left(\vert m_{\tau}  \vert^{2}+\vert m_{\mu}  \vert^{2}+\vert m_{e}  \vert^{2}-\vert a_{l}  \vert^{2}-R_{e}\right)-2\vert m_{\tau}\vert \vert m_{\mu}\vert \vert m_{e} \vert}{2\vert a_{l}\vert}};\nn\\
\vert c_{l}\vert&=&\sqrt{\frac{\vert a_{l}  \vert\left(\vert m_{\tau}  \vert^{2}+\vert m_{\mu}  \vert^{2}+\vert m_{e}  \vert^{2}-\vert a_{l}  \vert^{2}+R_{e}\right)-2\vert m_{\tau}\vert \vert m_{\mu}\vert \vert m_{e} \vert}{2\vert a_{l}\vert}};\nn\\
\vert d_{l}\vert&=&\sqrt{\frac{\vert m_{\tau}\vert \vert m_{\mu}\vert \vert m_{e} \vert}{\vert a_{l}\vert}}.
\end{eqnarray}
where $R_{e}=\sqrt{\left(\vert m_{\tau}  \vert^{2}+\vert m_{\mu}  \vert^{2}+\vert m_{e}  \vert^{2}-\vert a_{l}  \vert^{2}\right)^{2}-4\left[\vert m_{e}  \vert^{2} \left(\vert m_{\tau}  \vert^{2}+\vert m_{\mu}  \vert^{2}\right)+\vert m_{\tau}  \vert^{2}\vert m_{\mu} \vert^{2}-2\vert m_{\tau} \vert \vert m_{\mu} \vert\vert m_{e} \vert \vert a_{l} \vert\right]}$.
Having fixed three parameters, the $\mathbf{O}_{l}$ real and orthogonal matrix is parametrized as
\begin{equation}
\mathbf{O}_{l}= \left( X_{1}\quad X_{2} \quad X_{3}
\right),
\end{equation}
where the eigenvectors are written explicitly 
\begin{eqnarray}
X_{1}&=& \left( 
\begin{array}{c}
-\sqrt{\frac{2\vert m_{e}\vert\left(\vert m_{\tau}\vert\vert m_{\mu}\vert-\vert a_{l}\vert\vert m_{e}\vert\right)^{2}\left[\vert a_{l}\vert\left(\vert m_{\tau}  \vert^{2}+\vert m_{\mu}  \vert^{2}+\vert m_{e}  \vert^{2}-\vert a_{l}  \vert^{2}+R_{e}\right)-2\vert m_{\tau}  \vert \vert m_{\mu}  \vert \vert m_{e}  \vert\right]}{D_{e}}}\\
\sqrt{\frac{4\vert m_{\tau}\vert \vert m_{\mu}\vert \left(\vert m_{\tau}\vert\vert m_{\mu}\vert-\vert a_{l}\vert\vert m_{e}\vert\right) \left(\vert m_{\mu}\vert\vert a_{l}\vert-\vert m_{\tau}\vert\vert m_{e}\vert\right) \left(\vert m_{\tau}\vert\vert a_{l}\vert-\vert m_{\mu}\vert\vert m_{e}\vert\right)}{D_{e}}}\\
\sqrt{\frac{\vert a_{l}\vert \vert m_{e}\vert \left[2 \vert m_{\tau}\vert \vert m_{\mu}\vert \vert a_{l}\vert-\vert m_{e}\vert\left(\vert m_{\tau}  \vert^{2}+\vert m_{\mu}  \vert^{2}-\vert m_{e}  \vert^{2}+\vert a_{l}  \vert^{2}-R_{e}\right)\right]^{2}}{D_{e}}}
\end{array}%
\right);\nn\\
X_{2}&=&\left( 
\begin{array}{c}
\sqrt{\frac{\vert m_{\mu}\vert\left(\vert m_{\mu}\vert\vert a_{l}\vert-\vert m_{\tau}\vert\vert m_{e}\vert\right)\left(\vert m_{\tau}  \vert^{2}-\vert m_{\mu}  \vert^{2}+\vert m_{e}  \vert^{2}-\vert a_{l}  \vert^{2}+R_{e}\right)}{D_{\mu}}}\\  \sqrt{\frac{\vert m_{\tau}\vert \vert m_{e}\vert \left[ \vert m_{\mu}\vert\left(\vert m_{\tau}  \vert^{2}-\vert m_{\mu}  \vert^{2}+\vert m_{e}  \vert^{2}+\vert a_{l}  \vert^{2}+R_{e}\right) -  2 \vert m_{\tau}\vert \vert m_{e}\vert \vert a_{l}\vert\right]}{D_{\mu}}}\\ 
-\sqrt{\frac{\vert a_{l}\vert \vert m_{\mu}\vert \left[\vert m_{\mu}\vert\left(\vert m_{\tau}  \vert^{2}-\vert m_{\mu}  \vert^{2}+\vert m_{e}  \vert^{2}+\vert a_{l}  \vert^{2}-R_{e}\right) -  2 \vert m_{\tau}\vert \vert m_{e}\vert \vert a_{l}\vert\right]}{D_{\mu}}}
\end{array}%
\right);\nn\\
X_{3}&=&\left( 
\begin{array}{c}
\sqrt{\frac{2\vert m_{\tau}\vert\left(\vert m_{\tau}\vert\vert a_{l}\vert-\vert m_{\mu}\vert\vert m_{e}\vert\right)^{2}\left[\vert a_{l}\vert\left(\vert m_{\tau}  \vert^{2}+\vert m_{\mu}  \vert^{2}+\vert m_{e}  \vert^{2}-\vert a_{l}  \vert^{2}+R_{e}\right)-2\vert m_{\tau}  \vert \vert m_{\mu}  \vert \vert m_{e}  \vert\right]}{D_{\tau}}}\\  \sqrt{\frac{4\vert m_{\mu}\vert \vert m_{e}\vert \left(\vert m_{\tau}\vert\vert m_{\mu}\vert-\vert a_{l}\vert\vert m_{e}\vert\right) \left(\vert m_{\mu}\vert\vert a_{l}\vert-\vert m_{\tau}\vert\vert m_{e}\vert\right) \left(\vert m_{\tau}\vert\vert a_{l}\vert-\vert m_{\mu}\vert\vert m_{e}\vert\right)}{D_{\tau}}}\\ 
\sqrt{\frac{\vert a_{l}\vert \vert m_{\tau}\vert \left[\vert m_{\tau}\vert\left(\vert m_{\tau}  \vert^{2}-\vert m_{\mu}  \vert^{2}-\vert m_{e}  \vert^{2}-\vert a_{l}  \vert^{2}+R_{e}\right) + 2 \vert m_{\mu}\vert \vert m_{e}\vert \vert a_{l}\vert\right]^{2}}{D_{\tau}}}
\end{array}%
\right).
\end{eqnarray}
with
\begin{eqnarray}
D_{e}&=&2\vert a_{l}\vert \left(\vert m_{\tau}  \vert^{2}-\vert m_{e}  \vert^{2}\right)\left(\vert m_{\mu}  \vert^{2}-\vert m_{e}  \vert^{2}\right)\left[2 \vert m_{\tau}\vert \vert m_{\mu}\vert \vert a_{l}\vert-\vert m_{e}\vert\left(\vert m_{\tau}  \vert^{2}+\vert m_{\mu}  \vert^{2}-\vert m_{e}  \vert^{2}+\vert a_{l}  \vert^{2}-R_{e}\right)\right];\nn\\ 
D_{\mu}&=&2\vert a_{l}\vert \left(\vert m_{\tau}  \vert^{2}-\vert m_{\mu}  \vert^{2}\right)\left(\vert m_{\mu}  \vert^{2}-\vert m_{e}  \vert^{2}\right);\nn\\ D_{\tau}&=&2\vert a_{l}\vert \left(\vert m_{\tau}  \vert^{2}-\vert m_{e}  \vert^{2}\right)\left(\vert m_{\tau}  \vert^{2}-\vert m_{\mu}  \vert^{2}\right)\left[\vert m_{\tau}\vert\left(\vert m_{\tau}  \vert^{2}-\vert m_{\mu}  \vert^{2}-\vert m_{e}  \vert^{2}-\vert a_{l}  \vert^{2}+R_{e}\right) + 2 \vert m_{\mu}\vert \vert m_{e}\vert \vert a_{l}\vert\right].
\end{eqnarray}
The $\vert a_{l}\vert$ free parameter is constrained in the region 
$\vert m_{\tau}\vert>\vert a_{l}\vert>(\vert m_{\tau}\vert/\vert m_{\mu}\vert)\vert m_{e}\vert$. Nonetheless, the correct charged lepton masses are getting with $\vert a_{l}\vert\approx (\vert m_{\tau}\vert/\vert m_{\mu}\vert)\vert m_{e}\vert$, in this case, the $\mathbf{U}_{l L}=\mathbf{P}_{l}\mathbf{O}_{l}$ is close the identity matrix. Therefore, in this scenario, the PMNS mixing matrix is controlled by the Cobimaximal pattern that comes from the neutrino sector.

Next, let us show you briefly a limit case where $\mathbf{U}_{l L}\approx \mathbf{1}$. To do this, if $\vert a_{l}\vert=(\vert m_{\tau}\vert/\vert m_{\mu}\vert)\vert m_{e}\vert$, one would obtain
\begin{equation}
R_{e}=\frac{\left(\vert m_{\tau}  \vert^{2}-\vert m_{\mu}  \vert^{2}\right)\left(\vert m_{\mu}  \vert^{2}-\vert m_{e}  \vert^{2}\right)}{\vert m_{\mu}  \vert^{2}}, \quad \vert b_{l}\vert=0,\quad   \vert c_{l}\vert=\sqrt{\left(\vert m_{\tau}  \vert^{2}-\vert m_{\mu}  \vert^{2}\right)\left(1-\frac{\vert m_{e}\vert^{2}}{\vert m_{\mu}\vert^{2}}\right)},\quad \vert d_{l}\vert= \vert m_{\mu}\vert. 
\end{equation}
In consequence
\begin{equation*}
\mathbf{O}_{l}=\left( 
\begin{array}{ccc}
-\sqrt{\frac{\vert m_{\tau}\vert^{2}\left(\vert m_{\mu}  \vert^{2}-\vert m_{e}  \vert^{2}\right)}{\vert m_{\mu}\vert^{2}\left(\vert m_{\tau}  \vert^{2}-\vert m_{e}  \vert^{2}\right)}} & 0 & \sqrt{\frac{\vert m_{e}\vert^{2}\left(\vert m_{\tau}  \vert^{2}-\vert m_{\mu}  \vert^{2}\right)}{\vert m_{\mu}\vert^{2}\left(\vert m_{\tau}  \vert^{2}-\vert m_{e}  \vert^{2}\right)}} \\ 
0 & 1 & 0 \\ 
\sqrt{\frac{\vert m_{e}\vert^{2}\left(\vert m_{\tau}  \vert^{2}-\vert m_{\mu}  \vert^{2}\right)}{\vert m_{\mu}\vert^{2}\left(\vert m_{\tau}  \vert^{2}-\vert m_{e}  \vert^{2}\right)}} & 0 & \sqrt{\frac{\vert m_{\tau}\vert^{2}\left(\vert m_{\mu}  \vert^{2}-\vert m_{e}  \vert^{2}\right)}{\vert m_{\mu}\vert^{2}\left(\vert m_{\tau}  \vert^{2}-\vert m_{e}  \vert^{2}\right)}}%
\end{array}%
\right),
\end{equation*}
then $\mathbf{U}_{l L}\approx \mathbf{1}$ so that our statement is correct.

\subsection{Neutrino sector}
According to the neutrino section, the effective mass matrix possesses the cobimaximal pattern, this is
\begin{equation}
\mathbf{M}_{\nu }=\left( 
\begin{array}{ccc}
A_{\nu } & \tilde{B}_{\nu } & \tilde{B}_{\nu }^{\ast } \\ 
\tilde{B}_{\nu } & \tilde{C}_{\nu }^{\ast } & D_{\nu } \\ 
\tilde{B}_{\nu }^{\ast } & D_{\nu } & \tilde{C}_{\nu }%
\end{array}%
\right).
\end{equation}%

This kind of pattern was proposed many years ago and amazing predictions on the mixing angles and Majorana phases are notable. As it has been shown, $\mathbf{M}_{\nu }$
is diagonalized by the mixing matrix $\mathbf{U}%
_{\nu }$, this is, $\mathbf{U}_{\nu }^{\dagger }\mathbf{M}_{\nu }\mathbf{U}%
_{\nu }^{\ast }=\hat{\mathbf{M}}_{\nu }$ with $\hat{\mathbf{M}}_{\nu }=\text{%
Diag.}(|m_{1}|,|m_{2}|,|m_{3}|)$. The neutrino mixing
matrix is parametrized by $\mathbf{U}_{\nu }=\mathbf{U}_{\alpha }\mathbf{O}%
_{23}\mathbf{O}_{13}\mathbf{O}_{12}\mathbf{U}_{\beta }$. Explicitly, we have 
\begin{eqnarray}
\mathbf{U}_{\alpha } &=&%
\begin{pmatrix}
e^{i\alpha _{1}} & 0 & 0 \\ 
0 & e^{i\alpha _{2}} & 0 \\ 
0 & 0 & e^{i\alpha _{3}}%
\end{pmatrix}%
,\qquad \mathbf{U}_{\beta }=%
\begin{pmatrix}
1 & 0 & 0 \\ 
0 & e^{i\beta _{1}} & 0 \\ 
0 & 0 & e^{i\beta _{2}}%
\end{pmatrix}
\notag \\
\mathbf{O}_{23} &=&%
\begin{pmatrix}
1 & 0 & 0 \\ 
0 & \cos {\gamma }_{23} & \sin {\gamma }_{23} \\ 
0 & -\sin {\gamma }_{23} & \cos {\gamma }_{23}%
\end{pmatrix}%
,\quad \mathbf{O}_{13}=%
\begin{pmatrix}
\cos {\gamma }_{13} & 0 & \sin {\gamma }_{13}e^{-i\delta } \\ 
0 & 1 & 0 \\ 
-\sin {\gamma }_{13}e^{i\delta } & 0 & \cos {\gamma }_{13}%
\end{pmatrix}%
,\quad \mathbf{O}_{12}=%
\begin{pmatrix}
\cos {\gamma }_{12} & \sin {\gamma }_{12} & 0 \\ 
-\sin {\gamma }_{12} & \cos {\gamma }_{12} & 0 \\ 
0 & 0 & 1%
\end{pmatrix}%
.
\end{eqnarray}%
In the above matrices, $\alpha _{i}$ ($i=1,2,3$) are unphysical phases; $%
\beta _{j}$ ($j=1,2$) stands for the Majorana phases. In addition, there are
three angles and one phase that parameterize the rotations.

As one can verify, the $\alpha _{i}$ and $\beta _{j}$ phases are not
arbitrary since they can be fixed by inverting the expression, $\mathbf{U}%
_{\nu }^{\dagger }\mathbf{M}_{\nu }\mathbf{U}_{\nu }^{\ast }=\hat{\mathbf{M}}%
_{\nu }$ to obtain the effective mass matrix. This means explicitly, $%
\mathbf{M}_{\nu }=\mathbf{U}_{\nu }\hat{\mathbf{M}}_{\nu }\mathbf{U}_{\nu
}^{T}$, then, we obtain 
\begin{eqnarray}
A_{\nu } &=&\cos ^{2}{\gamma _{13}}\left( |m_{1}|\cos ^{2}{\gamma _{12}}%
+|m_{2}|\sin ^{2}{\gamma _{12}}\right) +|m_{3}|\sin ^{2}{\gamma _{13}}; 
\notag \\
B_{\nu } &=&\frac{\cos {\gamma _{13}}}{\sqrt{2}}\left[|m_{2}|\sin {\gamma _{12}}\left( \cos {\gamma _{12}}+i\sin {%
\gamma _{12}}\sin {\gamma _{13}}\right)-|m_{1}|\cos {%
\gamma _{12}}\left( \sin {\gamma _{12}}-i\cos {\gamma _{12}}\sin {\gamma
_{13}}\right)  -i|m_{3}|\sin {\gamma _{13}}\right] ;
\notag \\
\tilde{C}_{\nu } &=&\frac{1}{2}\left[ |m_{1}|\left( \sin {\gamma _{12}}%
+i\cos {\gamma _{12}}\sin {\gamma _{13}}\right) ^{2}+|m_{2}|\left( \cos {%
\gamma _{12}}-i\sin {\gamma _{12}}\sin {\gamma _{13}}\right)
^{2}-|m_{3}|\cos ^{2}{\gamma _{13}}\right]  \notag \\
D_{\nu } &=&\frac{1}{2}\left[ |m_{1}|\left( \sin ^{2}{\gamma _{12}}+\cos ^{2}%
{\gamma _{12}}\sin ^{2}{\gamma _{13}}\right) +|m_{2}|\left( \cos ^{2}{\gamma
_{12}}+\sin ^{2}{\gamma _{12}}\sin ^{2}{\gamma _{13}}\right) +|m_{3}|\cos
^{2}{\gamma _{13}}\right] .
\end{eqnarray}%
These matrix elements are obtained with $\alpha _{1}=\alpha _{2}=0$ and $%
\alpha _{3}=\pi $; $\beta _{1}=0$ and $\beta _{2}=\pi /2$. Along with these, 
$\gamma _{23}=\pi /4$ and $\delta =-\pi /2$.

Having given the above conditions, let us write explicitly the neutrino
mixing matrix 
\begin{eqnarray}
\mathbf{U}_{\nu}=%
\begin{pmatrix}
\cos{\gamma_{12}}\cos{\gamma_{13}} & \sin{\gamma_{12}}\cos{\gamma_{13}} & 
-\sin{\gamma_{13}} \\ 
-\frac{1}{\sqrt{2}}\left(\sin{\gamma_{12}}-i\cos{\gamma_{12}}\sin{\gamma_{13}}%
\right) & \frac{1}{\sqrt{2}}\left(\cos{\gamma_{12}}+i\sin{\gamma_{12}}\sin{%
\gamma_{13}}\right) & \frac{i\cos{\gamma_{13}}}{\sqrt{2}} \\ 
-\frac{1}{\sqrt{2}}\left(\sin{\gamma_{12}}+i\cos{\gamma_{12}}\sin{\gamma_{13}}%
\right) & \frac{1}{\sqrt{2}}\left(\cos{\gamma_{12}}-i\sin{\gamma_{12}}\sin{%
\gamma_{13}}\right) & -\frac{i\cos{\gamma_{13}}}{\sqrt{2}}%
\end{pmatrix}%
\end{eqnarray}

Finally, the PMNS mixing matrix is given by $\mathbf{U}=\mathbf{U}^{\dagger}_{l}\mathbf{U}_{\nu}=\mathbf{O}^{T}_{l}\mathbf{P}_{l}\mathbf{U}_{\nu}$. Consequently, the reactor, solar and atmospheric angles are give as follows
\begin{eqnarray}
\sin^{2}{\theta}_{13}&=& \big| \left(\mathbf{U}\right)_{13}\vert^{2}=\vert \left(\mathbf{O}_{l}\right)_{11}\left(\mathbf{U}_{\nu}\right)_{13}+\left(\mathbf{O}_{l}\right)_{21}\left(\mathbf{U}_{\nu}\right)_{23}e^{-i\eta_{\mu}}+\left(\mathbf{O}_{l}\right)_{31}\left(\mathbf{U}_{\nu}\right)_{33}e^{-i\eta_{\tau}}\big|^{2};\nn\\
\sin^{2}{\theta}_{12}&=&\frac{\big| \left(\mathbf{U}\right)_{12}\vert^{2}}{1-\vert \left(\mathbf{U}\right)_{13}\big|^{2}}=\frac{\vert \left(\mathbf{O}_{l}\right)_{11}\left(\mathbf{U}_{\nu}\right)_{12}+\left(\mathbf{O}_{l}\right)_{21}\left(\mathbf{U}_{\nu}\right)_{22}e^{-i\eta_{\mu}}+\left(\mathbf{O}_{l}\right)_{31}\left(\mathbf{U}_{\nu}\right)_{32}e^{-i\eta_{\tau}}\vert^{2}}{1-\vert \left(\mathbf{U}\right)_{13}\vert^{2}};\nn\\
\sin^{2}{\theta}_{23}&=&\frac{\big| \left(\mathbf{U}\right)_{12}\vert^{2}}{1-\vert \left(\mathbf{U}\right)_{13}\vert^{2}}=\frac{\vert \left(\mathbf{O}_{l}\right)_{12}\left(\mathbf{U}_{\nu}\right)_{13}+\left(\mathbf{O}_{l}\right)_{22}\left(\mathbf{U}_{\nu}\right)_{23}e^{-i\eta_{\mu}}+\left(\mathbf{O}_{l}\right)_{32}\left(\mathbf{U}_{\nu}\right)_{33}e^{-i\eta_{\tau}}\vert^{2}}{1-\vert \left(\mathbf{U}\right)_{13}\big|^{2}}.
\end{eqnarray}
Notice that there are still  free parameters in the PMNS matrix, these are $\gamma_{12}$, $\gamma_{13}$ and $\vert a_{l} \vert$. In addition to those, two phases $\eta_{\mu}$ and $\eta_{\tau}$.

\section{Scalar potential}\label{ap:scalar}
After the spontaneous breaking of the $Q_6$ discrete symmetry, the scalar potential takes the form:
\begin{eqnarray}\label{eq:potencial}
V &=& 
-\mu_2^2 \left(H_2^{\dagger}H_2\right) -\mu_3^2 \left(H_3^{\dagger}H_3\right) -\mu_{13}^2 \left(H_3^{\dagger}H_1+ H_1^{\dagger}H_3\right)-\mu_{23}^2\left(H_2^{\dagger}H_3 + H_3^{\dagger}H_2\right)-\mu_{12}^2 \left(H_1^{\dagger}H_2+ H_2^{\dagger}H_1\right) -\mu_4^2 \left(\sigma^*\sigma\right) \notag\\
&&-\mu_5^2 \left(\xi_1^*\xi_1\right)-\mu_6^2 \left(\xi_2^*\xi_2\right) -\mu_7^2 \left(\rho^*\rho\right) +\lambda_1 \left(H_1^{\dagger}H_2-H_2^{\dagger}H_1\right)^2 +\lambda_2 \left(H_3^{\dagger}H_3\right)^2 +\lambda_3 \left(\sigma^*\sigma\right)^2 +\lambda_4 \left(\xi_1^*\xi_2-\xi_2^*\xi_1\right)^2 \notag\\
&&+\lambda_5 \left(\rho^*\rho\right)^2 + \lambda_6 \left(H_1^{\dagger}H_2-H_2^{\dagger}H_1\right)\left(H_3^{\dagger}H_3\right) + \lambda_7 \left(H_1^{\dagger}H_2-H_2^{\dagger}H_1\right)\left(\sigma^*\sigma\right)+ \lambda_8 \left(H_1^{\dagger}H_2-H_2^{\dagger}H_1\right)\left(\rho^*\rho\right) \notag\\
&&+ \lambda_9 \left(\xi_1^*\xi_2-\xi_2^*\xi_1\right)\left(H_3^{\dagger}H_3\right) + \lambda_{10} \left(\xi_1^*\xi_2-\xi_2^*\xi_1\right)\left(\sigma^*\sigma\right)+ \lambda_{11} \left(\xi_1^*\xi_2-\xi_2^*\xi_1\right)\left(\rho^*\rho\right) \\
&&+\lambda_{12} \left(H_1^{\dagger}H_2-H_2^{\dagger}H_1\right)\left(\xi_1^*\xi_2-\xi_2^*\xi_1\right)
+\lambda_{13}\left(H_1^{\dagger}H_2+H_2^{\dagger}H_1\right)\left(\xi_1^*\xi_2+\xi_2^*\xi_1\right)+\lambda_{14}\left(H_1^{\dagger}H_2+H_2^{\dagger}H_1\right)\left(\sigma^*\rho\right)\notag\\
&&+\lambda_{15}\left(\xi_1^*\xi_2+\xi_2^*\xi_1\right)\left(\sigma^*\rho\right)
+\lambda_{16} \left(H_3^{\dagger}H_3\right)\left(\sigma^*\sigma\right) +\lambda_{17} \left(H_3^{\dagger}H_3\right)\left(\rho^*\rho\right)+\lambda_{18} \left(\sigma^*\sigma\right)\left(\rho^*\rho\right)+h.c.\notag
\end{eqnarray}
where $\lambda_6=\lambda_7=\lambda_8=\lambda_{12}=0$ as required by CP conservation.

The scalar mass matrices of the CP-even neutral, CP-odd neutral and electrically charged fields, considering the VEV of section \ref{quarmassesandmixings}, are given by:
\begin{align}
\mathbf{M}_{CP-\text{even}}^{2}&=\begin{pmatrix}
A_{3\times 3} & B_{3\times 4} \\
B_{4\times 3}^T & C_{4\times 4}
\end{pmatrix},\qquad \mathbf{M}_{CP-\text{odd}}^{2}=\begin{pmatrix}
X_{3\times 3} & 0_{3\times 4} \\
0_{4\times 3} & Y_{4\times 4}
\end{pmatrix},\\[5pt]
\mathbf{M}_{\text{charged}}^{2}&=
\begin{pmatrix}
-\mu_1^2 & \frac{\mu _{13}^2 v_3}{v_2} & -\mu _{13}^2 \\
 \frac{\mu _{13}^2 v_3}{v_2} & \frac{\mu _{23}^2 v_3}{v_2}
   & -\mu _{23}^2 \\
 -\mu _{13}^2 & -\mu _{23}^2 & \frac{\mu _{23}^2 v_2}{v_3}
\end{pmatrix},
\end{align}
where:
\begin{align}
A_{3\times 3}&= \begin{pmatrix}
-\mu_1^2 & \frac{\mu _{13}^2 v_3}{v_2} & -\mu _{13}^2 \\
 \frac{\mu _{13}^2 v_3}{v_2} & \frac{\mu _{23}^2 v_3}{v_2}
   & -\mu _{23}^2 \\
 -\mu _{13}^2 & -\mu _{23}^2 & 2 \lambda _2
   v_3^2+\frac{\mu _{23}^2 v_2}{v_3}
\end{pmatrix}, \notag\\
B_{3\times 4}&= \begin{pmatrix}
\lambda _{13} v_2 \cos (\theta ) v_{\xi } & \lambda _{13}
   v_2 \cos (\theta ) v_{\xi } & \frac{1}{2} \lambda _{14}
   v_2 v_{\rho } & \frac{1}{2} \lambda _{14} v_2 v_{\sigma
   } \\
 0 & 0 & 0 & 0 \\
 -i \lambda _9 v_3 \sin (\theta ) v_{\xi } & -i \lambda _9
   v_3 \sin (\theta ) v_{\xi } & \lambda _{16} v_3
   v_{\sigma } & 0
\end{pmatrix}, \\
C_{4\times 4}&=
\frac{1}{2}
\scalebox{0.75}{$
\begin{pmatrix}
-4 \lambda _4 \sin ^2(\theta ) v_{\xi }^2-\lambda _{15}
   \sec (2 \theta ) v_{\rho } v_{\sigma } & \lambda _{15}
   v_{\rho } v_{\sigma }-4 \lambda _4 \sin ^2(\theta )
   v_{\xi }^2 & v_{\xi } \left(\lambda _{15} \cos (\theta
   ) v_{\rho }-2 i \lambda _{10} \sin (\theta ) v_{\sigma
   }\right) & v_{\xi } \left(\lambda _{15} \cos (\theta )
   v_{\sigma }-2 i \lambda _{11} \sin (\theta ) v_{\rho
   }\right) \\
 \lambda _{15} v_{\rho } v_{\sigma }-4 \lambda _4 \sin
   ^2(\theta ) v_{\xi }^2 & -4 \lambda _4 \sin ^2(\theta )
   v_{\xi }^2-\lambda _{15} \sec (2 \theta ) v_{\rho }
   v_{\sigma } & v_{\xi } \left(\lambda _{15} \cos (\theta
   ) v_{\rho }-2 i \lambda _{10} \sin (\theta ) v_{\sigma
   }\right) & v_{\xi } \left(\lambda _{15} \cos (\theta )
   v_{\sigma }-2 i \lambda _{11} \sin (\theta ) v_{\rho
   }\right) \\
 v_{\xi } \left(\lambda _{15} \cos (\theta ) v_{\rho }-2 i
   \lambda _{10} \sin (\theta ) v_{\sigma }\right) &
   v_{\xi } \left(\lambda _{15} \cos (\theta ) v_{\rho }-2
   i \lambda _{10} \sin (\theta ) v_{\sigma }\right) & 4
   \lambda _3 v_{\sigma }^2-\frac{\lambda _{15} \cos (2
   \theta ) v_{\xi }^2 v_{\rho }}{v_{\sigma }} & \lambda
   _{15} \cos (2 \theta ) v_{\xi }^2+2 \lambda _{18}
   v_{\rho } v_{\sigma } \\
 v_{\xi } \left(\lambda _{15} \cos (\theta ) v_{\sigma }-2
   i \lambda _{11} \sin (\theta ) v_{\rho }\right) &
   v_{\xi } \left(\lambda _{15} \cos (\theta ) v_{\sigma
   }-2 i \lambda _{11} \sin (\theta ) v_{\rho }\right) &
   \lambda _{15} \cos (2 \theta ) v_{\xi }^2+2 \lambda
   _{18} v_{\rho } v_{\sigma } & 4 \lambda _5 v_{\rho
   }^2-\frac{\lambda _{15} \cos (2 \theta ) v_{\xi }^2
   v_{\sigma }}{v_{\rho }}
\end{pmatrix}\notag
$}.\\
X_{3\times 3}&=\begin{pmatrix}
-2 \lambda _1 v_2^2-\mu_1^2 & \frac{\mu _{13}^2 v_3}{v_2} & -\mu
   _{13}^2 \\
 \frac{\mu _{13}^2 v_3}{v_2} & \frac{\mu _{23}^2 v_3}{v_2}
   & -\mu _{23}^2 \\
 -\mu _{13}^2 & -\mu _{23}^2 & \frac{\mu _{23}^2 v_2}{v_3}
\end{pmatrix}, \\
Y_{4\times 4}&= \frac{1}{4}
\scalebox{0.85}{$
\begin{pmatrix}
-2 \left(4 \lambda _4 \cos ^2(\theta ) v_{\xi }^2+\lambda
   _{15} \sec (2 \theta ) v_{\rho } v_{\sigma }\right) & 8
   \lambda _4 \cos ^2(\theta ) v_{\xi }^2+2 \lambda _{15}
   v_{\rho } v_{\sigma } & 2 i \lambda _{15} \sin (\theta
   ) v_{\xi } v_{\rho } & -2 i \lambda _{15} \sin (\theta
   ) v_{\xi } v_{\sigma } \\
 8 \lambda _4 \cos ^2(\theta ) v_{\xi }^2+2 \lambda _{15}
   v_{\rho } v_{\sigma } & -2 \left(4 \lambda _4 \cos
   ^2(\theta ) v_{\xi }^2+\lambda _{15} \sec (2 \theta )
   v_{\rho } v_{\sigma }\right) & -2 i \lambda _{15} \sin
   (\theta ) v_{\xi } v_{\rho } & 2 i \lambda _{15} \sin
   (\theta ) v_{\xi } v_{\sigma } \\
 2 i \lambda _{15} \sin (\theta ) v_{\xi } v_{\rho } & -2
   i \lambda _{15} \sin (\theta ) v_{\xi } v_{\rho } &
   -\frac{2 \lambda _{15} \cos (2 \theta ) v_{\xi }^2
   v_{\rho }}{v_{\sigma }} & 2 \lambda _{15} \cos (2
   \theta ) v_{\xi }^2 \\
 -2 i \lambda _{15} \sin (\theta ) v_{\xi } v_{\sigma } &
   2 i \lambda _{15} \sin (\theta ) v_{\xi } v_{\sigma } &
   2 \lambda _{15} \cos (2 \theta ) v_{\xi }^2 & -\frac{2
   \lambda _{15} \cos (2 \theta ) v_{\xi }^2 v_{\sigma
   }}{v_{\rho }}
\end{pmatrix}
$}.\notag
\end{align}
%
%
%
%
%
%
%
%
%
%

\bibliographystyle{utphys}
\bibliography{BiblioQ6.bib}

\end{document}